\documentclass[fleqn,usenatbib]{mnras}

\usepackage{newtxtext,newtxmath}

\usepackage[T1]{fontenc}

\DeclareRobustCommand{\VAN}[3]{#2}
\let\VANthebibliography\thebibliography
\def\thebibliography{\DeclareRobustCommand{\VAN}[3]{##3}\VANthebibliography}

\usepackage{graphicx}	
\usepackage{amsmath}	
\usepackage{subcaption}
\usepackage{svg}
\usepackage{graphicx}
\usepackage{multirow} 
\usepackage{caption}
\setcitestyle{notesep={ }}
\usepackage{caption}
\DeclareCaptionFont{myfont}{\fontsize{14pt}{11pt}\selectfont}
\usepackage{cprotect}
\usepackage{algorithm}
\usepackage{algpseudocode}
\usepackage{tablefootnote}
\usepackage{comment}
\usepackage{anyfontsize}

\title[Merger tree accuracy and consequences]{On the accuracy of dark matter halo merger trees and the consequences for semi-analytic models of galaxy formation}

\author[\'Angel Chandro-G\'omez et al.]{
\parbox[t]{\textwidth}{
\vspace{-0.5cm}
\'Angel Chandro-G\'omez$^{1,2}$\thanks{E-mail: angel.chandrogomez@research.uwa.edu.au}, Claudia del P. Lagos$^{1,2}$, Chris Power$^{1,2}$, Victor J. Forouhar Moreno$^{3}$, John C. Helly$^{4}$, Cedric G. Lacey$^{4}$, Robert J. McGibbon$^{3}$, Matthieu Schaller$^{5,3}$, Joop Schaye$^{3}$}
\vspace*{6pt} \\
$^{1}$ International Centre for Radio Astronomy Research, The University of Western Australia, 35 Stirling Highway, Crawley, Western Australia 6009, Australia\\
$^{2}$ ARC Centre for All-Sky Astrophysics in 3 Dimensions (ASTRO 3D)\\
$^{3}$ Leiden Observatory, Leiden University, PO Box 9513, 2300 RA Leiden, The Netherlands\\
$^{4}$ Institute for Computational Cosmology, Department of Physics, University of Durham, South Road, Durham, DH1 3LE, UK\\
$^{5}$ Lorentz Institute for Theoretical Physics, Leiden University, PO Box 9506, NL-2300 RA Leiden, The Netherlands\\
\vspace*{-0.5cm}}

\date{Accepted XXX. Received YYY; in original form ZZZ}

\pubyear{2025}

\begin{document}
\label{firstpage}
\pagerange{\pageref{firstpage}--\pageref{lastpage}}
\maketitle

\begin{abstract}
Galaxy formation and evolution models, such as semi-analytic models, are powerful theoretical tools for predicting how galaxies evolve across cosmic time. These models follow the evolution of galaxies based on the halo assembly histories inferred from large $N$-body cosmological simulations. This process requires codes to identify haloes (\textit{``halo finder''}) and to track their time evolution (\textit{``tree builder''}). While these codes generally perform well, they encounter numerical issues when handling dense environments. In this paper, we present how relevant these issues are in state-of-the-art cosmological simulations. We characterise two major numerical artefacts in halo assembly histories: (i) the non-physical swapping of large amounts of mass between subhaloes, and (ii) the sudden formation of already massive subhaloes at late cosmic times. We quantify these artefacts for different combinations of halo finder ({\sc Subfind}, {\sc VELOCIraptor}, {\sc HBT-HERONS}) and tree builder codes ({\sc D-Trees}+{\sc DHalo}, {\sc TreeFrog}, {\sc HBT-HERONS}), finding that in general more than $50\%$ ($80\%$) of the more massive subhaloes with $>10^{3}$ ($>10^{4}$) particles at $z=0$ inherit them in most cases. However, {\sc HBT-HERONS}, which explicitly incorporates temporal information, effectively reduces the occurrence of these artefacts to $5\%$ ($10\%$). We then use the semi-analytic models {\sc Shark} and {\sc Galform} to explore how these artefacts impact galaxy formation predictions. We demonstrate that the issues above lead to non-physical predictions in galaxies hosted by affected haloes, particularly in {\sc Shark} where the modelling of baryons relies on subhalo information. Finally, we propose and implement fixes for the numerical artefacts at the semi-analytic model level, and use {\sc Shark} to show the improvements, especially at the high-mass end, after applying them.

\end{abstract}




\section{Introduction}
\label{sec:intro} 

In the $\Lambda$ Cold Dark Matter ($\Lambda$CDM) cosmological model, structures in the universe form hierarchically, following a ``bottom-up'' scenario. Small matter over-densities merge over time with each other, giving rise to larger structures \citep{lacey93}. This process applies primarily to the dominant dark-matter (DM) density field whose cold, collisionless nature is governed solely by gravity. However, baryons behave differently as they can dissipate energy through radiative processes. Despite this, DM plays a crucial role in how baryons evolve as the evolution of DM gravitationally-bound structures, or \textit{``haloes''}, directly influences the formation and evolution of galaxies \citep{white91}. Galaxies form within these haloes and their hierarchical merging scenario also leads to galaxy mergers.

Large-volume cosmological simulations can model the formation of hierarchical structures. These simulations numerically trace the evolution of the DM density field (in $N$-body/DM-only simulations) or jointly with baryons (in hydrodynamical simulations) by solving both the fluid and gravity equations \citep[see][for a review]{somerville15}. State-of-the-art simulations such as {\sc FLAMINGO} \citep{flamingo}, {\sc IllustrisTNG} \citep{tng_b}, {\sc EAGLE} \citep{eagle}, {\sc Magneticum} \citep{magneticum_b} or {\sc Horizon-AGN} \citep{horizonagn_b} offer powerful tools for making theoretical predictions for galaxies across cosmic time. They can generate mock catalogues that emulate galaxy surveys, which can be compared directly to observations. Moreover, they allow testing different subgrid physics prescriptions for baryonic processes that remain not completely understood, such as feedback mechanisms, star formation or gas cooling. 

Despite their success, cosmological (either $N$-body or hydrodynamical) simulations simply model the evolution of DM (or in addition gas/stellar) particles or cells, which are tracers of the underlying density fields. To enhance their predictive power and enable direct comparison with the observable Universe, raw simulation data need to be processed to track the evolution of gravitationally bound structures, both haloes and galaxies. Post-processing analysis tools, such as group-finding algorithms, termed as \textit{``halo/subhalo finder''} codes, identify gravitationally bound particles, thus detecting haloes and substructures/galaxies within them at different epochs, as well as allowing for their properties to be measured. Additionally, tree-building algorithms, known as \textit{``merger tree builder''} codes, link these structures across time, constructing their individual evolutionary histories. A huge diversity of these post-processing codes is available, based on different methodologies (see \citealp{knebe11,knebe13b,knebe13a,onions12} for halo/subhalo-finding and \citealp{srisawat13} for tree-building tools).

The use of halo/subhalo finder and tree builder codes is essential from the point of view of galaxy formation and evolution since their predictions depend on them. Every hydrodynamical simulation requires these post-processing tools to evaluate the individual growth of galaxies and haloes/subhaloes in them — e.g. codes such as {\sc VELOCIraptor} \citep{velociraptor}, {\sc D-Trees}+{\sc DHalo} \citep{dhalos} or {\sc HBT-HERONS} \citep{hbtp,hbt-herons} have been used in {\sc FLAMINGO}; {\sc Subfind} \citep{subfind,subfind_b} and {\sc SubLink} \citep{sublink} in {\sc IllustrisTNG}, {\sc EAGLE} and {\sc Magneticum} \citep{dolag23} — there is {\sc LHaloTree} \citep{lhalotree} data for {\sc IllustrisTNG} as well —; or {\sc AdaptaHOP} \citep{adaptahop} in {\sc Horizon-AGN}. Plus certain subgrid physics recipes in hydrodynamical schemes rely on halo properties. For instance, supermassive black holes (SMBHs) are typically seeded at the centre of haloes \citep{dimatteo08}.

Beyond hydrodynamical simulations (which are the most computationally expensive way of predicting galaxy populations), other approaches model galaxies on top of DM-only runs using as a backbone the output from the halo/subhalo finder or merger tree builder codes. A detailed description of these models can be found in \citet{wechsler18}, which vary in their assumptions, predictive power and computational cost. Models relying on less physical assumptions require the halo/subhalo-finding step to match halo properties to galaxy properties. Halo occupation distribution (HOD) models, based on a probability distribution conditioned on the halo mass, populate them with a certain number of galaxies meeting some stellar mass or luminosity criteria \citep{benson00}; abundance matching models directly connect stellar masses or luminosities of galaxies to haloes by matching the abundance of haloes with the abundance of galaxies of a certain mass or internal velocity \citep{wechsler98}; or models that parametrise the stellar-to-halo mass relation as well \citep{moster10}. Halo mass accretion histories from the tree-building step are also crucial for empirical models that constrain the galaxy-halo connection at different epochs \citep{behroozi19} and for the most physical and more computationally expensive semi-analytic models (SAMs), which solve baryonic prescriptions based on individual halo growth histories \citep{white91}. 

All galaxy formation models — from the most physical hydrodynamical simulations to simpler empirical models —  require the use of these simulation post-processing tools. Moreover, these models are complementary to one another: physical models inform assumptions used in more empirical techniques, while empirical models can help constrain parameters in more physical approaches. Accurate halo-finding and tree-building algorithms are therefore critical for producing reliable predictions across all types of galaxy formation models. Consequently, new methods are continually being developed \citep{poole17,canas18,bahe19,bose22,mansfield23}.

While the numerous available halo/subhalo finder and merger tree builder codes usually work well, difficulties arise in specific circumstances, as has been widely reported in workshops such as the ``Haloes going MAD'' \citep{knebe11,knebe13b}, ``Subhaloes Going Notts'' \citep{onions12,onions13,elahi13,hoffmann14,behroozi15} and ``Sussing Merger Trees'' \citep{srisawat13,avila14,lee14}. Common challenges include tracing haloes with few particles \citep{onions12}, tracking substructures inside haloes in very dense regions \citep{muldrew10,vandenbosch18,mansfield23}, dealing with tidal streams and shells arising from substructures being stripped \citep{elahi13}, tracing structures through major mergers \citep{behroozi15} or defining halo boundaries \citep{diemer21}. The academic studies from these workshops analyse the performance of multiple algorithms and provide recommendations to improve them. However, despite identifying potential numerical issues, there is still a lack of systematic quantification of how different codes perform and how their limitations affect predictions from galaxy formation models \citep{gomez21}. As a result, this paper aims to address these gaps by characterizing and quantifying the most relevant numerical issues in merger tree post-processed data and evaluating their impact on models of galaxy formation. 

The analysis focuses particularly on SAM predictions, as these models offer a powerful and efficient tool for large-volume studies while testing subgrid prescriptions for different baryonic processes \citep[see a review in][]{baugh06,benson10,somerville15}. SAMs have been used to study a range of galaxy populations at different cosmic times for the dual purpose of making predictions about observable properties of galaxies and to help design galaxy surveys. For instance, they have been used to infer cosmological parameters in the local universe \citep[e.g.][]{2dfgrs,sdss}, to predict the atomic hydrogen (HI) properties of galaxies supporting surveys with the Square Kilometre Array (SKA) and its pathfinders \citep{ska}, and to help understand how well halo masses can be derived from spectroscopic surveys \citep{robotham11}. Additionally, given that SAMs depend on the evolution of DM haloes, they provide an ideal framework for checking how numerical artefacts in both halo finder and tree builder codes propagate into such galaxy formation models. While some studies have explored how different post-processing codes impact SAMs, their findings are contradictory. For example, some conclude that modelled galaxy properties remain consistent despite variations in the underlying merger tree catalogues \citep{gomez21}, whereas others highlight substantial discrepancies \citep{lee14}. Nevertheless, there is little explanation of how the numerical artefacts that post-processing tools produce propagate to the predicted galaxy properties. This paper analyses this impact and develops strategies to improve the reliability of SAM predictions, focusing on the robustness of theoretical models rather than merely reproducing observables.

We begin in \S~\ref{sec:methods} with a description of the simulation data used for this analysis, giving details on the halo-finding and tree-building algorithms employed. In \S~\ref{sec:merger-trees-artefacts} we categorise two key numerical problems and quantify their frequency as a function of halo mass and cosmic time, while we examine their impact on SAMs. In \S~\ref{sec:fix-galaxy-formation-models} we propose solutions to minimise the impact of numerical problems and produce more reliable predictions, understanding how these predictions change. We finally summarise our results in \S~\ref{sec:conclusions}.

\section{Merger tree data: algorithm description}
\label{sec:methods} 


For this analysis, we use four state-of-the-art large-volume cosmological simulations: (i) the fiducial DM-only run ``L1\_m8\_DMO'' and (ii) the fiducial hydrodynamical run ``L1\_m8'' both from the {\sc FLAMINGO} project \citep{flamingo,flamingo_b}; as well as (iii) {\sc PMill} \citep{pmill} and (iv) {\sc medi-SURFS} \citep{surfs}. Each simulation has slightly different cosmological parameters: (i)--(ii) $h=0.681$, $\rm \Omega_{m}=0.306$, $\rm \Omega_{\Lambda}=0.694$ and $\rm \Omega_{b}=0.0486$; (iii) $h=0.6777$, $\rm \Omega_{m}=0.307$, $\rm \Omega_{\Lambda}=0.693$ and $\rm \Omega_{b}=0.04825$; and (iv) $h=0.6751$, $\rm \Omega_{m}=0.3121$, $\rm \Omega_{\Lambda}=0.6879$ and $\rm \Omega_{b}=0.0488$. These variations involve a mass difference of less than $1$\% driven by the $h$ differences, which is negligible for our purposes. Since we focus on relative mass changes and particle counts in each simulation, the cosmological parameters do not play a significant role in this study and are henceforth ignored. Other relevant parameters, such as the simulation box size, the DM particle mass or the number of snapshots are detailed in Table~\ref{tab:sim}. The simulations also employ different gravity and hydrodynamics solver codes — {\sc SWIFT} \citep{swift} for (i)--(ii) and {\sc GADGET-2} \citep{gadget2} for (iii)--(iv) — but these differences do not matter for our results.


Ultimately, we evaluate six merger tree catalogues, since the {\sc FLAMINGO} runs have been processed twice using different sets of halo/subhalo-finding and merger tree-building codes. The key features for all the catalogues are presented in Table~\ref{tab:cat}. The main difference for our purposes is that the catalogues are post-processed using various halo/subhalo finder and tree builder codes. This approach allows us to study the frequency of potential numerical artefacts and any dependence on the code employed. In addition, we use two SAMs to explore how these numerical artefacts impact models that apply different descriptions of baryon physics.

\begin{table}
	\centering
	\caption{Simulations used. $L$: the periodic box size in comoving Mpc; $N_{\rm part}$: the number of DM particles in the volume; $m_{\rm part}$: the DM particle mass in M$_{\odot}$; $N_{\rm out}$: the number of simulation outputs/snapshots; and the type of simulation (DM-only or hydro).}
	\label{tab:sim}
	\begin{tabular}{ccccccc}
		\hline
		simulation & $L$/Mpc & $N_{\rm part}$ & $m_{\rm part}$/M$_{\odot}$ & $N_{\rm out}$ & type\\
		\hline
        {\sc FLAM-DM} & 1000 & 3600$^3$ & 8.40$\times$10$^8$ & 79 & DM-only\\
        {\sc FLAM-Hydro} & 1000 & 3600$^3$ & 7.06$\times$10$^8$ & 79 & hydro\\
        {\sc PMill} & 800 & 5040$^3$ & 1.56$\times$10$^8$ & 271 & DM-only\\
        {\sc medi-SURFS} & 311.07 & 1536$^3$ & 3.27$\times$10$^8$ & 200 & DM-only\\
		\hline
	\end{tabular}
\end{table}

\begin{table*}
	\centering
	\caption{Merger tree catalogues used. The halo finder code; the merger tree builder code; the halo finder approach; the merger tree builder approach; and $N^{\rm min}_{\rm part,subhalo}$: the minimum particle number for a subhalo to be detected in the simulation.}
	\label{tab:cat}
    \begin{tabular}{cccccc}
		\hline
		catalogue & halo finder & tree builder & halo finder approach & tree builder approach & $N^{\rm min}_{\rm part,subhalo}$\\
		\hline
        {\sc FLAM-DM-VR} & {\sc VELOCIraptor} & {\sc D-Trees}+{\sc DHalo} & \textit{phase-space} & \textit{adjacent snapshots} & 20\\
        {\sc FLAM-Hydro-VR} & {\sc VELOCIraptor} & {\sc D-Trees}+{\sc DHalo} & \textit{phase-space} & \textit{adjacent snapshots} & 20\footnotemark[1]\\
        {\sc FLAM-DM-HBT} & {\sc HBT-HERONS} & {\sc HBT-HERONS} & \textit{history-space} & \textit{history-space} & 20\\
        {\sc FLAM-Hydro-HBT} & {\sc HBT-HERONS} & {\sc HBT-HERONS} & \textit{history-space} & \textit{history-space} & 20\footnotemark[1]\\
        {\sc PMill} & {\sc Subfind} & {\sc D-Trees}+{\sc DHalo} & \textit{configuration-space} & \textit{adjacent snapshots} & 15\\
        {\sc medi-SURFS} & {\sc VELOCIraptor} & {\sc TreeFrog}(+{\sc DHalo}) & \textit{phase-space} & \textit{adjacent snapshots} & 20\\
		\hline
	\end{tabular}
\end{table*}

Below we briefly summarise the halo finder and merger tree builder codes used in these catalogues in \S~\ref{ssec:halo-finding} and \S~\ref{ssec:tree-finding}, respectively, while the SAMs employed for the analysis are discussed in \S~\ref{ssec:sam}.


\subsection{Halo/subhalo-finding}
\label{ssec:halo-finding}
Halo/subhalo finders are designed to identify large, isolated over-densities in cosmological simulations, classifying them as \textit{``host/main haloes''}. These large over-densities are typically identified using a Friends-Of-Friends (FOF) algorithm, which groups particles that are spatially close (3D FOF) or close in phase-space (6D FOF). The process involves defining a linking length — a threshold (spatial or phase-space) distance — below which particles are linked together. By applying this criterion to all particles in the simulation, groups of clustered particles, named FOF groups, are formed based on the chosen distance metric.

Within the host haloes identified by the FOF algorithm, there can be smaller over-densities referred to as \textit{``subhaloes''}. Various methods exist to detect these self-bound groups of particles. Regardless of the method, there is always a dominant subhalo determined by a finder-specific metric, such as the most massive subhalo within a host halo. The dominant structure is termed \textit{``central subhalo''}; while the remaining smaller over-densities are called \textit{``satellite subhaloes''}. A host halo can consist of a single subhalo (the central subhalo) or multiple subhaloes (a central subhalo with its satellites). These definitions are visually illustrated in Fig.~\ref{fig:mb-diagram}. Below we summarise the finder codes used in the analysed catalogues, each based on a distinct approach to identify DM (sub)haloes across simulation snapshots. 

\subsubsection{{\sc Subfind} (configuration-space)}
\label{sssec:subfind}
{\sc Subfind} \citep{subfind} is a configuration-based finder. It starts identifying host haloes using a 3D FOF algorithm with a standard linking length of $b=0.2$ times the mean interparticle distance, without incorporating velocity information. Within each host halo, subhaloes are detected as local over-densities, defined as self-bound groups of particles. To identify density peaks (subhalo candidates) within the FOF group, the code calculates local densities by applying an adaptive kernel interpolation to the $N_{\rm dens}$ nearest neighbours for each particle in the group.

Starting with the densest particle and proceeding in decreasing order, the code selects its $N_{\rm ngb}$ nearest particles and isolates the subset with higher density estimates. The algorithm then examines up to the two closest neighbouring particles from this subset, which may include none, one or two particles: (i) if such particles are not found, a local density maximum is identified, forming a new subhalo; (ii) if one or two particles belonging to the same subhalo are found, they are added to that existing subhalo; and (iii) if two particles belonging to different subhaloes are found, these subhaloes are marked as subhalo candidates, as a saddle point between two separate over-dense regions is located. Once subhalo candidates are identified, an unbinding procedure is subsequently performed. Particles with positive total energy are removed iteratively, defining the subhalo's centre as the most bound particle and its velocity centre as the centre-of-mass velocity of the particles in the group. 

{\sc Subfind} assigns particles to the smaller over-densities first, ensuring each particle is associated with only one subhalo. This approach establishes an exclusive mass definition. Finally, the code rechecks the self-bound nature of the disjoint subhaloes, defining them as satellites. Particles belonging to the background of the FOF group are also subjected to the unbinding procedure, forming the largest subhalo in the group or the central subhalo. Resolved subhaloes are retained if their particle number exceeds a predefined threshold $N_{\rm ngb}$.

\subsubsection{{\sc VELOCIraptor} (phase-space)}
\label{sssec:velociraptor}
{\sc VELOCIraptor} \citep{velociraptor} is a phase-space finder. It identifies host haloes using either a 3D or 6D FOF approach, taking into account a particle's position and velocity information. In this study, the code was applied with a simple 3D FOF approach (with the standard linking length of $b=0.2$ times the mean interparticle distance) for the {\sc FLAMINGO}-related runs. For {\sc medi-SURFS}, an additional 6D FOF algorithm was employed (with a velocity-space linking length proportional to the velocity dispersion of the 3D FOF groups), which helps separate structures that might be incorrectly linked by spurious particle bridges.

After host haloes are identified, {\sc VELOCIraptor} locates substructures within them based on phase-space information. This method leverages the distinct local velocity distributions of substructure particles compared to the background host halo. For each particle in a FOF group, following the methodology of the original STructure Finder ({\sc STF}) paper \citep{stf}, its local velocity density is estimated based on its $N_{\rm v}$ nearest neighbours in velocity space, which are selected from its $N_{\rm se}$ nearest spatial neighbours ($N_{\rm v}\leq N_{\rm se}$). The background velocity density, on the other hand, is calculated by dividing the group into cells, where the number of particles per cell, $N_{\rm cell}$, is chosen to strike a balance between statistical reliability and minimal density variations between cells. Each particle's background velocity density is then interpolated using the cell it occupies and its six neighbouring cells. Particles whose local velocity density exceeds the background density by a predefined threshold are grouped into dynamically distinct substructures if they satisfy a secondary 3D FOF criterion and also share similar velocities. 
An iterative unbinding procedure removes unbound particles from each substructure. The remaining bound particles are assigned exclusively to a single substructure, which is retained only if it meets a minimum particle count threshold. 

Moreover, remnants of mergers or ``cores'' are identified afterwards within the remaining background particles using a conservative 6D FOF, in which the linking length is progressively reduced to again isolate over-densities distinct from the background distribution. These over-densities are reconstructed as satellite subhaloes as well after 
performing an unbinding procedure. Although {\sc VELOCIraptor} is capable of identifying tidal streams, this feature is not used in this analysis. 

After all these steps, the remaining self-bound background over-density is defined as the central subhalo, while substructures identified through the velocity distribution and merger remnants are classified as satellite subhaloes. For hydrodynamical simulations, the initial FOF links that determine host haloes are based solely on DM particles, with baryons subsequently assigned to the FOF group and subhalo corresponding to the closest DM particle in phase space \citep[see][for an application to the {\sc Horizon-AGN} hydrodynamical simulations]{canas18,canas20}. 

\subsubsection{{\sc HBT-HERONS} (history-space)}
\label{sssec:hbt}
{\sc HBT-HERONS} \citep{hbt-herons}  differs from both configuration- and phase-space-based codes, employing a unique methodology, which can be described as a history-space finder. The code is an improved version of {\sc HBT+} \citep{hbtp}, featuring bug fixes and enhancements in substructure identification, self-consistent merger tree building and computational efficiency. It begins with a 3D FOF catalogue using a linking length of $b=0.2$ times the mean interparticle distance. Subhaloes are then identified by cross-matching particle memberships as detailed in \citet{hbt}, based on the principle that every satellite subhalo was once the only central subhalo hosted by its halo. This approach effectively separates heavily overlapping subhaloes. For hydrodynamical runs, the FOF groups are identified using merely DM particles, with baryons consequently linked to them.

The core of the cross-matching algorithm involves assigning a set of particles, referred to as the ``source'', to every subhalo. The source is then tracked across snapshots to detect descendant subhaloes. Self-bound subhaloes with $N_{\rm bound}$ particles are subsequently identified by applying an unbinding procedure to the source. To avoid instabilities, the code limits the number of particles in the source ($N_{\rm source}\leq 3N_{\rm bound}$), updating it after unbinding.

Starting from the earliest snapshot, the code locates isolated FOF groups in the simulation. These groups determine the source, with its self-bound particles defining a newly created central subhalo. As the simulation progresses, a weighted estimate of the 10 most bound tracer particles from the source is tracked forward to identify which host halo the descendant structure belongs to. In hydrodynamical simulations, only collisionless and time persistent particles (DM or stellar) are used to track descendant subhaloes, as gas or BH particles could cause sudden mass loss due to ejection/consumption or mergers, respectively. When multiple previously central subhaloes are found within a single FOF group, this indicates a merger event.

Central candidates are the central subhaloes at the latest snapshot whose previous bound mass is greater than some fraction of the largest bound mass among the candidates ($f_{\rm major}\mathrm{max}(M_{\rm bound})$). If more than one candidate exists, the subhalo with the lowest specific orbital kinetic energy (computed in the centre of mass reference frame of the FOF group) is designated as central. The central subhalo acquires the diffuse mass accreted by the host halo, while the remaining subhaloes are classified as satellites, preventing the accretion of particles that belong to the background of the FOF group. The sources for central subhaloes are updated to the current host halo, excluding particles belonging to any satellite sources; while the sources for the satellites continue to be traced back to their progenitors.

The unbinding algorithm operates recursively from the deepest nested structures to the shallowest, leveraging the code's record of the satellite-of-satellite hierarchy. Stripped particles from satellites are transferred to the central subhalo source to ensure exclusive mass definitions. For each source, unbound particles are removed until the mass converges within a specified tolerance, 
using the centre of mass of all bound particles as a reference to compute kinetic energies. This is applied to both DM and hydrodynamical simulations, treating all particle types equally. In hydrodynamical runs, at least 10 collisionless particles must be present for any resolved subhalo.

{\sc HBT-HERONS} also tracks the evolution of subhaloes whose mass (temporarily) drops below the resolution limit. These ``orphan'' subhaloes are tracked using the most bound particle (tracer particle) and reappear if that particle is hosted by a FOF group with no resolved subhaloes, resulting in zero masses at the intermediate snapshots. The code also includes a prescription for merging subhaloes affected by dynamical friction, coalescing structures that overlap in phase-space. This is assessed using the mass-weighted position and velocity of the 10 most bound particle tracers, except for orphan structures, where simply the tracer particle is used. After the merger, all source particles from the deeper structure are transferred to the shallower one. Further details about other implementations are provided in \citet{hbt-herons}, which not only introduces the code but also compares various subhalo finders, ultimately concluding that {\sc HBT-HERONS} is the most reliable choice for subhalo identification.

\footnotetext[1]{accounting for baryonic particles as well}

\begin{figure*}
\centering
\includegraphics[trim={0cm 1.5cm 8.9cm 1.5cm},clip,width=\textwidth]{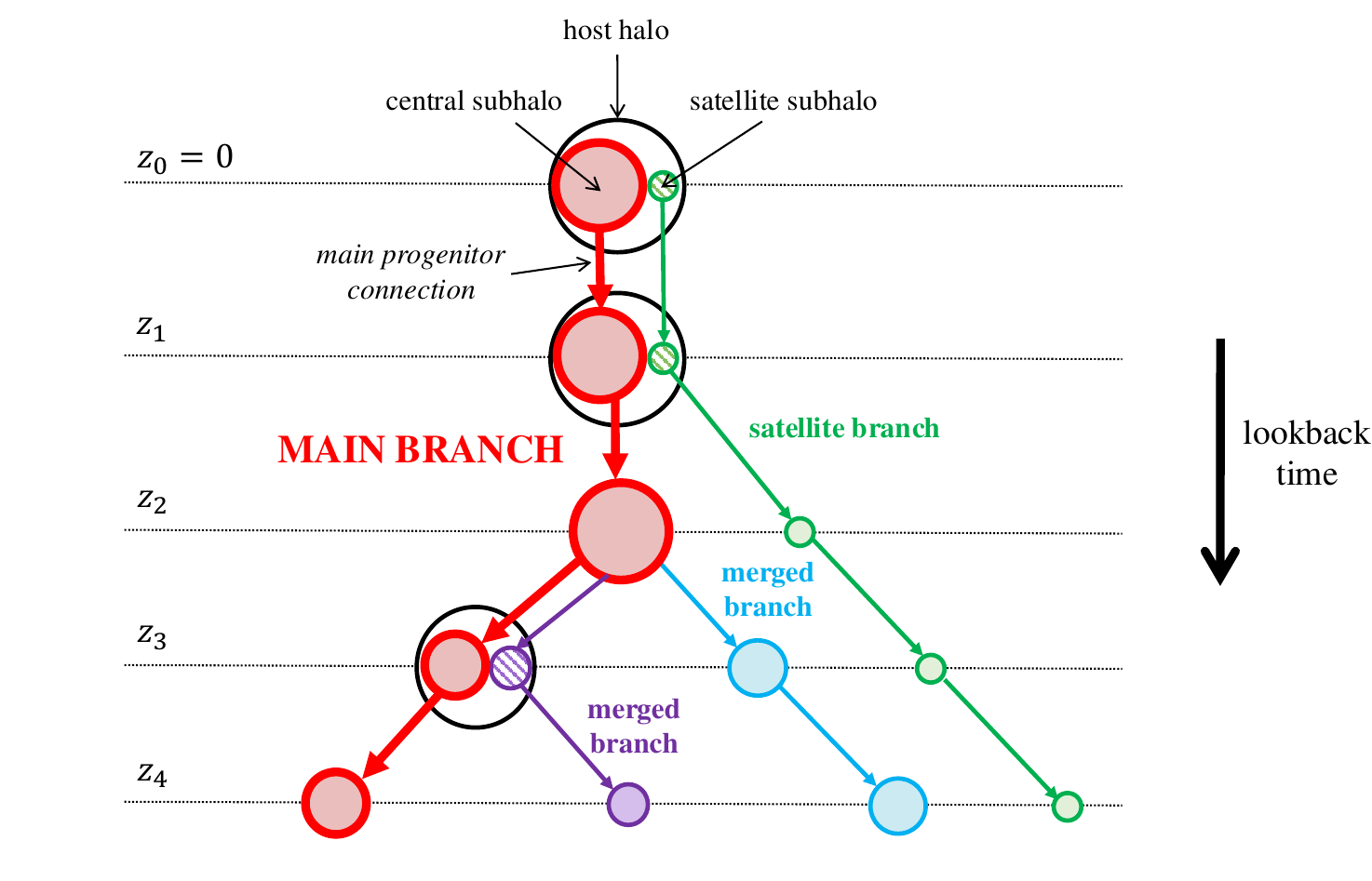}
\caption{Diagram showing the nomenclature used for halo/subhalo finder and merger tree builder codes. Each colour corresponds to a different halo/subhalo, with the circle size indicating roughly their mass — larger circles represent more massive systems. Solid-coloured circles designate central subhaloes within the host halo, while hatched patterns denote subhaloes classified as satellites. When a host halo contains more than one subhalo, it is outlined in black. The main branch stands out in bold red, displaying the evolution of the central subhalo from $z=0$ backwards in time, while other colours show merged/satellite branches.} 
\label{fig:mb-diagram}
\end{figure*}

\subsection{Merger tree-building}
\label{ssec:tree-finding}
Once the structure hierarchy is established, merger tree builders link these structures across the snapshots of the simulation, making a progenitor-subhalo/descendant connection. The links are typically made by cross-matching particles shared by the structures between snapshots. This process is guided by a merit function, which evaluates the particle information considered in the algorithm (e.g. the number of shared particles between structures or the binding energy of the particles). The most likely progenitor/descendant is selected as the one with the highest merit function value. 

Although a subhalo can have multiple progenitors in a merger event (as several subhaloes merge), one is always designated as the main progenitor. The designation can be determined in various ways, such as identifying the subhalo with the highest merit function value (e.g., the one that shares the most particles or the most bound particles with the structure at the latest snapshot) or the most massive one. The main progenitor links are indicated by arrows in Fig.~\ref{fig:mb-diagram}. Below, we outline the approach each tree builder code follows.

\subsubsection{{\sc D-Trees}+{\sc DHalo} (adjacent snapshots)}
\label{sssec:dhalos}
{\sc D-Trees} \citep[explained in appendix~A2 in][]{dhalos} is a subhalo merger tree builder that identifies the progenitor/descendant for each subhalo. The code links structures between snapshots by identifying particles that subhaloes have in common. It maximises a merit function \citep[see equation~A1 in][]{dhalos}, focusing on the most bound particles of each subhalo, known as ``core'' particles. These core particles
help connect subhaloes irrespective of any lost mass as well as identify the subhalo that survives in a merger event.

For each snapshot in the simulation, first {\sc D-Trees} identifies the main progenitor of each subhalo by determining which subhalo in the preceding snapshot contains most of the descendant's core particles (regardless of whether they are part of the main progenitor's core). Secondly, it makes descendant links based on which subhalo at the later snapshot has the highest number of core particles from the earlier subhalo (again, regardless of whether they are part of the descendant's core). Furthermore, the code can connect subhaloes that may be temporarily lost when passing near the dense regions of a larger one, enabling to distinguish between real mergers and those caused by the finder code temporarily not detecting the subhalo. This capability allows the search of main progenitor-descendant connections across multiple snapshots, $N_{\rm step}$.
If snapshots are skipped due to the best link being between non-consecutive snapshots, {\sc D-Trees} inserts ``interpolated'' subhaloes to allow for a continuous merger tree. The properties of these interpolated subhaloes are assigned the same values as those of the subhalo at the earlier time between the two interpolated snapshots.

After establishing these links, the code {\sc DHalo} \citep[explained in appendix~A3 in][]{dhalos} processes the central-satellite hierarchy and groups the subhalo merger tree into host haloes that define the largest gravitationally bound structures. This is based on the idea that once a subhalo becomes part of a host halo, it remains attached to it as long as it is resolved. {\sc DHalo} applies the following conditions: (i) a newly formed subhalo that does not lie inside 
a predefined factor, $R_{\rm factor}$, of the half mass radius of any other subhalo is considered to create a new separate host halo, preventing artificial mergers caused by tenuous bridges of particles; (ii) satellite subhaloes are designated as separate host haloes if they have not lost enough mass, $M_{\rm frac}$, upon becoming satellites, ensuring artificially linked FOF groups are split; and (iii) once conditions (i)--(ii) are applied, a subhalo can not change the host halo to which it belongs. Additionally, a central identifier for each group is assigned to its most massive subhalo. Note, however, that the {\sc Shark} and {\sc Galform} SAMs make their own definitions of the central subhalo, as described later in \S~\ref{sssec:shark}~and~\S~\ref{sssec:galform}, and do not use this central identifier.

The host haloes defined by {\sc DHalo} are suitable for modelling galaxy formation with masses generally growing monotonically \citep{gomez21}. Consequently, both SAMs discussed in \S~\ref{ssec:sam} require {\sc DHalo} to process merger tree data before incorporating any baryonic physics. The code is therefore used in most of the merger tree catalogues listed in Table~\ref{tab:cat}, with four of them employing the {\sc D-Trees}+{\sc DHalo} combination. However, {\sc DHalo} is not restricted to outputs from {\sc D-Trees} and can be applied independently to outputs generated by other tree builders, ensuring compatibility with the input format required by the SAMs. In such a way, the {\sc medi-SURFS} catalogue uses {\sc TreeFrog} (explained in \S~\ref{sssec:treefrog}) to identify progenitor-descendant connections, while {\sc DHalo} redefines host haloes according to conditions (i)--(iii).

\subsubsection{{\sc TreeFrog} (adjacent snapshots)}
\label{sssec:treefrog}
{\sc TreeFrog} \citep{treefrog} operates similarly to {\sc D-Trees}, performing a cross-matching of particles to maximise a merit function \citep[see equation~3 in][]{treefrog}. However, its merit function accounts for the total number of shared particles between subhaloes while weighting them based on the number of equally well-bound shared particles. The weighting scheme is calculated via the rank of the particles, based on their positions in an array sorted by binding energy. This merit function is computed for a fraction $f_{\rm TF}$ of the most bound particles of each subhalo. 

For each snapshot with an available structure catalogue, the code first examines the previous snapshot to identify multiple progenitors (structures with a merit function value above some predefined threshold, $\rm \mathcal{M}_{lim}$, when looking back in time) and the main progenitor (the structure with the highest merit function value). It then analyses the future snapshot to select a unique descendant (the structure with the highest merit function value when looking ahead above the predefined threshold). This ensures that each structure has a single descendant, as tree building can become complex due to mass loss and tidal disruption. The code can also search across multiple snapshots, $\Delta_{\rm s}$, for progenitor-descendant links if no progenitor or descendant is found at the subsequent earlier or later snapshot due to the merit function values being below the threshold. 

After processing the initial merger tree, {\sc TreeFrog} attempts to correct truncation events arising from misidentifications by the halo/subhalo finder. It adjusts links between structures without a main progenitor and structures without a descendant if they are in close proximity in terms of merger tree history and merit function values. Post-processing also considers phase-space position and number of particles to create new links.

\subsubsection{{\sc HBT-HERONS} (history-space)}
\label{sssec:hbt2}
{\sc HBT-HERONS} \citep{hbtp,hbt-herons}, as detailed in \S~\ref{sssec:hbt}, constructs the merger tree branches simultaneously with the subhalo-finding process in a history-space approach. In this method, particles associated with a subhalo, along with additional unbound particles (collectively referred to as the ``source''), are tracked in consecutive snapshots. The descendant is then selected as the bound structure that inherits this set of particles, leading to a more robust and efficient merger tree catalogue.

\subsection{Semi-analytic models of galaxy formation}
\label{ssec:sam}

SAMs solve for the evolution of galaxies by relying on the DM structures and merger histories provided by tree-building algorithms. They treat the different physical processes affecting baryons using a set of analytic equations, generally incorporating various mass reservoirs: hot halo gas, cold gas and stars in a disc or bulge, SMBHs and gas expelled from the halo, as a minimum. More mass reservoirs can potentially be tracked, such as several interstellar medium (ISM) or additional halo gas reservoirs.  A set of coupled parametrised differential equations aims to encapsulate all the physical processes relevant to galaxy formation and evolution, describing how the mass, metals, angular momentum and energy are exchanged between these reservoirs. The parameters involved in these equations are calibrated against a set of observations.

The physical processes represented by the set of equations are largely consistent across different codes \citep{baugh06,benson10}: (i) gas accretion onto DM host haloes, forming a hot halo gas reservoir through smooth accretion; (ii) cooling of this hot gas in the densest regions, settling down in the form of a disc; (iii) star formation within the disc; (iv) stellar feedback, which either returns gas to the hot halo or ejects it from the halo to the intergalactic medium; (v) chemical enrichment of stars and gas; (vi) disc instabilities transferring stars and/or gas to a bulge component; (vii) galaxy mergers, in some cases tracked using ``orphan'' satellite galaxies as satellite subhaloes are not resolved below a resolution limit, which can also alter the disc morphology; (viii) starbursts induced by mergers and/or disc instabilities; (ix) SMBH growth through mergers, starburst-driven cold gas or hot halo gas accretion; (x) active galactic nuclei (AGN) feedback; (xi) photoionisation feedback; and (xii) environmental processes, such as ram pressure stripping, tidal stripping, among others.

SAMs rely on smooth halo/subhalo properties to accurately model baryonic processes. Without this smoothness, non-physical effects, such as the triggering of artificial starbursts, can happen. However, the specific processing of halo/subhalo information can vary between different codes. Below, we provide an overview of how the SAMs used in the analysis handle this information, focusing on how the halo/subhalo properties from tree builder codes are processed given the paper's aim. In our case, both SAMs are run over the same merger tree catalogue in the {\sc DHalo} format, therefore ensuring the input data share the same definitions. 

The central-satellite hierarchy of the subhaloes plays a key role in the accurate treatment of galaxies, as central galaxies are assigned to central subhaloes, while satellite galaxies to satellite subhaloes. The modelling of galaxy properties depends on this hierarchy; for instance, satellite galaxies are subject to environmental processes. Initially, SAMs assume all galaxies originate as central galaxies, corresponding to central subhaloes that appear in isolation as the only subhalo within a host halo. During mergers, galaxies can either retain their status as central objects or transition to satellites based on whether their associated subhalo becomes a central or a satellite within the remnant host halo. Furthermore, galaxies hosted by subhaloes that are no longer resolved due to a low particle number are also classified as satellites.

\subsubsection{{\sc Shark} (subhalo-based)}
\label{sssec:shark}

{\sc Shark} was first presented in \citet{sharkv1}, where prescriptions for the physical processes (i)--(xii) are outlined. A key feature of {\sc Shark} is its flexibility in exploring a wide range of models for the same baryonic processes, allowing for a more thorough exploration of their effect on the predictions of galaxy populations. The code was updated to its latest version in \citet{sharkv2} to model massive galaxies consistently with observations over time. The most significant improvement over the first version of {\sc Shark} is the implementation of a new AGN feedback model that includes jet and wind feedback modes; while other new elements include the exchange of angular momentum between the ISM and stars, a more refined dynamical friction time-scale for mergers \citep{poulton21}, a treatment for the black hole (BH) spin, improved modelling of environmental processes for satellite galaxies, as well as a package that allows for automatic parameter exploration. For this paper, we run the latest version of {\sc Shark}, which has been recalibrated to fit the stellar mass function (SMF) at $z=0$ for the {\sc FLAM-DM-HBT} merger tree catalogue. Details of this recalibration are provided in Appendix~\ref{asec:shark-calibration}.

In {\sc Shark}, when a central subhalo forms, a hot halo gas reservoir at the virial temperature is assigned to its host halo. This hot gas grows in tandem with the host halo and cools down within a cooling radius as per \citet{croton06}. Cold gas assembles into a disc, whose size is governed by the halo gas specific angular momentum and the subhalo maximum circular velocity, becoming available to form stars at a rate computed by integrating the surface density of molecular gas according to \citet{br06} (a.k.a. star formation in the disc). Major and wet (i.e. gas-rich) minor mergers drive the cold gas in the disc to inflow to the bulge component, triggering starbursts (a.k.a. star formation via mergers and disc instabilities) that are more efficient than the disc star formation mode. BHs are seeded with $10^{4}\mathrm{M}_{\odot}/h$ in subhaloes with $10^{10}\mathrm{M}_{\odot}/h$ and can accrete starburst-driven cold gas (starburst mode) or hot halo gas (hot halo mode), as described in \citet{kh00} and \citet{croton16}, respectively. Stellar feedback due to winds or supernova (SN) explosions \citep[assuming a universal][initial mass function (IMF)]{chabrier03} is modelled as in \citet{lagos13}, whereas AGN feedback from winds or jets follows \citet{sharkv2}.

The model pre-processes the {\sc DHalo} merger tree skeleton provided as input before introducing baryonic physics \citep[a full description is in section~4.1 in][]{sharkv1}. This step generates a consistent subhalo hierarchy: selecting at the final simulated snapshot the most massive structure in each group as the central subhalo and tracing its {\sc DHalo} main progenitors (defined by the most bound particles; see \S\ref{sssec:dhalos}) backwards in time, marking them as central subhaloes at earlier times to ensure that the central flag remains unchanged. This prevents artefacts from propagating into the predicted evolution and properties of the galaxy population. Central subhaloes without progenitors are populated with a central galaxy, whose properties — current and future — are calculated making use of either host halo or subhalo information. The code primarily relies on subhalo properties to compute these various galaxy properties, such as the disk size, BH accretion rates or the mass outflow due to SN feedback, among others, aiming for a more accurate physical description of the involved baryonic processes. Additionally, host haloes are constrained to grow in mass consistently with a hierarchical $\Lambda$CDM scenario, thus ensuring the mass of a host halo is equal to or greater than the mass of its progenitors, controlling a physical gas accretion for galaxies.

\subsubsection{{\sc Galform} (host halo-based)}
\label{sssec:galform}

{\sc Galform} was first introduced in \citet{cole00}, with the latest major update presented in \citet{lacey16}, which includes implementations for the physical processes (i)--(xii). The flavour we employ in this paper is GP19 \citep{gp19}, which uses a universal \citep{chabrier03} IMF for star formation in both the disc and the bulge and features an updated model for the BH growth and spin evolution \citep{griffin19}, gradual ram pressure stripping of gas for satellite galaxies \citep{font08} and a merging scheme that follows satellite orbits introduced in \citet{gonzalezperez18,pmill}. The free parameters are calibrated using the {\sc MS-W7} simulation, with a box of $500 \mathrm{Mpc}/h$ \citep{guo13} and a WMAP7 cosmology \citep{komatsu11}. The different cosmology does not affect our analysis, which focuses on merger tree artefact propagation rather than statistical predictions, and the variations involve negligible effects on the studied properties. The calibration is set to reproduce the $z=0$ luminosity function in the $b_{\rm J}$ and $K$ bands, the BH-bulge mass relation and the local passive galaxy fraction. 

The gas treatment for haloes in {\sc Galform} is similar to {\sc Shark}, with hot gas settling into a spherically symmetric distribution and cooling down slowly (hot accretion) or rapidly (cold accretion) inside a cooling radius as explained in \citet{cole00}. The disc has an exponential surface density profile, whose half-mass radius is determined by the conservation of specific angular momentum and the centrifugal equilibrium in the combined gravitational potential of the halo, disc and bulge. Star formation in the disc follows the \citet{br06} prescription as described in \citet{lagos11}. Gas transfer to the bulge due to major, wet minor mergers and disc instabilities and its resulting size is modelled as in {\sc Shark}, but the star formation model for starbursts is different. Here, the star formation rate depends on the dynamical time-scale of the bulge \citep{kenicutt98}. BHs, seeded with $ 10\mathrm{M}_{\odot}/h$ once a galaxy is formed, accrete cold gas from the bulge during starbursts (starburst mode), whose quantity is a fraction of the stellar mass formed, or from the hot gas (hot halo mode). Stellar feedback depends on the galaxy's circular velocity as in \citet{cole00}, and AGN feedback operates only on galaxies whose host 
haloes are under hydrostatic equilibrium, such that the gas cooling flow is shut off if the cooling luminosity is similar to or smaller than the Eddington luminosity.

{\sc Galform} also manipulates the merger tree skeleton provided by {\sc DHalo} to avoid artefact propagation. While its treatment shares some similarities with {\sc Shark} — both impose that host halo masses cannot decrease with time and both pre-process the central-satellite subhalo hierarchy —, there are significant differences. In {\sc Galform}, central flags for each host halo are assigned to the subhalo with the highest cumulative mass throughout its past merger tree. All subhaloes along the main progenitor branch of this subhalo are set to be the central subhalo for their respective host haloes. However, {\sc Galform} depends on the host halo merger tree rather than the subhalo merger tree. As a result, the model prioritises the main progenitor of a host halo, defined as the host containing the main progenitor subhalo of its central subhalo. If no central subhalo exists, the main progenitor is assigned to the host containing the main progenitor of the most massive satellite subhalo.

Furthermore, the overall processing for the evolution of DM structures differs. The code defines ``halo formation'' events when host haloes grow by a factor of 2 in mass, ensuring that host halo properties (circular velocity $V_{\rm vir}$, halo concentration $c$ and spin $\lambda$), except for the host halo mass $\rm M_{halo}$ and radius $r_{\rm vir}$, remain constant between such events. Then, all baryonic processes rely primarily on these properties of the host halo, as they provide a smoother evolution \citep{gomez21}. The only exception is the use of subhalo information to compute the dynamical friction time-scale of a satellite galaxy merging with its central galaxy once the satellite subhalo is no longer resolved. During host halo mergers, the central galaxy in the main progenitor becomes the central galaxy in the subsequent snapshot.

\section{Defining and quantifying merger tree artefacts}
\label{sec:merger-trees-artefacts} 

We analyse the merger tree catalogues described in \S~\ref{sec:methods}. We focus on the central subhaloes at $z=0$ in the simulations and track them backwards by following their main progenitors in what is called the \textit{``main branch''} (see the bold red colour lines in Fig.~\ref{fig:mb-diagram}). Other branches merge with the main branch (blue and purple colours in Fig.~\ref{fig:mb-diagram}) or become satellite subhaloes (green colour). We focus on the main branches because: 1. they characterise the growth of the DM-bound structures, representing the main backbone for SAMs, which assume that central galaxies populate these central subhaloes; and 2. main branches exhibit a wide variety of growth histories that reflect the diversity seen in all subhaloes, and by only analysing those we lower computational costs.

\begin{figure}
\centering
\begin{subfigure}[b]{0.45\textwidth}
   \includegraphics[width=\textwidth]{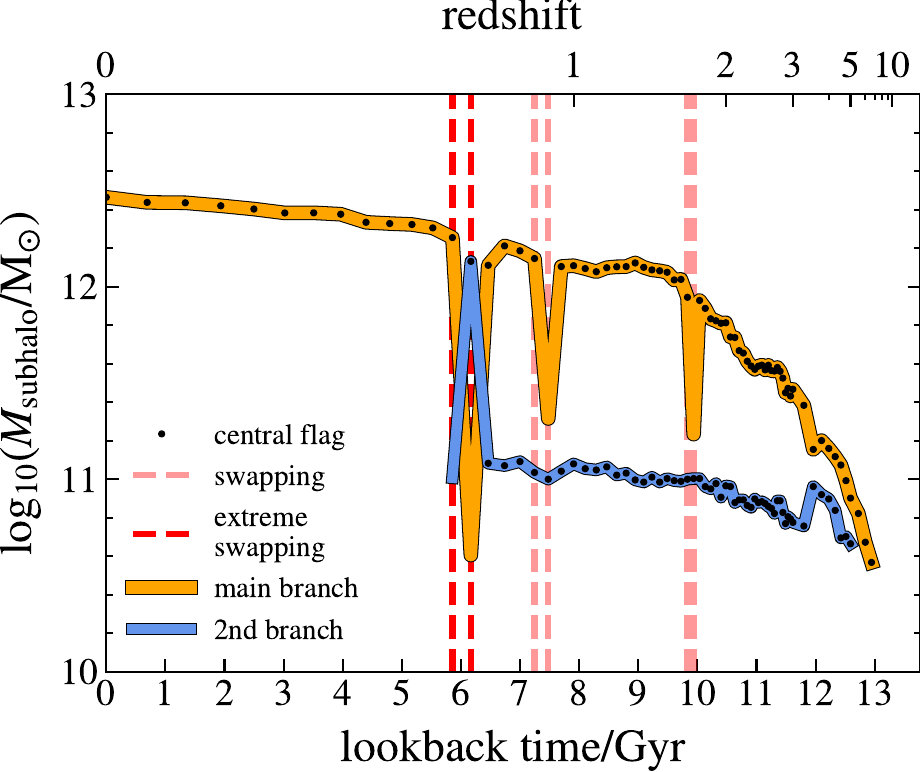}
\end{subfigure}
\begin{subfigure}[b]{0.5\textwidth}
   \includegraphics[trim={1.5cm 3.5cm 15.5cm 1.5cm},clip,width=\textwidth]{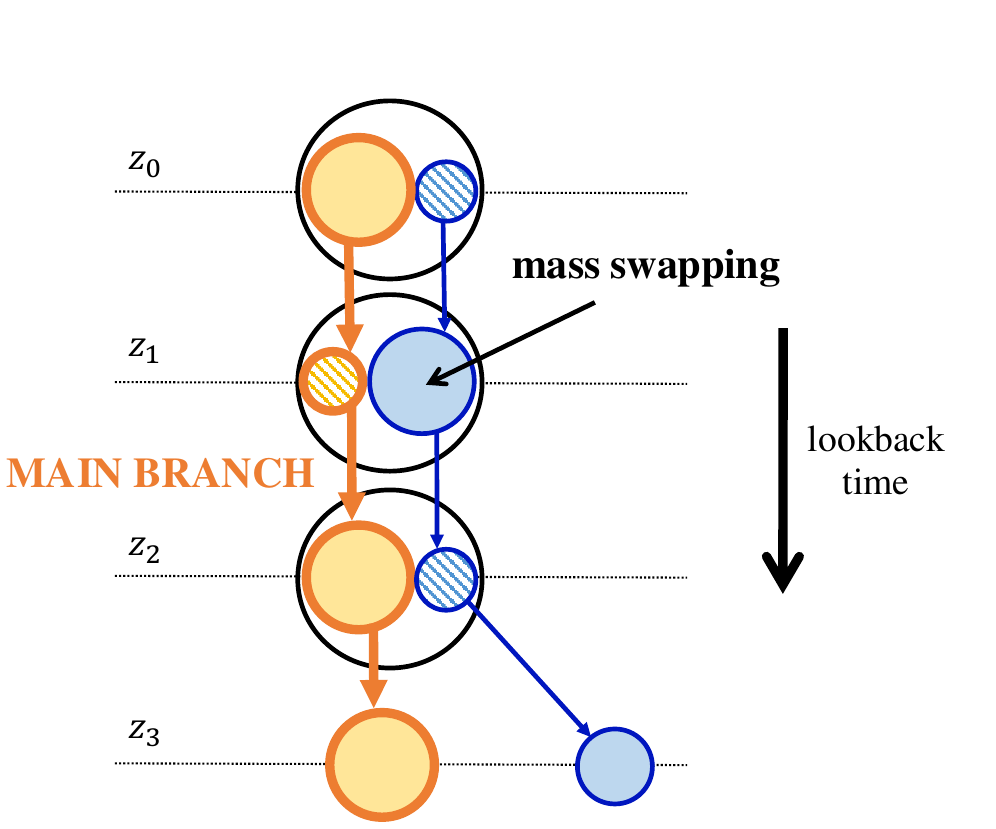}
\end{subfigure}
\caption{Example of a main branch undergoing several mass-swapping events. \textit{Top panel}: MAH (extracted from the {\sc FLAM-DM-VR} catalogue) of a central subhalo (orange line) with several large decreases in mass produced by mass-swapping events marked by red-dashed vertical lines, with the most extreme cases in a more opaque red. The abrupt mass loss around lookback times $\rm \approx6~Gyr$ (the largest of the mass drops in this main branch) is due to the subhalo (blue line) swapping its hierarchy. The other two dips are swapping events with subhaloes that are not shown. The black dots show the snapshots at which each structure is defined as being central. In this instance, the subhalo represented by the blue line is incorrectly identified as the central subhalo without first being classified as a satellite. \textit{Bottom panel}: schematic diagram where each colour refers to a different structure and the circle size represents qualitatively mass, using the same colours as in the top panel to ease the connection between the two. The figure shows that at $z_1$, the central subhalo (solid-coloured circles) is misidentified as being a satellite (hatched pattern) for that one snapshot, and the satellite subhalo in blue is momentarily identified as being central.} 
\label{fig:ms-diagram} 
\end{figure}

We examine the mass accretion histories (MAH) for the main progenitor subhalo, as defined by each merger tree code (\S~\ref{ssec:tree-finding}), of all main branches. Some SAMs use these histories that describe the growth of DM within subhaloes to model galaxy properties accordingly (see the {\sc Shark} description in \S~\ref{sssec:shark} for details). By examining the MAHs generated by different combinations of halo/subhalo finder and merger tree algorithms, we categorise the numerical artefacts that affect them into two main types: \textit{``mass-swapping''} and \textit{``massive-transient''} events. These artefacts are described and their frequencies analysed in \S~\ref{ssec:mass-swapping}~and~\ref{ssec:massive-transients}, respectively. In \S~\ref{ssec:impact-galaxy-formation-models}, we provide examples of how these numerical artefacts in merger trees directly impact the predicted evolution of galaxies in SAMs. We initially visualised the MAHs for both main and satellite branches at $z=0$ and found no artefact exclusively affecting the satellite population, supporting our decision to focus solely on the main branches in this study.

\subsection{Mass-swapping events}
\label{ssec:mass-swapping} 

\subsubsection{Definition}
\label{sssec:mass-swapping-def} 

The first artefact affecting merger trees that we identify is the instantaneous drops in mass during their assembly, which contrasts with the expectations of central MAHs within a hierarchical structure formation. A central subhalo should grow in mass exclusively through smooth accretion or mergers with other haloes/subhaloes over time. An example of this artefact is shown in the top panel in Fig.~\ref{fig:ms-diagram}, where the evolution of a main branch (in orange) displays several significant mass drops, highlighted by red dashed vertical lines. The largest mass drop, a change of almost $2$~dex, is marked with a more intense red line.

These dips in the central subhalo mass are numerical issues arising from the misidentification of structures by halo/subhalo finder and merger tree builder codes, particularly when dealing with massive structures undergoing merger events. This type of artefact is well-known in the literature and has been reported in works such as \citet{srisawat13} and \citet{poole17}. 

The diagram in the bottom panel in Fig.~\ref{fig:ms-diagram} illustrates the general phenomenon: a central subhalo and a satellite subhalo swap their hierarchy, resulting in the material on the outskirts of the halo being exchanged between them, as those particles are typically assigned to whichever subhalo is identified as central. This can be observed in the example in the top panel, focusing on the most significant swapping event. Here, the evolution of the satellite that inherits these particles is shown in blue. This satellite subhalo becomes the central, indicated by the black dots, at the specific snapshot where there is a noticeable jump in mass around lookback times $\rm \approx6~Gyr$, but subsequently reverts to being a satellite. When this occurs, all the mass on the outskirts is re-assigned to what had been the central subhalo before the mass-swapping event. In this particular example, the blue subhalo does not initially possess a satellite hierarchy before the swapping event, but the underlying phenomenon remains the same. The remaining
two dips in Fig.~\ref{fig:ms-diagram} at roughly lookback times $\rm \approx7.5~and~10~Gyr$ are swapping events with other subhaloes that are not shown.

Events where the subhalo mass undergoes large mass increases (not produced by mergers) or decreases are considered non-physical and should be classified as mass-swapping events. To isolate them, we introduce a physically reasonable definition and follow \citet{wechsler02}, where the authors assume the MAHs recovered from $\rm \Lambda CDM$ cosmological simulations can be universally described by exponential profiles of the form:

\begin{equation}
    M(z)=M_{0}\ e^{-\alpha z}
    \label{eq:1}
\end{equation}

\noindent where $M_{0}$ is the final mass of the subhalo, $z$ the redshift and $\alpha$ a free parameter.

When the tree builder codes link the main progenitor and descendant subhaloes, we compute the mass ratio that arises following equation~(\ref{eq:1}) as a function of the redshift:

\begin{equation}
    \frac{M(z_{\mathrm{i+1}})}{M(z_{\mathrm{i}})}=\frac{M_{\mathrm{subhalo}}}{M_{\mathrm{main}\ \mathrm{progenitor}}}=e^{\alpha \Delta z}
    \label{eq:2}
\end{equation}

\noindent where $\Delta z = \lvert z_{\mathrm{i+1}}-z_{\mathrm{i}} \rvert$ is the redshift separation between consecutive snapshots. Thus, we can compute the expected change in mass according to these exponential profiles independently of the snapshot cadence of each simulation. This definition helps avoid events being misclassified as mass-swapping at early times due to the longer redshift span between snapshots.

With all these ingredients we can give a proper definition for the mass-swapping events. Typical values for $\alpha$ lie in the range $\approx[0.3-2]$ \citep[see fig. 3 in][for structures spanning a range of masses and concentration parameters]{wechsler02}. However, for our purposes, we select (realistic) extreme values considering always matter accretion (i.e. increases in mass) at varying rates: an extremely rapid rate $\alpha_{\mathrm{u}}=3$ as the upper limit and an extremely slow $\alpha_{\mathrm{l}}=0.1$ as the lower limit. Additionally, instead of defining unrealistic $\alpha$ values (outside the range of values set above) to account for non-physical mass increases or decreases, we introduce extreme offsets in equation~(\ref{eq:2}) that imply in the case of no mass change ($\alpha=0$) an increase larger than a factor of two ($O_{\mathrm{u}}=1$) or a decrease more significant than a factor of half ($O_{\mathrm{l}}=-0.5$) in the original mass. The final equation we use to classify mass-swapping events takes the form:

\begin{equation}
    \frac{M_{\mathrm{subhalo}}}{M_{\mathrm{main}\ \mathrm{progenitor}}}=\begin{cases}
        e^{\alpha_{\mathrm{u}} \Delta z}+O_{\mathrm{u}} = e^{3\Delta z}+1 & \text{(upper limit)}\\
        e^{\alpha_{\mathrm{l}} \Delta z}+O_{\mathrm{l}} = e^{0.1\Delta z}-0.5 & \text{(lower limit)}
    \end{cases}
    \label{eq:3}
\end{equation}

We classify as physically well-defined MAHs those whose mass changes lie between the upper and lower limits in equation~(\ref{eq:3}). If the relative mass gained or lost in a subhalo-main progenitor connection exceeds these imposed thresholds, that connection is considered numerically incorrect. Since the parameters chosen for equation~(\ref{eq:3}) correspond to extreme mass-swapping events, we are confident our selection isolates only numerically incorrect events.

\subsubsection{Characterisation}
\label{sssec:mass-swapping-calc}

\begin{figure*}
\vspace{0.5cm}
\centering 
\includegraphics[trim={0.1cm 0.1cm 0.1cm 0.1cm},clip,width=0.5\linewidth]{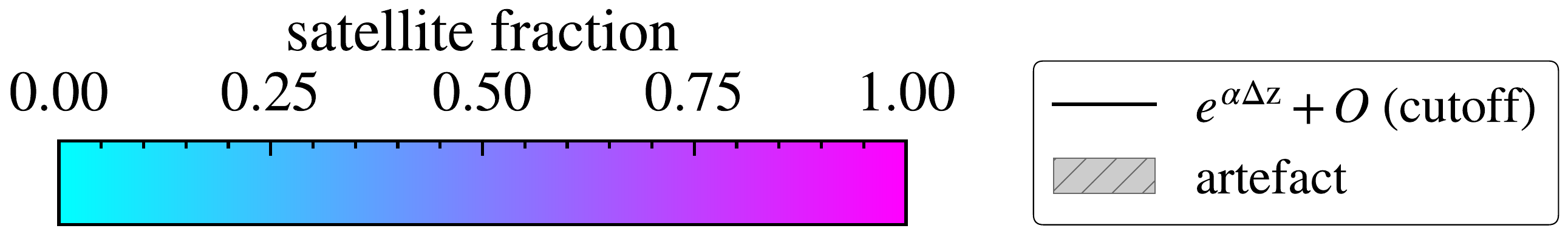}
\\[0.3cm]
\medskip
\captionsetup[subfigure]{labelfont=bf,font=large}
\begin{subfigure}{0.3635\textwidth}
  \includegraphics[trim={0.1cm 0.1cm 0.1cm 0.1cm},clip,width=\linewidth]{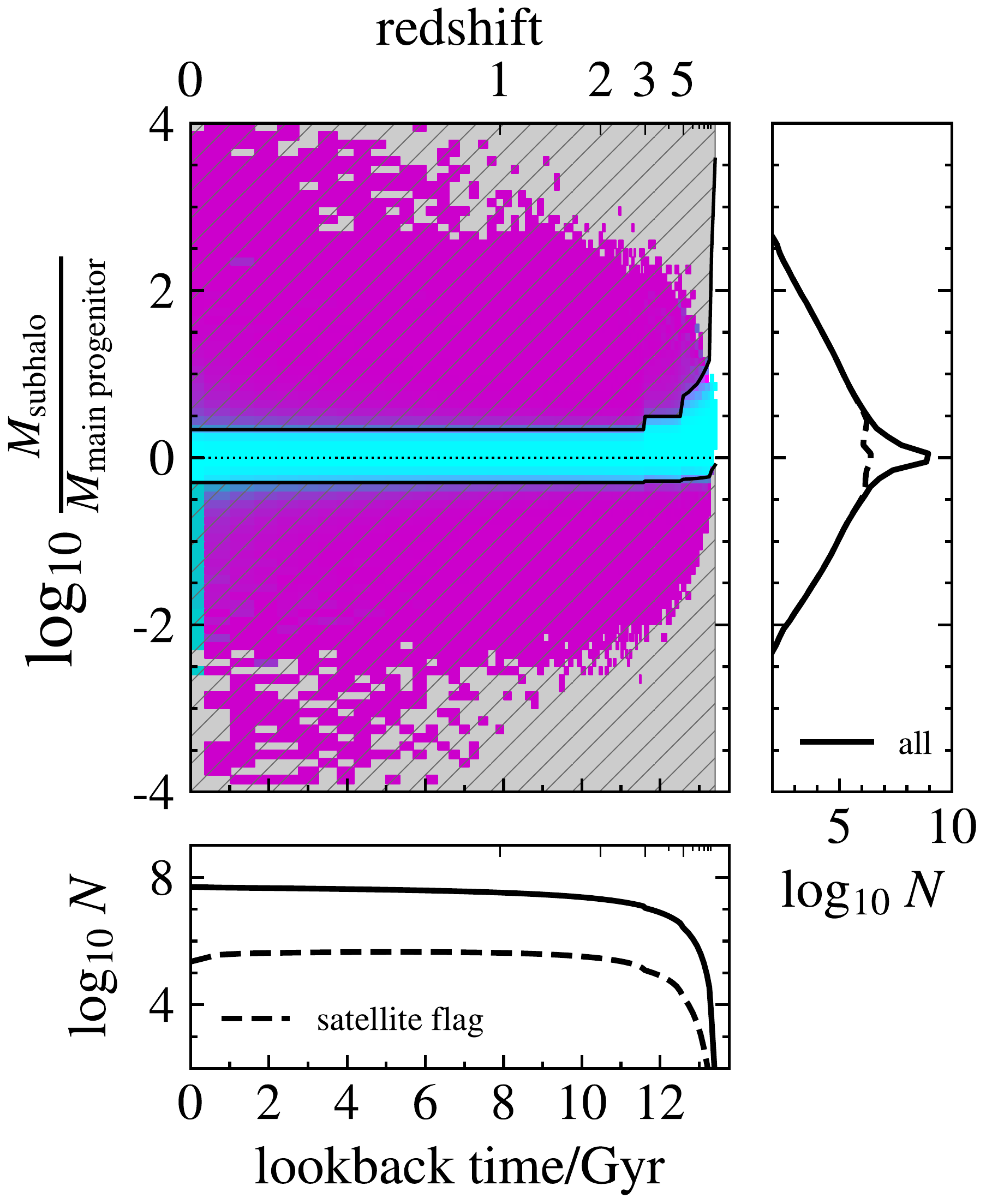}
  \caption{\textbf{\textsc{FLAM-DM-VR}}}
  \label{fig:msdef1}
\end{subfigure}\hfil 
\begin{subfigure}{0.3\textwidth}
  \includegraphics[trim={0.1cm 0 0 0},clip,width=\linewidth]{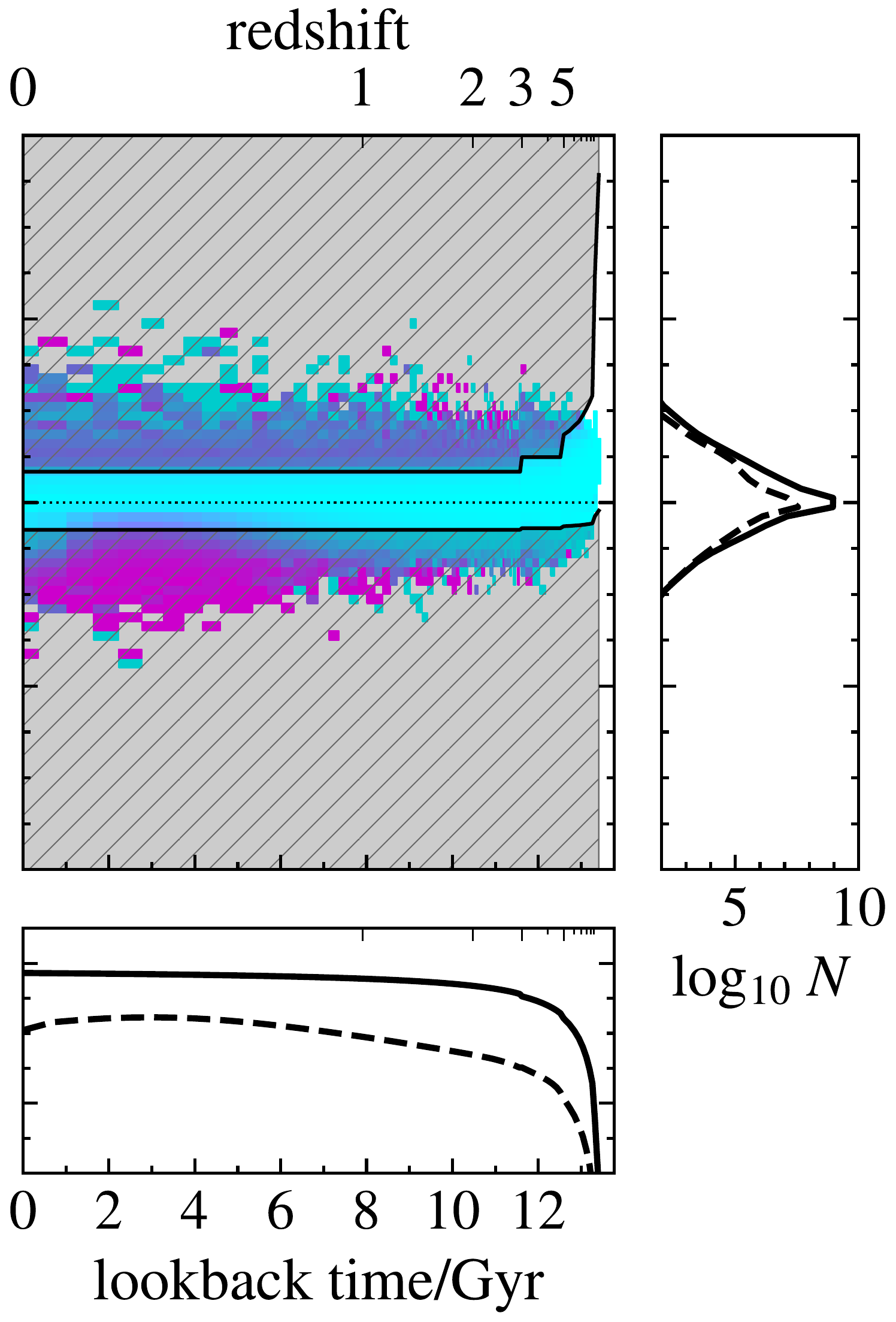}
  \caption{\textbf{\textsc{FLAM-DM-HBT}}}
  \label{fig:msdef2}
\end{subfigure}\hfil 
\begin{subfigure}{0.3\textwidth}
  \includegraphics[trim={0.1cm 0 0 0},clip,width=\linewidth]{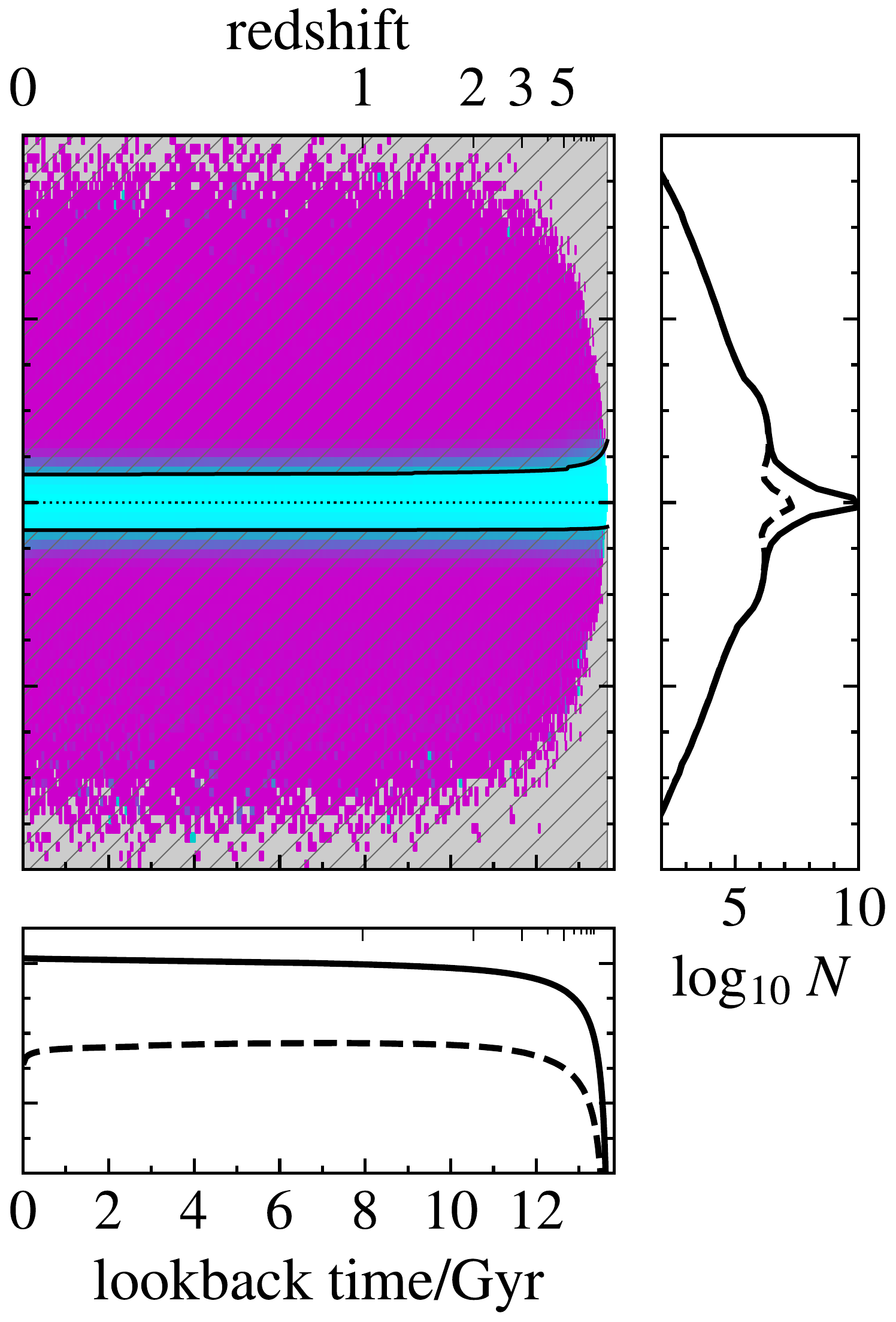}
  \caption{\textbf{\textsc{PMill}}}
  \label{fig:msdef3}
\end{subfigure}
\\[0.3cm]
\medskip
\begin{subfigure}{0.3635\textwidth}
  \includegraphics[trim={0.1cm 0.1cm 0.1cm 0.1cm},clip,width=\linewidth]{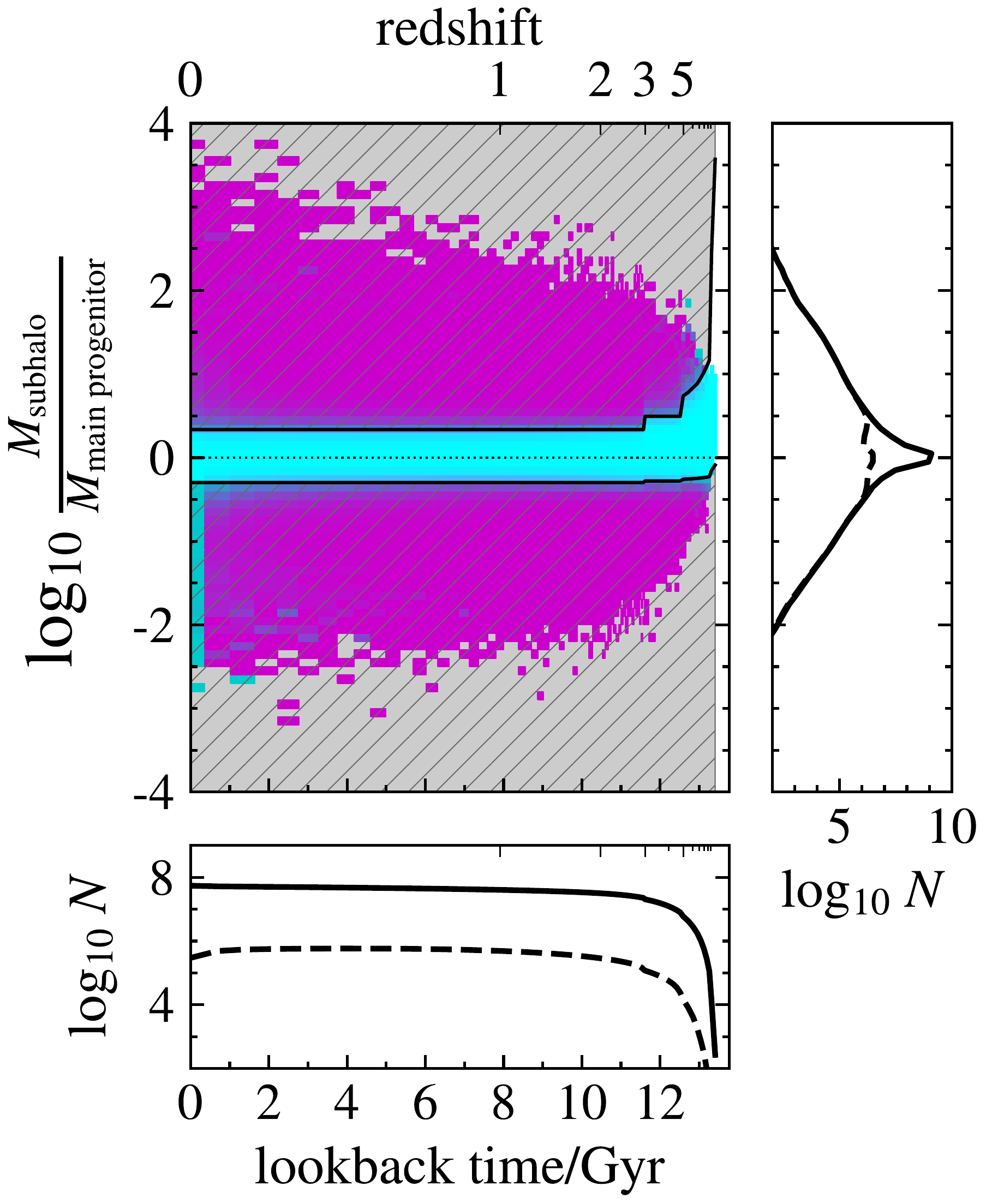}
  \caption{\textbf{\textsc{FLAM-Hydro-VR}}}
  \label{fig:msdef4}
\end{subfigure}\hfil 
\begin{subfigure}{0.3\textwidth}
  \includegraphics[trim={0.1cm 0 0 0},clip,width=\linewidth]{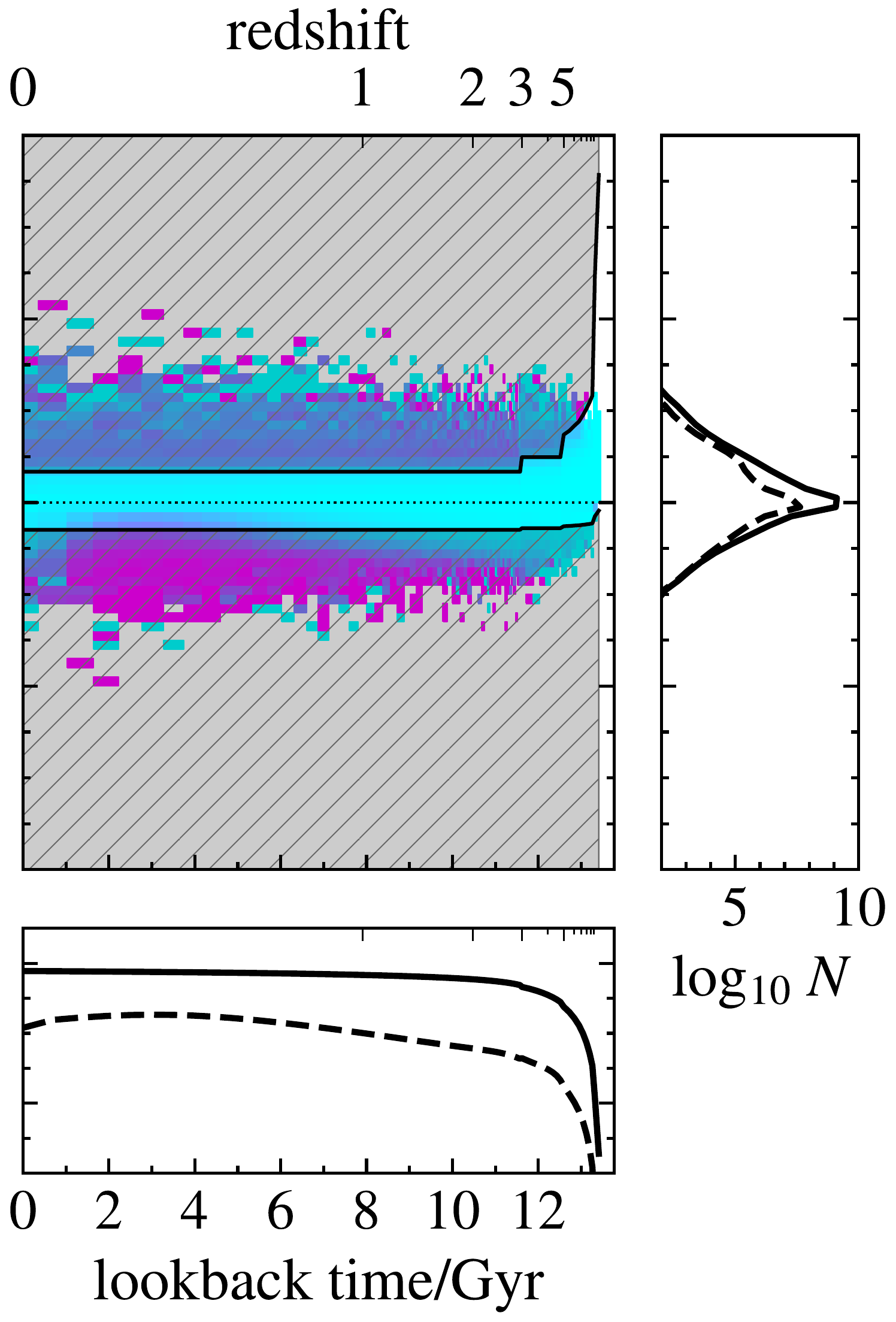}
  \caption{\textbf{\textsc{FLAM-Hydro-HBT}}}
  \label{fig:msdef5}
\end{subfigure}\hfil 
\begin{subfigure}{0.3\textwidth}
  \includegraphics[trim={0.1cm 0 0 0},clip,width=\linewidth]{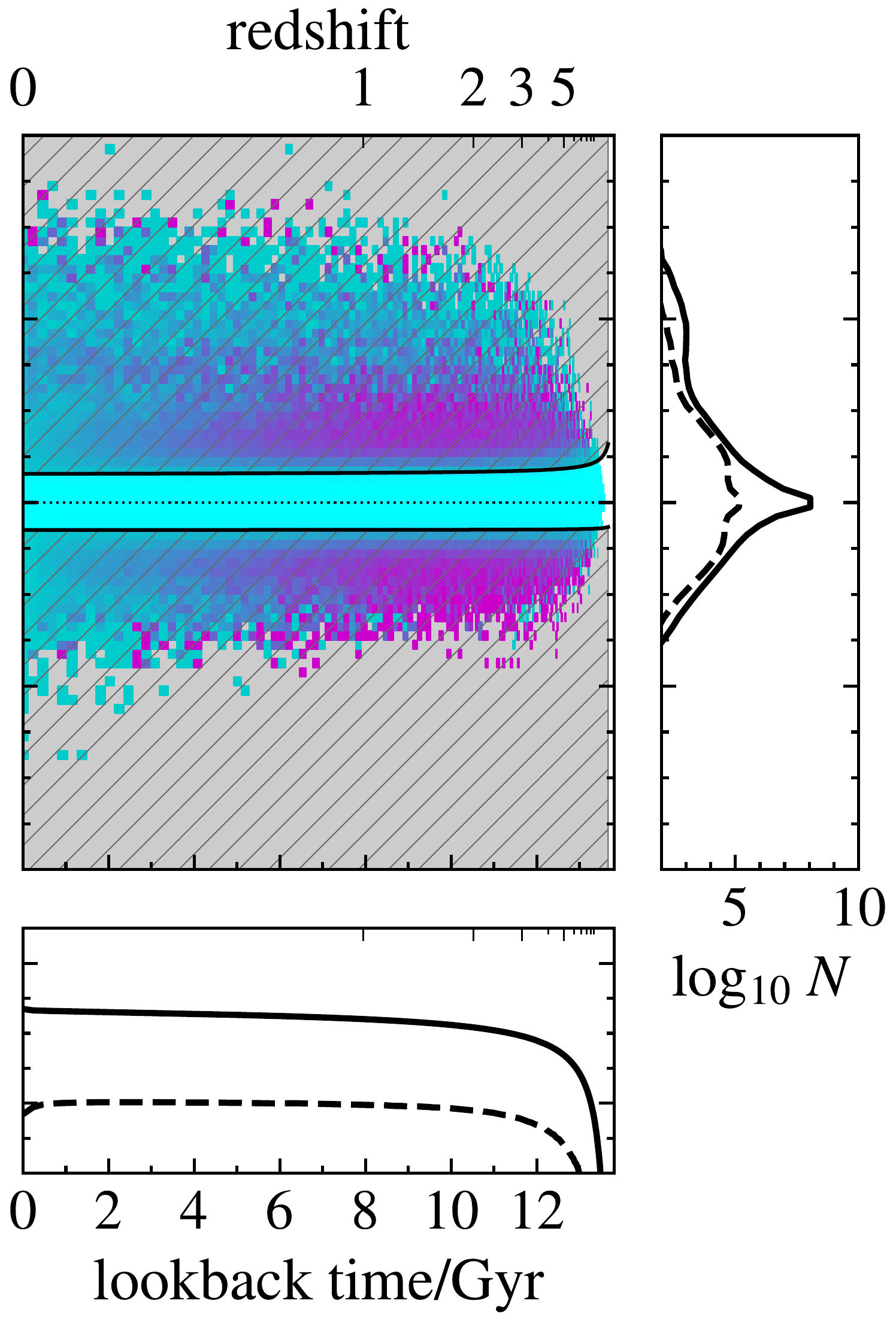}
  \caption{\textbf{\textsc{medi-SURFS}}}
  \label{fig:msdef6}
\end{subfigure}
\\[0.3cm]
\caption{2D maps of the relative change in mass when there is a connection between a main progenitor and a descendant subhalo for a main branch as a function of the lookback time/redshift at which that event is happening, for the 6 merger tree catalogues described in Table~\ref{tab:cat}. The main panel is colour-coded by the fraction of events within each 2D bin where the main progenitor or the descendant are satellites, following the colour bar at the top. The shaded area marks the numerical artefacts based on the thresholds defined in equation~(\ref{eq:3}) (solid line). A dotted line represents cases where there is no mass change ($M_{\mathrm{subhalo}}=M_{\mathrm{main}\ \mathrm{progenitor}}$). On the right and bottom panels, we show the 1D projected histograms for the mass ratio and the lookback time/redshift, respectively, for all the connections (solid line) and the satellite-related ones (dashed line).}
\label{fig:ms-def-panel}
\end{figure*}

Fig.~\ref{fig:ms-def-panel} shows 2D maps of the relative mass change as a function of the lookback time/redshift for all the main branch connections for each of the merger tree catalogues. Throughout this paper, for {\sc HBT-HERONS} subhaloes that become ``orphans'' (i.e., with just the most bound particle) across a range of snapshots because their particle number fluctuates around the minimum to detect them and then reappear, their mass is assigned the same values as the mass of the subhalo at the earlier snapshot where it is well-defined. Thus, these instances are not flagged as mass-swapping events. The shaded areas indicate regions considered mass-swapping artefacts (i.e. large mass increases, that are not caused by mergers, at positive values of the y-axis, and huge mass decreases at negative values of the y-axis). Additionally, there are 1D projected histograms on the right and bottom panels to aid in the analysis.

The most striking difference across the catalogues is how populated the artefact area (shaded region) is. The {\sc PMill} catalogue spans the largest range in parameter space with mass changes reaching extreme upper and lower values as large as $\rm log_{10}(\textit{M}_{subhalo}/\textit{M}_{main\ progenitor})\approx[-4,4]$. Similarly, the {\sc FLAM-DM-VR} and {\sc FLAM-Hydro-VR} catalogues also cover a wide area around $\rm log_{10}(\textit{M}_{subhalo}/\textit{M}_{main\ progenitor})\approx[-3,3]$, though they are not as broad as {\sc PMill}, particularly at a high lookback time/redshift. This broad distribution can result from {\sc D-Trees} tracking only the most bound cores of subhaloes while disregarding the bulk of the subhalo mass. The {\sc medi-SURFS} catalogue produces slightly better results, possibly due to its smaller box size and {\sc TreeFrog} post-processing. Configuration-based finders, such as {\sc Subfind}, are less effective at accurately recovering subhalo masses in mergers events — characteristic of these artefact areas — compared to phase-space-based finders like {\sc VELOCIraptor} — since the former ones struggle to resolve structures at nearly identical positions. However, mass fluctuations still exist due to the phase differences between the orbiting core and the merging mass \citep{behroozi15}.

The {\sc HBT}-related catalogues show the most accurate results, with the smallest population of artefacts and a well-defined peak with narrower wings in the 1D projected histogram in the right panel around $\rm log_{10}(\textit{M}_{subhalo}/\textit{M}_{main\ progenitor})\approx0$. Temporal finders provide more stable merger trees since they do not truncate branches early, they create smoother mass variations and better tracking of other properties as well \citep{srisawat13,avila14,behroozi15,wang16}.

The map in Fig.~\ref{fig:ms-def-panel} is colour-coded by the fraction of main branches in each 2D bin that exhibit a satellite hierarchy in one of the two snapshots where the mass change is calculated. Main branches should maintain a central hierarchy over time. Therefore, regions of the parameter space dominated by satellite events (shown in more purple colours) are not physically realistic. As these purple areas are well-captured in the shaded prohibited areas, the cutoffs defined in equation~(\ref{eq:3}) effectively isolate temporal changes in hierarchy. We do not directly flag cases with a change in the hierarchy as numerical artefacts because this could overlook swapping events that do not involve such a huge mass exchange, e.g. fly-by or splashback cases. Conversely, there may be instances where the mass-swapping is not linked to the typical satellite-central type swapping.

All examples that employ directly {\sc D-Trees} to build the subhalo trees exhibit a satellite fraction close to unity in the artefact area, even when combined with {\sc DHalo}, which aims to define this hierarchy consistently to avoid these swapping events, as stated in \citet{gomez21}. In contrast, the {\sc HBT}-related cases and {\sc medi-SURFS} display many mass-swapping events where the subhalo retains its central status (more cyan colours in the shaded region). This indicates that mass-swapping does not always correspond to a hierarchy swap, particularly in the case of non-physical mass increases.
For {\sc medi-SURFS}, {\sc TreeFrog} appears to clean the satellite hierarchy in central branches.
Moreover, the last snapshot (at lookback times $\rm \approx0~Gyr$) for the {\sc FLAM-DM-VR} and {\sc FLAM-Hydro-VR} catalogues show interesting mass decreases without corresponding hierarchy change. 

When comparing the DM-only and hydrodynamical runs, the 2D maps in Fig.~\ref{fig:ms-def-panel} reveal fewer extreme mass-swapping events in the {\sc VELOCIraptor} runs, while for {\sc HBT-HERONS} the results remain quite similar. As gas particles tend to cluster more strongly than the DM counterparts \citep{knebe13b}, bound structures exhibit stronger gravitational potentials, which reduces the amount of swapped material on the outskirts and makes the mass-swapping events less extreme when these objects are identified with a phase-space method. For {\sc HBT-HERONS}, the history-based method of tracking which particles belong to which subhalo is minimally impacted by the inclusion of baryons thanks to the new additions made to the algorithm, as described in \citet{hbt-herons}. Plus, when baryons are included, the masses and numbers of particles for subhaloes involve all particles, regardless of their type (DM or baryons).


\subsubsection{Quantification}
\label{sssec:mass-swapping-quant} 

After properly defining the mass-swapping events, we identify the main branches affected by them and quantify how many main branches exhibit such behaviour. Table~\ref{tab:statistics} lists the number of main branches analysed for each catalogue, along with their median mass and particle number. We also report the fraction of main branches affected by non-physical mass changes in their accretion histories, as well as the median mass and particle number for the affected branches. 

In most catalogues, $\approx10\%$ of branches experience at least one mass-swapping event, whereas for the {\sc HBT}-related catalogues, this fraction is significantly lower ($\approx2\%$). This suggests, as previously seen in the 2D maps, that the distinct {\sc HBT-HERONS} algorithm effectively reduces the number of branches that are artificially affected, likely due to the employed temporal information and the advantage of building the branches simultaneously. Notably, branches affected by mass-swapping tend to have a larger median maximum particle number ($\approx200-300$) compared to the median maximum number of particles for all branches ($\approx 50-80$). The artefact may occur at any stage of the subhalo evolution, but the maximum values, presumably reached near $z=0$, are
used for consistency. Larger subhaloes with later formation times are more prone to undergo major mergers since they trace higher density peaks (denser environments) \citep{behroozi15}, making structure identification more challenging and increasing the likelihood of mass-swapping.

The left panel of Fig.~\ref{fig:diagnostics-part} presents the fraction of main branches that have suffered from mass-swapping events at any point of their evolution as a function of the maximum particle number along the main branch to remove the resolution dependence. For the two catalogues derived from hydrodynamical runs ({\sc FLAM-Hydro-VR} and {\sc FLAM-Hydro-HBT}), the maximum particle number is estimated by dividing the maximum subhalo mass (accounting for all particle types) by the DM particle mass, $N^{\rm max}_{\rm part,\ subhalo}\approx M^{\rm max}_{\rm part,\ subhalo}/m_{\rm part}$. This provides a lower limit that does not bias the results. Each colour represents a different merger tree catalogue, while the vertical lines mark the resolution limit for each of them. Plus, in Fig.~\ref{fig:diagnostics}, we show a version of Fig.~\ref{fig:diagnostics-part} that replaces the x-axis with the subhalo mass.

The most noticeable trend is that more massive structures are more prone to mass-swapping across all catalogues, especially those with maximum particle numbers above $10^{3}$. These high-particle-number structures may, however, have faced numerical issues when they were less massive, having grown in mass since then. Broadly, we find more than half of the branches more massive than $10^{3}$ particles are affected by mass-swapping events. The exception is {\sc HBT-HERONS}, which adopts a history-space approach based on tracking particles between snapshots, effectively minimising this artefact. Nevertheless, the peak of affected main branches for the {\sc HBT} catalogues still occurs at the high-mass end.

The general trend for the non-{\sc HBT} cases is similar, but the shapes of the curves differ for the different codes employed. The {\sc medi-SURFS} catalogue behaves similarly to the {\sc FLAM-DM-VR} and {\sc FLAM-Hydro-VR} catalogues because they share the same halo finder ({\sc VELOCIraptor}), which indicates the issues stem initially from the structure-finding step. This interpretation is further supported by the difference with {\sc PMill}, which follows a distinct trend as a result of using {\sc Subfind}. Configuration-based algorithms in principle struggle more to separate structures in dense environments, but in this case, they show fewer affected massive branches. This discrepancy may arise from the varying definitions of central subhaloes between the codes, with {\sc VELOCIraptor} also identifying remnants of mergers as satellite subhaloes, reallocating particles from the central subhalo accordingly.

The hydrodynamical runs, although having fewer extreme cases and fewer in number as stated before, show that the fraction of branches affected by mass-swapping is larger for the catalogues processed with both {\sc VELOCIraptor} and {\sc HBT-HERONS} than for DM-only in Fig.~\ref{fig:diagnostics-part}, and the median values correspond to fewer massive branches in Table~\ref{tab:statistics}. While DM-only runs may exhibit more mass-swapping events, these tend to occur repeatedly on the same branches due to less strong gravitational potentials. In contrast, branches in hydrodynamical runs are more susceptible to mass-swapping since the centre of the host haloes is denser due to the stronger clustering, making the particle assignment over different subhaloes more challenging.

\subsection{Massive transients}
\label{ssec:massive-transients} 

\subsubsection{Definition}
\label{sssec:massive-transients-def} 

\begin{figure}
\centering
\begin{subfigure}[b]{0.45\textwidth}
   \includegraphics[width=\textwidth]{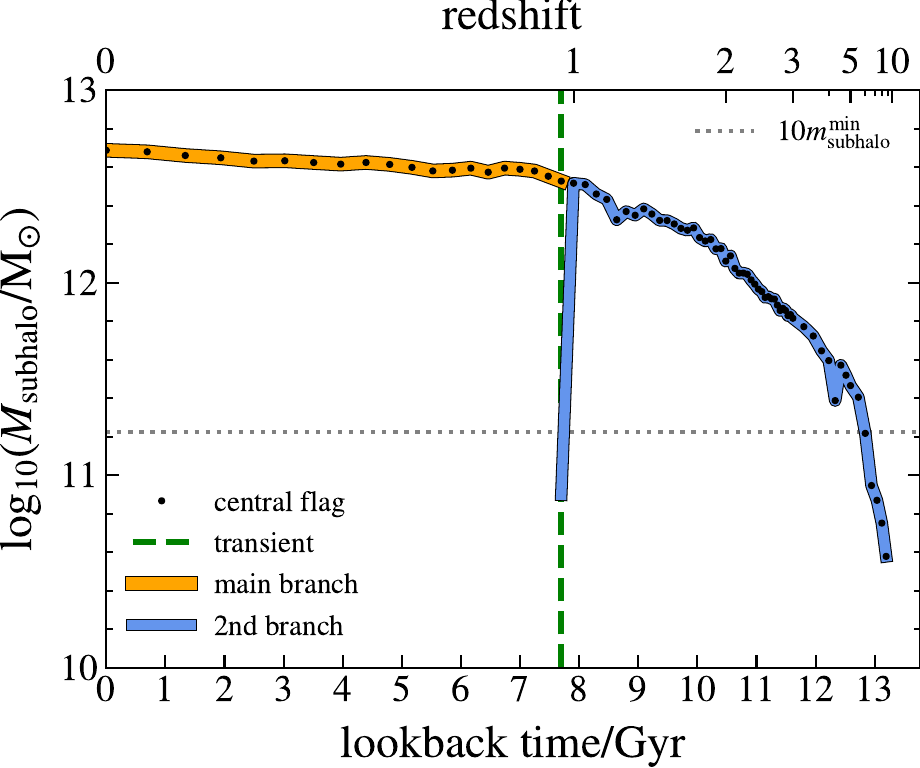}
\end{subfigure}
\begin{subfigure}[b]{0.5\textwidth}
   \includegraphics[trim={5cm 3cm 12.5cm 2cm},clip,width=\textwidth]{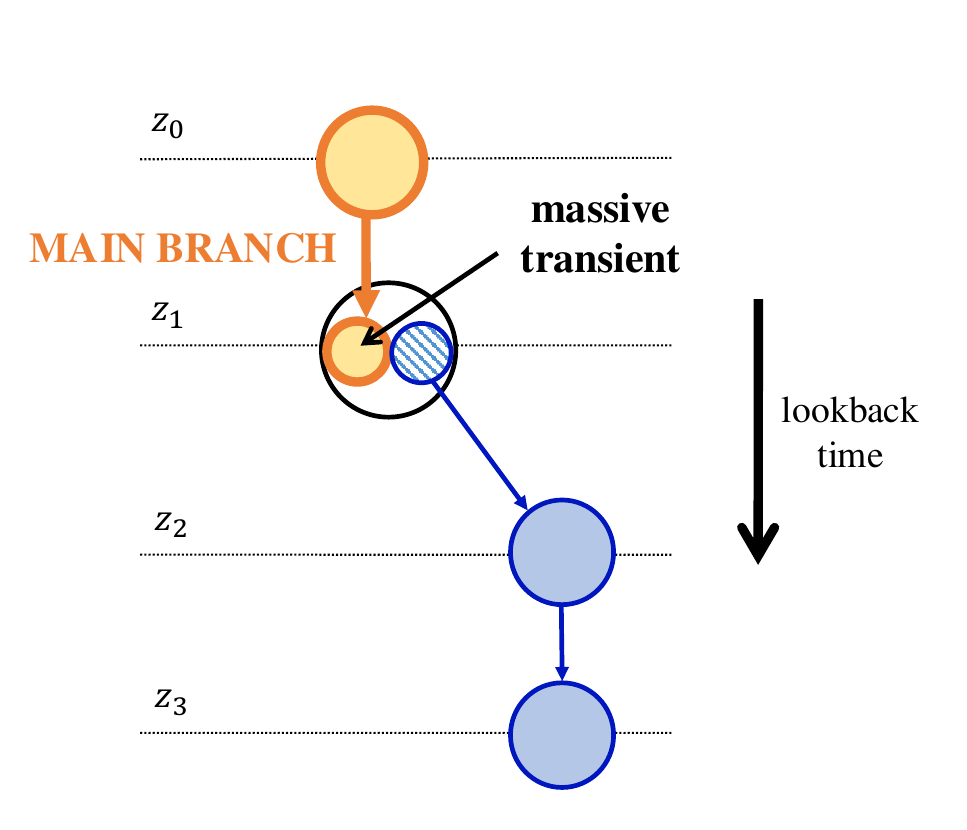}
\end{subfigure}
\caption{Example of a massive transient subhalo. \textit{Top panel}: MAH (extracted from the {\sc FLAM-DM-VR} catalogue) of a central subhalo (orange line) that unexpectedly emerges at a late cosmic time (lookback times $\rm \approx8~Gyr$), marked by a green dashed vertical line. Before this event, another subhalo (blue line) was growing steadily. These 2 structures should be linked in the tree-building process, however they are not. The black dots show the snapshots at which each structure is defined as being central. \textit{Bottom panel}: schematic diagram where each colour refers to a different structure and the circle size represents qualitatively mass, using the same colours as in the top panel to ease the connection between the two. The figure describes that at $z_{1}$ the blue subhalo, which was the main branch up to that cosmic time, is suddenly split into 2 distinct structures. A new central subhalo (solid-coloured circles) appears in orange and the former main branch is classified as a satellite (hatched pattern). Afterwards, the orange subhalo survives as the new main branch.} 
\label{fig:mt-diagram}
\end{figure}

The second type of artefact we observe in the MAHs of the main branches is the sudden appearance of subhaloes with an unexpectedly high number of particles. Typically, bound structures are first identified by the codes described in \S~\ref{ssec:halo-finding}, when they consist of around $20$ particles. However, there are cases in which structures are detected for the first time with hundreds or even thousands of particles. Throughout this paper, we refer to these structures as massive transients. An example of a massive transient is shown in orange in the top panel in Fig.~\ref{fig:mt-diagram}. Initially, there is a structure (depicted in blue) that grows in mass smoothly, but it abruptly disappears at a late cosmic time (in this case around lookback times $\rm \approx8~Gyr$) after being redefined as a satellite subhalo of a new central subhalo. This new subhalo branch emerges for the first time with a very high mass, $\gtrsim 10^{12.6}\,\rm M_{\odot}$, and corresponds to the final main branch. Such structures are not expected to appear in the current model of structure formation, thus they represent a non-physical numerical artefact resulting from the combination of halo-finding and tree-building processes, as has been previously reported in the literature \citep[e.g.][]{wang16,srisawat13}.

The threshold for the particle number at birth, which we consider indicative of numerical inaccuracies, is somewhat ambiguous and not straightforward. We decide to use the standard deviation of the particle number at birth for all main branches in the catalogue at each specific snapshot, $\sigma(N^{\mathrm{birth}}_{\mathrm{part,subhalo}},z)$, to define it. We flag as numerical artefacts subhaloes born with a particle number exceeding the threshold:

\begin{equation}
    N^{\mathrm{birth}}_{\mathrm{part,subhalo}}>\mathrm{max}(10N^{\mathrm{min}}_{\mathrm{part,subhalo}},3\sigma(N^{\mathrm{birth}}_{\mathrm{part,subhalo}},z))
\label{eq:4}
\end{equation}

\noindent which means the particle number must be either above three times the standard deviation at each snapshot or exceed one order of magnitude above the minimum particle number required to detect a subhalo. This approach helps avoid considering very low values for the standard deviation at high-$z$ when the Universe is not dense enough to form very massive objects. 

We tested other arbitrary threshold values for the particle number at birth, such as $100$, $200$, $1000$ or $3\sigma(N^{\mathrm{birth}}_{\mathrm{part,subhalo}})$ globally regardless of the snapshot. The results were practically the same in all these variations. Therefore, we prefer a $z$/time-dependant threshold that has a more meaningful physical interpretation.

\begin{figure*}
\vspace{0.5cm}
\centering 
\includegraphics[trim={0.1cm 0.1cm 0.1cm 0.1cm},clip,width=0.5\linewidth]{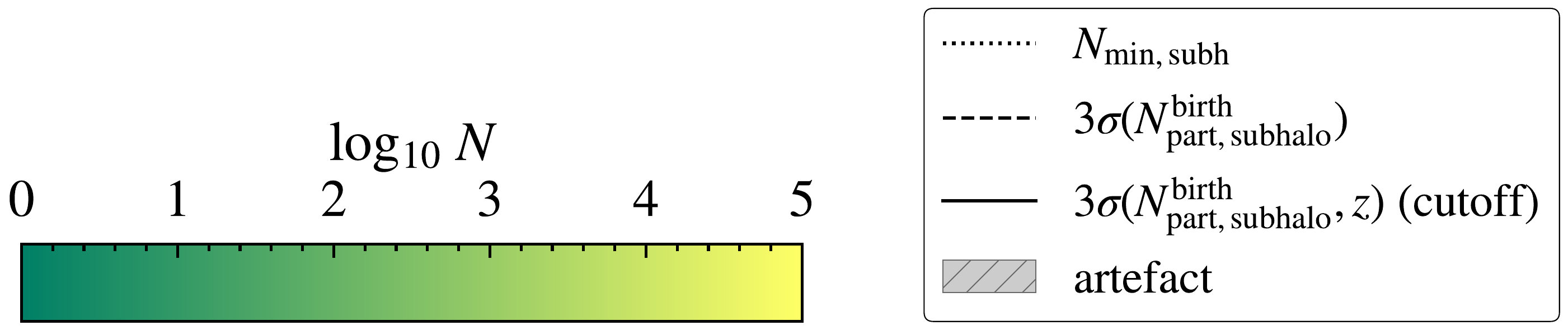}
\\[0.3cm]
\medskip
\captionsetup[subfigure]{labelfont=bf,font=large}
\begin{subfigure}{0.355\textwidth}
  \includegraphics[trim={0.1cm 0.1cm 0.1cm 0.1cm},clip,width=\linewidth]{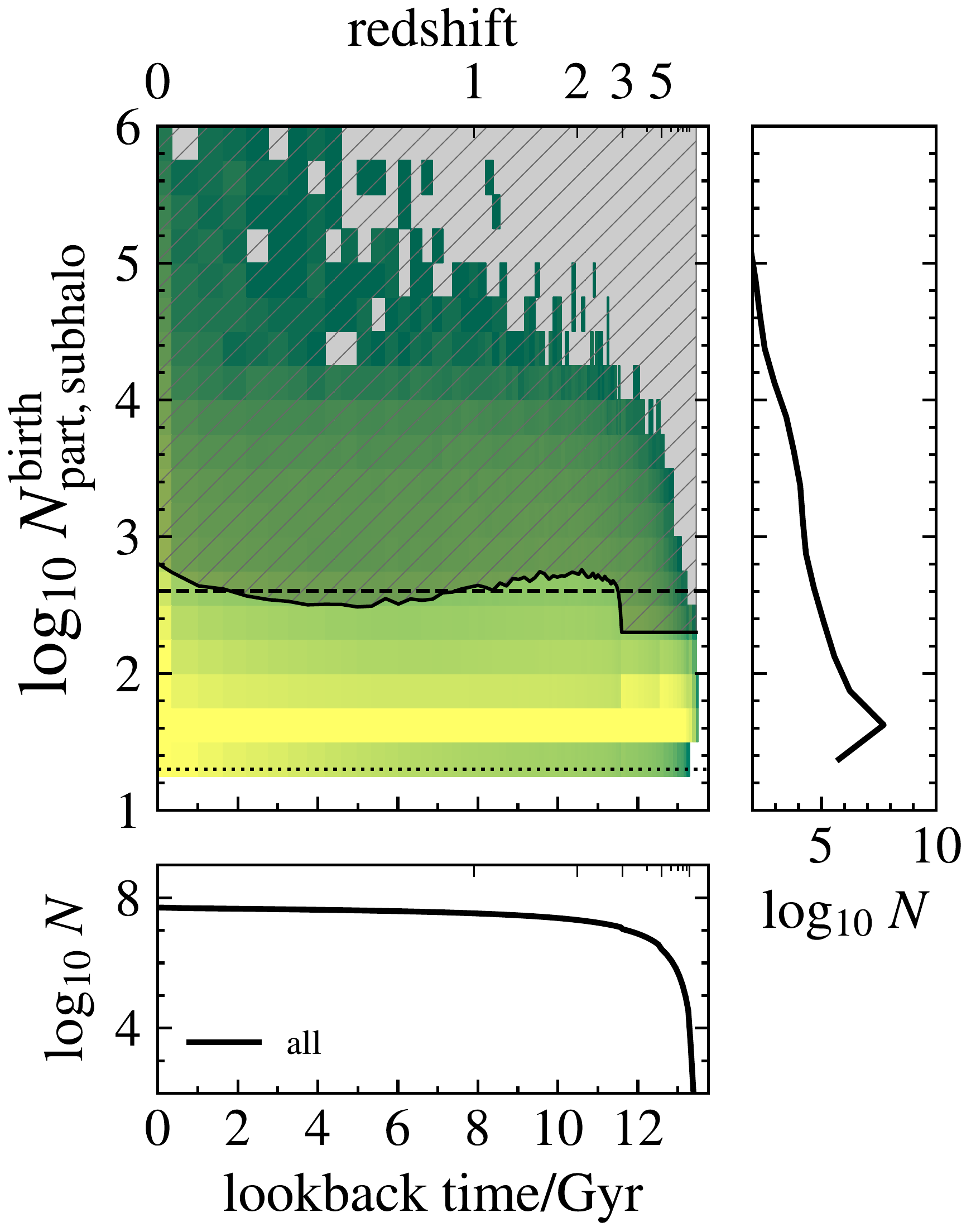}
  \caption{\textbf{\textsc{FLAM-DM-VR}}}
  \label{fig:mtdef1}
\end{subfigure}\hfil 
\begin{subfigure}{0.305\textwidth}
  \includegraphics[trim={0.1cm 0 0 0},clip,width=\linewidth]{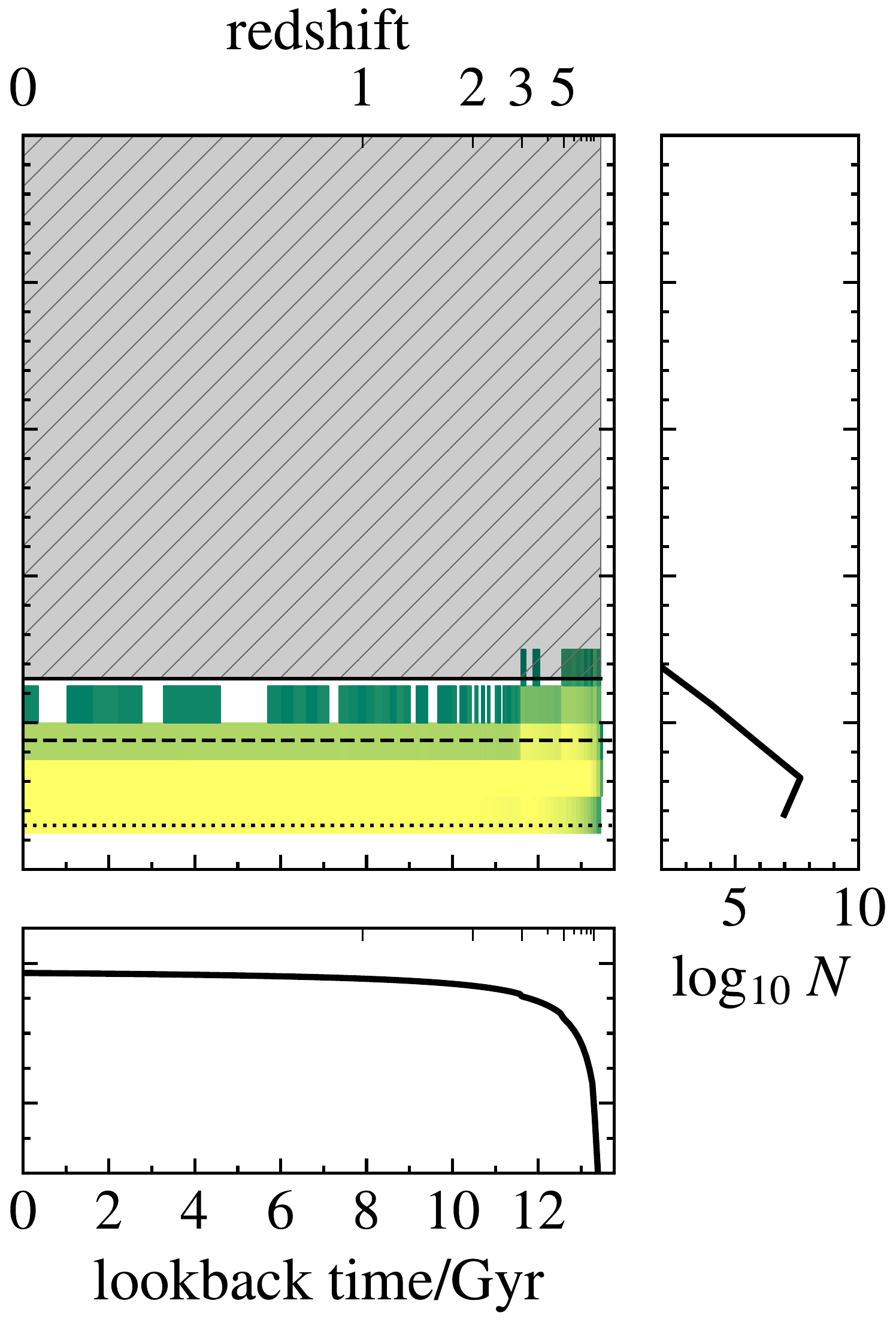}
  \caption{\textbf{\textsc{FLAM-DM-HBT}}}
  \label{fig:mtdef2}
\end{subfigure}\hfil 
\begin{subfigure}{0.305\textwidth}
  \includegraphics[trim={0.1cm 0 0 0},clip,width=\linewidth]{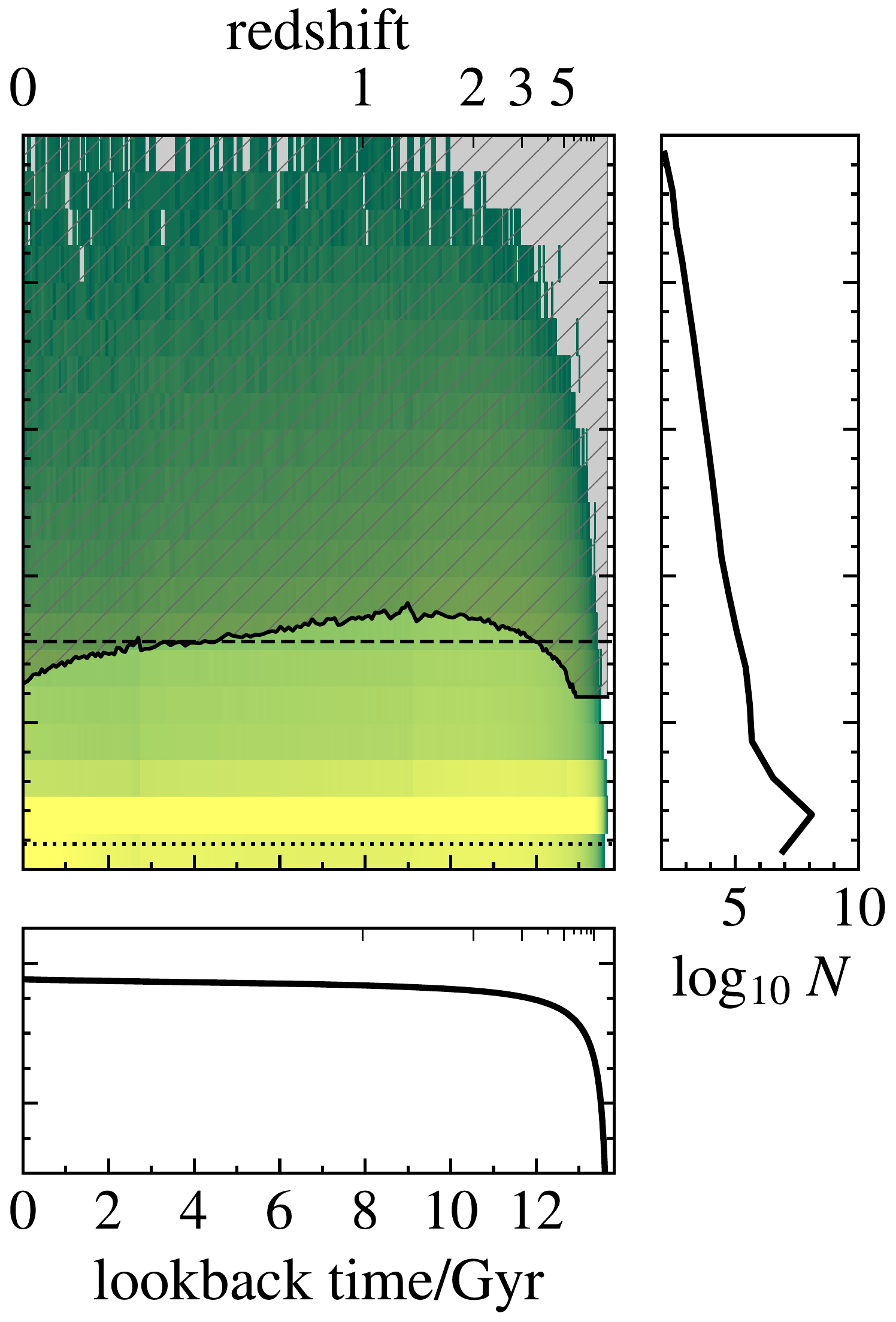}
  \caption{\textbf{\textsc{PMill}}}
  \label{fig:mtdef3}
\end{subfigure}
\\[0.3cm]
\medskip
\begin{subfigure}{0.355\textwidth}
  \includegraphics[trim={0.1cm 0.1cm 0.1cm 0.1cm},clip,width=\linewidth]{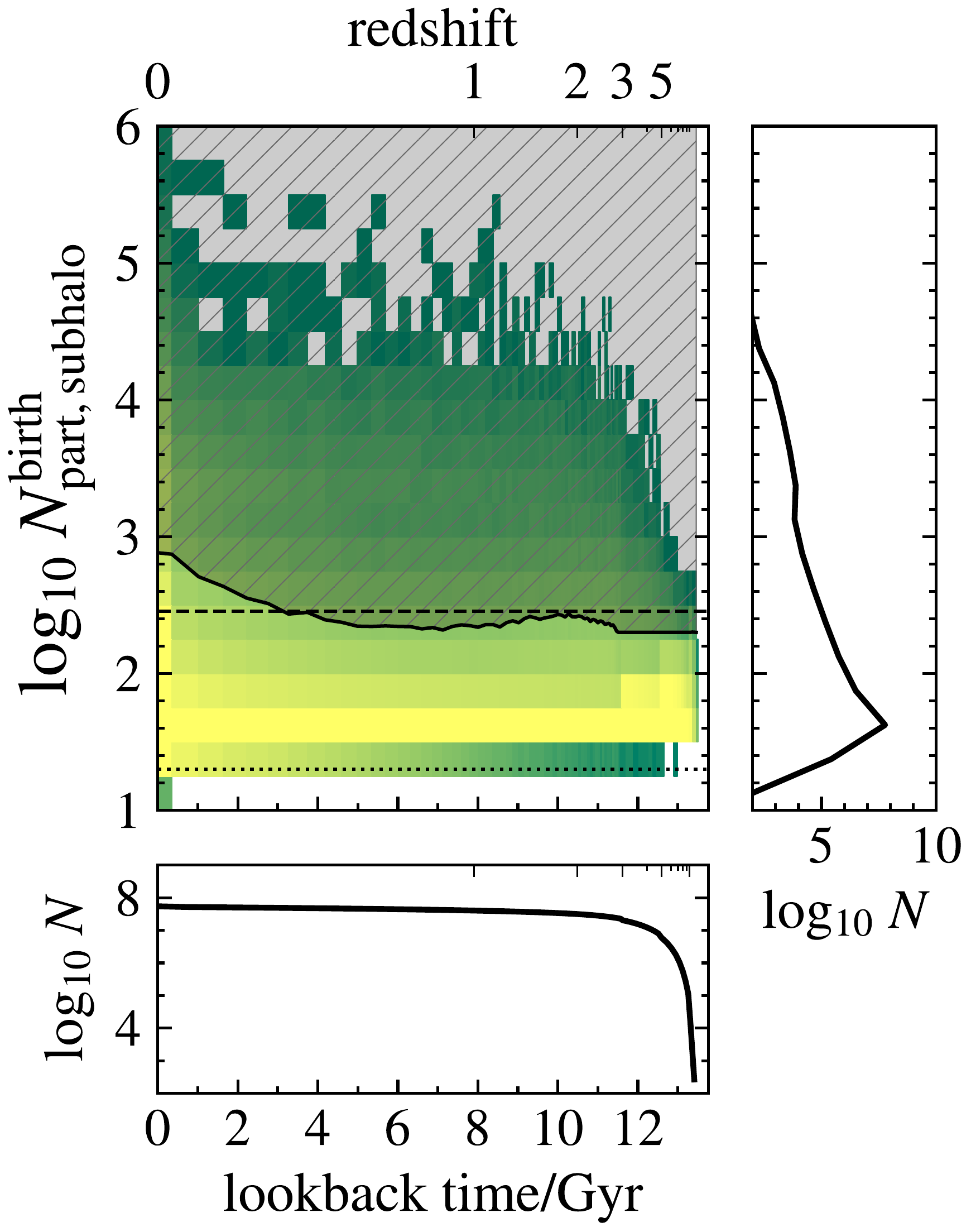}
  \caption{\textbf{\textsc{FLAM-Hydro-VR}}}
  \label{fig:mtdef4}
\end{subfigure}\hfil 
\begin{subfigure}{0.305\textwidth}
  \includegraphics[trim={0.1cm 0 0 0},clip,width=\linewidth]{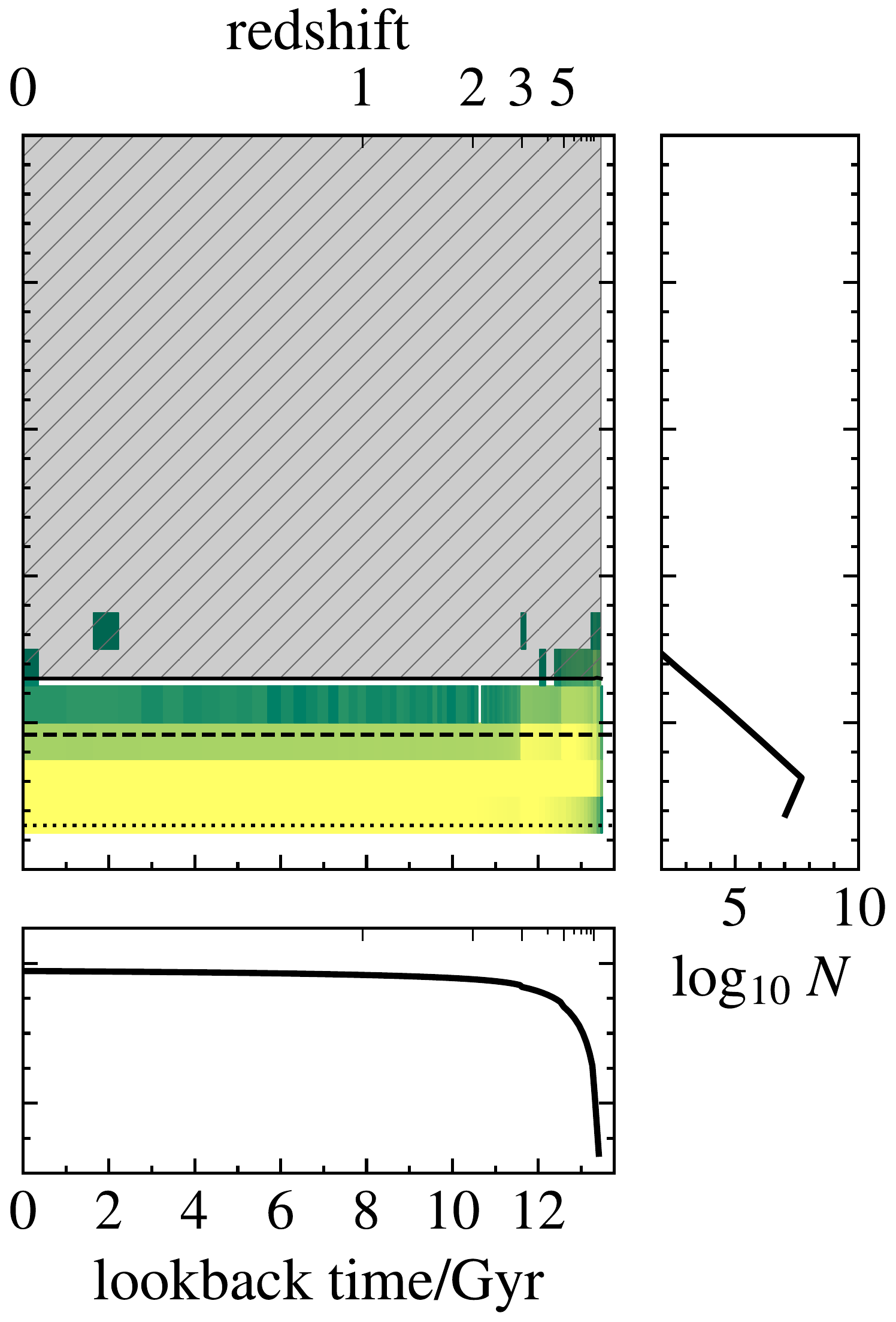}
  \caption{\textbf{\textsc{FLAM-Hydro-HBT}}}
  \label{fig:mtdef5}
\end{subfigure}\hfil 
\begin{subfigure}{0.305\textwidth}
  \includegraphics[trim={0.1cm 0 0 0},clip,width=\linewidth]{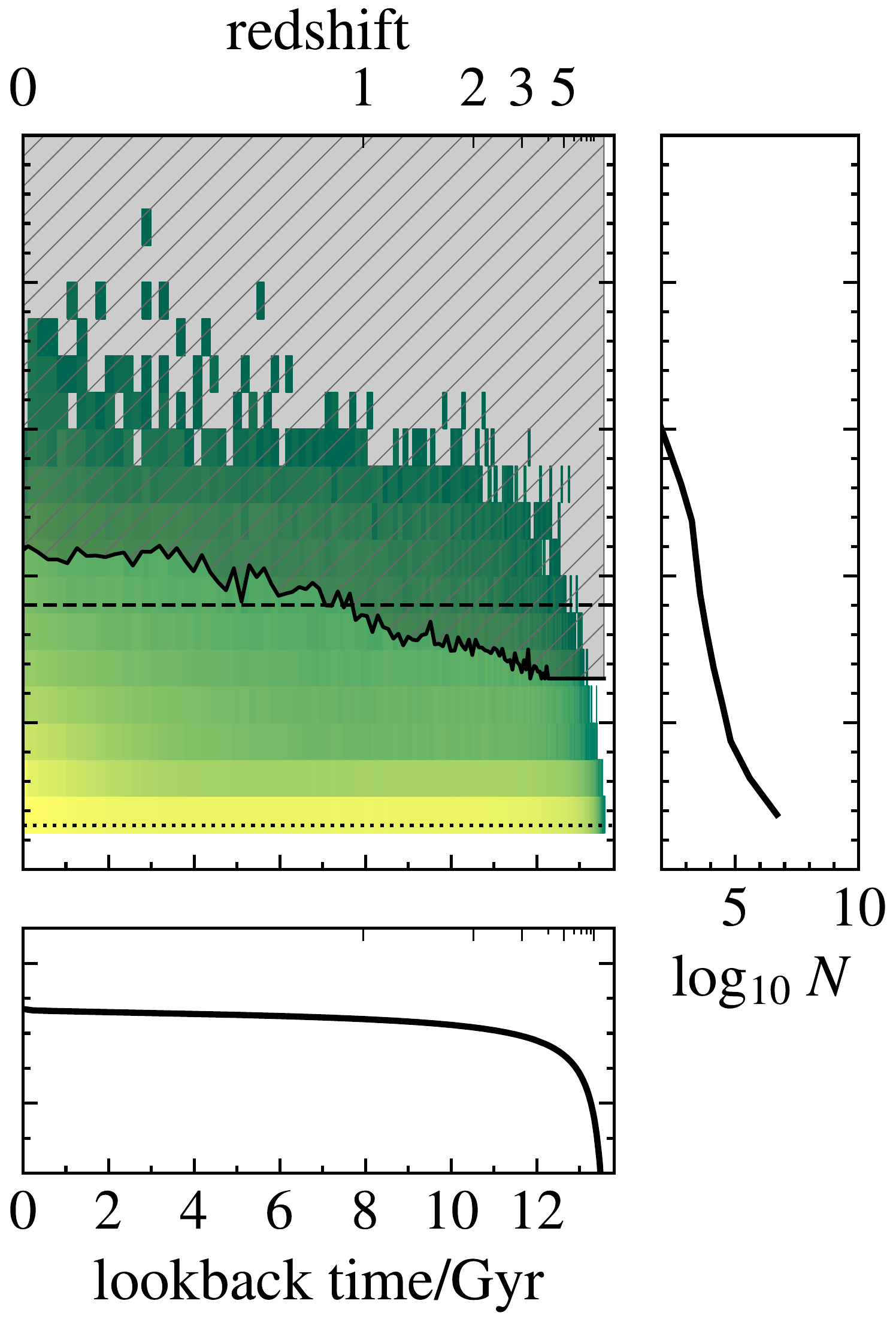}
  \caption{\textbf{\textsc{medi-SURFS}}}
  \label{fig:mtdef6}
\end{subfigure}
\\[0.3cm]
\caption{2D histograms of the particle number with which main branches are born in the simulation as a function of the lookback time/redshift at which the birth is happening, for the 6 merger tree catalogues described in Table~\ref{tab:cat}. The main panel is colour-coded by the number of counts within each 2D bin, following the colour bar at the top. The shaded area marks the numerical artefacts based on the threshold $3\sigma(N^{\mathrm{birth}}_{\mathrm{part,subhalo}},z)$ defined in equation~(\ref{eq:4}) (solid line). A dotted line represents the minimum particle number for a subhalo to be detected ($N^{\mathrm{min}}_{\mathrm{part,subhalo}}$), whereas a dotted line shows $3\sigma(N^{\mathrm{birth}}_{\mathrm{part,subhalo}})$ globally. On the right and bottom panels, we show the 1D projected histograms for the mass ratio and the lookback time/redshift, respectively.}
\label{fig:mt-def-panel}
\end{figure*}


\subsubsection{Characterisation}
\label{sssec:massive-transients-calc} 

Having established the transient definition, we examine their time distribution in Fig.~\ref{fig:mt-def-panel} for all catalogues. We generate a 2D histogram of the particle number at birth for all main branches in the catalogues as a function of the lookback time/redshift. Subhaloes that fall within the shaded area in this panel are considered transients according to our definition, with the solid black line indicating the boundary defined in equation~(\ref{eq:4}). For comparison, the black dashed line represents $3\sigma(N^{\mathrm{birth}}_{\mathrm{part,subhalo}})$ globally across the entire redshift range, dominated by the large number of subhaloes at low-$z$. Additionally, a dotted black line marks the minimum particle number required to detect a subhalo. 1D projected histograms for the particle number and the lookback time/redshift are included on the right and bottom panels, respectively. 

Across all catalogues, we find most subhaloes are born with $\approx 10-10^{2}$ particles and massive transient occurrences are more common at smaller lookback times, where there are very dense environments and the post-processing codes used to find and track bound systems encounter difficulties \citep{behroozi15}. The trends for the massive transients are similar to those observed for mass-swapping. More extreme cases are found in the {\sc PMill}, {\sc FLAM-DM-VR} and {\sc FLAM-Hydro-VR} catalogues. In {\sc PMill}, a significant number of subhaloes emerge with around $10^{6}$ particles or more at lookback times $\rm <10~Gyr$, while in {\sc FLAM-DM-VR} and {\sc FLAM-Hydro-VR} the most extreme transients have $~10^{4}$ particles, with a few cases with higher particle numbers, but they only appear at lookback times $\rm <8~Gyr$. {\sc Subfind} exhibits more transients than {\sc VELOCIraptor} as configuration-based codes struggle to distinguish very nearby objects. If subhaloes are temporarily split off from the bulk of the mass, {\sc D-Trees}, which track the most bound particles of subhaloes, may mistakenly identify a new massive subhalo with no progenitors.

The {\sc medi-SURFS} catalogue shows similar results, though less extreme: subhaloes appear with around $10^{3.5}$ particles in the worst cases. {\sc D-Trees} and {\sc TreeFrog} may skip snapshots to connect subhaloes across time when it is unnecessary. Interestingly, the dashed line indicating $3\sigma(N^{\mathrm{birth}}_{\mathrm{part,subhalo}})$ is higher than in other runs, suggesting that subhaloes tend to form with a larger variety of particle numbers in this catalogue.
The most notable result comes from the {\sc HBT}-related catalogues, where massive transient subhaloes are rare and flagged subhaloes have roughly $10^{2}$ particles. Most subhaloes appear with a reasonable particle number, highlighting that {\sc HBT-HERONS} effectively reduces this artefact thanks to its history-space algorithm, as pointed out in previous studies \citep{srisawat13,avila14,behroozi15,wang16}.

\begin{table*}
	\caption{Statistics of numerical artefacts. $N_{\rm mb}$: the number of main branches or central subhaloes at $z=0$ whose merger trees we track; $\tilde{x}$($M_{\rm subh}^{\rm max}$): median maximum mass for all main branches in M$_{\odot}$; $\tilde{x}$($N_{\rm p,subh}^{\rm max}$): median maximum particle number for all main branches; $f_{\rm s}$: fraction of main branches affected by mass-swapping in \%; $\tilde{x}$($M_{\rm subh,s}^{\rm max}$): median maximum mass for main branches affected by mass-swapping in M$_{\odot}$; $\tilde{x}$($N_{\rm p,subh,s}^{\rm max}$): median maximum particle number for main branches affected by mass-swapping; $f_{\rm t}$: fraction of main branches affected by massive transients in \%; $\tilde{x}$($M_{\rm subh,t}^{\rm max}$): median maximum mass for main branches affected by massive transients in M$_{\odot}$; $\tilde{x}$($N_{\rm p,subh,t}^{\rm max}$): median maximum particle number for main branches affected by massive transients; $f_{\rm art}$: fraction of main branches affected by either mass-swapping or massive transients in \%.}
	\label{tab:statistics}
	\begin{tabular}{ccccccccccc} 
		\hline
		\multirow{2}{*}{catalogue} & \multirow{2}{*}{$N_{\rm mb}$} & $\tilde{x}$($M_{\rm subh}^{\rm max}$) & \multirow{2}{*}{$\tilde{x}$($N_{\rm p,subh}^{\rm max}$)} & $f_{\rm s}$ & $\tilde{x}$($M_{\rm subh,s}^{\rm max}$) & \multirow{2}{*}{$\tilde{x}$($N_{\rm p,subh,s}^{\rm max}$)} & $f_{\rm t}$ & $\tilde{x}$($M_{\rm subh,t}^{\rm max}$) & \multirow{2}{*}{$\tilde{x}$($N_{\rm p,subh,t}^{\rm max}$)} & $f_{\rm art}$ \\
        & & /M$_{\odot}$ & & /\% & /M$_{\odot}$ & & /\% & /M$_{\odot}$ & & /\% \\
		\hline
        {\sc FLAM-DM-VR} & 54348687 & 6.6$\times$10$^{10}$ & 78 & 9.7 & 3.2$\times$10$^{11}$ & 383 & 0.13 & 4.0$\times$10$^{12}$ & 4725 & 9.8\\
        {\sc FLAM-Hydro-VR} & 63442932 & 5.1$\times$10$^{10}$ & 72 & 10 & 1.9$\times$10$^{11}$ & 266 & 0.10 & 1.3$\times$10$^{12}$ & 1864 & 10\\
        {\sc FLAM-DM-HBT} & 53941906 & 5.3$\times$10$^{10}$ & 63 & 1.8 & 1.4$\times$10$^{11}$ & 162 & 7.4$\times$10$^{-5}$ & 6.1$\times$10$^{13}$ & 72301 & 1.8\\
        {\sc FLAM-Hydro-HBT} & 61092496 & 4.0$\times$10$^{10}$ & 48 & 2.5 & 8.8$\times$10$^{10}$ & 104 & 3.5$\times$10$^{-4}$ & 4.3$\times$10$^{13}$ & 51640 & 2.5\\
        {\sc PMill} & 4295458 & 7.8$\times$10$^{9}$ & 50 & 12 & 3.3$\times$10$^{10}$ & 211 & 0.15 & 5.5$\times$10$^{11}$ & 3545 & 12\\
        {\sc medi-SURFS} & 5311603 & 1.8$\times$10$^{10}$ & 55 & 12 & 5.4$\times$10$^{10}$ & 166 & 0.14 & 1.5$\times$10$^{12}$ & 4589 & 12\\
		\hline
	\end{tabular}
\end{table*}

The thresholds defined in equation~(\ref{eq:4}) generally show larger values at lower lookback times. However, the {\sc PMill} catalogue deviates from this pattern, as there is a significant population of subhaloes born with a high number of particles around lookback times $\rm \approx 8-10~Gyr$, which then decreases up to the present day. A similar population is observed in {\sc FLAM-DM-VR}, though it is overshadowed by the presence of massive structures born at very low-$z$. This may indicate an increase in merging activity around $z\approx 1-2$, which agrees with results from other simulations showing that merger activity peaks at that cosmic time \citep{bertone09}. Besides, looking at the 1D histograms on the right side, it is interesting that the preferred particle number at birth for subhaloes usually falls in the second least massive bin rather than the smallest, except for {\sc medi-SURFS}.

In the catalogues involving hydrodynamics, the 2D maps in Fig.~\ref{fig:mt-def-panel} look very similar to the DM-only cases, with the exception that the {\sc FLAM-DM-VR} case has slightly more extreme transients than {\sc FLAM-Hydro-VR}. The discrepancy may arise from the weaker gravitational potentials in the DM-only case, which can cause a structure to split between the core and its outskirts \citep[ejected cores in fig.~1 in][]{poole17}. 

\subsubsection{Quantification}
\label{sssec:massive-transients-quant} 

As a result of defining this numerical artefact, we can quantify how transients impact each of the catalogues. Table~\ref{tab:statistics} summarises the number fraction of main branches affected by this issue: values are around $0.1\%$ for all catalogues, except for the {\sc HBT}-related catalogues, which exhibit significantly better performance with fractions several orders of magnitude lower. While overall the problem of massive transients is less severe than that of mass-swapping, the main branches impacted by transient events tend to be the most massive ones, as indicated by the median maximum particle number for the affected branches (above $10^{3}$ particles). Thus, it is crucial to consider them when analysing massive structures. These most massive branches are likely impacted because numerical misidentification occurs in the densest environments, which are more susceptible to major merger events \citep{behroozi15}. 

Less massive subhaloes, which are more often affected by mass-swapping, seem to avoid this issue. However, there is a bias by construction introduced by the definition of transients as massive structures. The most extreme case is found in the {\sc HBT}-related catalogues, where the median maximum particle number along the main branch evolution exceeds $10^{4}$ particles. Most of these {\sc HBT-HERONS} subhaloes initially emerge in the simulation with a relatively low particle number ($\approx200$), as shown in the Fig.~\ref{fig:mt-def-panel} middle column panels at $z>3$. Over time, however, they grow in mass accordingly, eventually reaching a much higher particle number. 

\begin{figure*}
\centering
\includegraphics[width=1\linewidth]{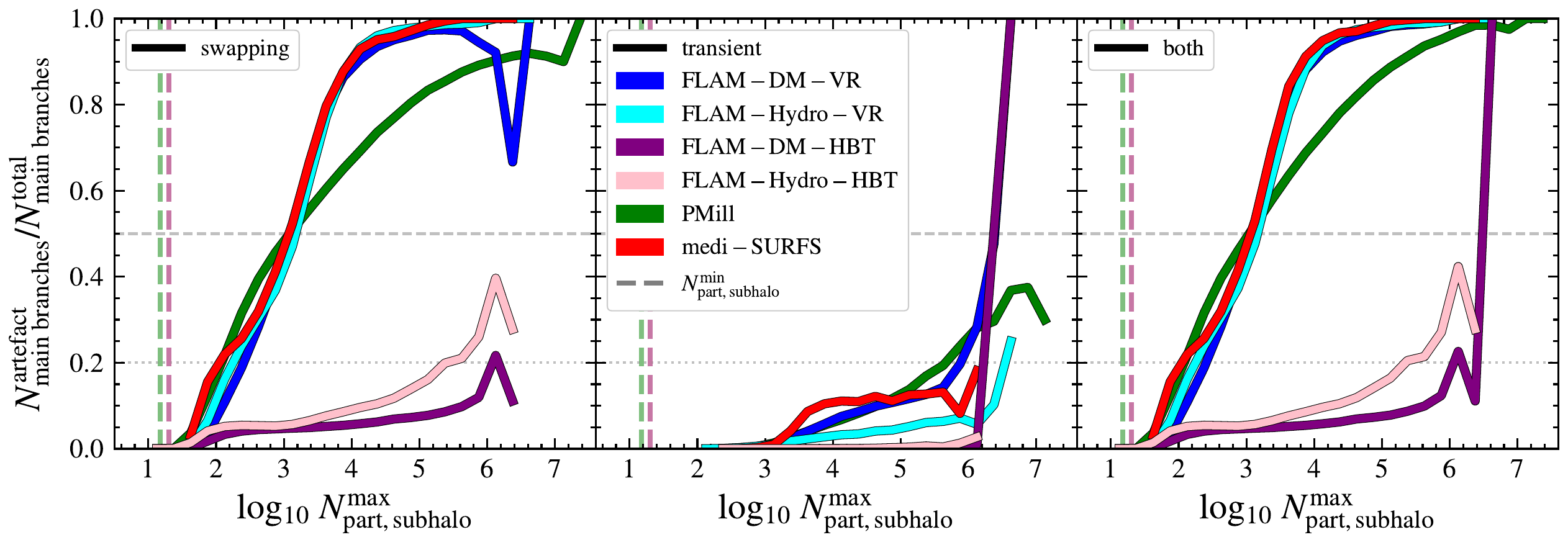}
\caption[caption]{\textit{Left panel}: fraction of main branches affected at least by 1 mass-swapping event during their lifetime as a function of the maximum subhalo particle number along the main branch \footnotemark. Each colour represents a different merger tree catalogue, as labelled, while the vertical lines mark the minimum particle number for a structure in each catalogue (usually around $20$ particles). \textit{Middle panel}: fraction of main branches defined as massive transients as a function of the maximum subhalo particle number along the main branch. \textit{Right panel}: fraction of main branches affected either by mass-swapping or massive transients as a function of the maximum subhalo particle number along the main branch.} 
\label{fig:diagnostics-part} 
\end{figure*}

The central panel of Fig.~\ref{fig:diagnostics-part} describes the number fraction of main branches affected by transients as a function of the maximum number of particles along the main branch. The same comparison is illustrated in Fig.~\ref{fig:diagnostics} as a function of the subhalo mass. Again, it is evident that this second type of artefact is less severe than the mass-swapping, but remains significant for the more massive subhaloes. It affects around $20\%$ of the main branches at the high-mass end for the {\sc PMill}, {\sc FLAM-DM-VR} and {\sc medi-SURFS} catalogues; while it has a little effect on the {\sc HBT}-related catalogues. The abrupt jump in the {\sc FLAM-DM-HBT} curve around $10^{6}$ particles 
is due to the small number of statistics — only one subhalo falls within that bin (this subhalo emerges early in time at $z\approx10$ with about $200$ particles, just above the defined threshold, and then grows in mass accordingly). The results for {\sc FLAM-Hydro-VR} fall between these 2 regimes.

The {\sc PMill} and {\sc FLAM-DM-VR} catalogues exhibit similar curves, both making use of the {\sc D-Trees}+{\sc DHalo} tree-building codes \citep[transients have been reported for this combination of codes in fig.~8 in][]{wang16}. This could suggest that the tree builder primarily addresses this issue. Algorithms with patching mechanisms which link structures separated by more than one snapshot in time alleviate these transient cases \citep{srisawat13}, although they are still present. Plus, the {\sc medi-SURFS} catalogue demonstrates a distinct performance, likely due to its different tree builder (the halo finder used is {\sc VELOCIraptor} as in {\sc FLAM-DM-VR}). Overall, it is clear that {\sc HBT-HERONS} manages these structures more effectively.

Examining the diagnostics plots in Fig.~\ref{fig:diagnostics-part} or the statistics in Table~\ref{tab:statistics}, it is apparent that comparing the {\sc FLAM-DM-VR} catalogue to the {\sc FLAM-Hydro-VR} one, the hydrodynamical run suffers less from this second artefact. The increased clustering of particles produces a deeper gravitational potential and less dispersed material that can be ejected from the core of the central subhalo. In that case, it appears that the transients are improved when adding baryons. Conversely, the {\sc HBT}-related catalogues have the opposite behaviour, with the hydrodynamical run being more affected. This is, in principle, related to its history-based method of tracking particles.

Table~\ref{tab:statistics} also presents the fraction of main branches affected by either of the two artefacts investigated — mass-swapping and massive transients —, while the right panel of Fig.~\ref{fig:diagnostics-part} exhibits the fraction of main branches affected by either artefact as a function of the maximum subhalo particle number. The results closely resemble those for mass-swapping, as transients affect a very small number of branches. In summary, these numerical issues affect about 10\% of all the main branches in non-history-based codes, whereas the fraction is reduced to about 2\% for {\sc HBT}-related catalogues. The fraction increases significantly at the high-mass end, reaching over 50\% for subhaloes with more than $10^{3}$ particles in the former and around 20\% for the most massive subhaloes in {\sc HBT-HERONS}. It is important to note that these two types of artefacts are connected, as the results do not change substantially when considering them together. Main branches flagged as problematic are often affected by both of them simultaneously. As noted in previous works \citep[e.g.][]{srisawat13,avila14}, the numerical artefacts studied in this paper can not be attributed to a single post-processing step, rather they result from the combination of halo/subhalo-finding and tree-building.

In general, resolution plays a critical role in all of the results presented. Artefacts typically arise due to the difficulty of assigning particles to structures within very dense environments. In such a way, increasing the particle resolution would push these issues to earlier snapshots in time, since these higher-density conditions would form earlier, challenging post-processing codes even further. That is the reason why we present our results as a function of the number of particles instead of mass to remove this dependency. In Appendix~\ref{assec:small-box-mass} we analyse small cosmological boxes to prove that the numerical issues arise in subhaloes with above $10^{3}$ particles, independently of the mass resolution and the number of snapshots. This shows that our analysis is not biased by the use of simulations with varying features. Furthermore, in Appendix~\ref{assec:small-box-analysis}, we use these same test boxes to show that numerical artefacts are more likely to arise in structures situated in denser environments. By analysing environmental and merger history metrics, we demonstrate that affected subhaloes reside in denser regions and have more complex merger histories.

\footnotetext{For the hydrodynamical runs $N^{\rm max}_{\rm part,\ subhalo}\approx M^{\rm max}_{\rm part,\ subhalo}/m_{\rm part}$, where the maximum subhalo mass is accounting for all the particles independently of their nature (DM or baryons), while $m_{\rm part}$ is the DM particle mass. In such a way, $N^{\rm max}_{\rm part,\ subhalo}$ is an approximate estimate that works as a lower limit since the baryonic particles are less massive.}

\subsection{Impact on galaxy formation models}
\label{ssec:impact-galaxy-formation-models}

After characterising and quantifying the numerical artefacts affecting merger trees in a range of catalogues, we evaluate the impact they have on the predictions made by SAMs. To do this, we run both {\sc Shark} and {\sc Galform} on the highly artificially affected {\sc FLAM-DM-VR} catalogue to determine whether these artefacts propagate into the galaxy predictions. Since each code computes galaxy properties based on different merger tree information ({\sc Shark} uses primarily subhalo data, while {\sc Galform} relies on host halo information), this analysis allows us to evaluate how varying baryonic treatments influence the propagation of these artefacts. We focus on main branches that had been flagged as suffering from mass-swapping or transient events and, for simplicity and efficiency, we select the central galaxies hosted by the problematic main branches.

Central galaxies are identified at the final snapshot of the simulation, as hosted by the corresponding same {\sc DHalo} affected central subhaloes. Then, we track their IDs backwards in time using the respective main branch definitions of each SAM. For {\sc Shark}, this involves selecting the most massive subhalo at the final snapshot and imposing central hierarchy across all {\sc DHalo} main progenitors (\S~\ref{sssec:shark}). In contrast, {\sc Galform} designates as central the subhalo with the highest cumulative mass in its merger tree and assigns its main progenitors as the host haloes main progenitors (\S~\ref{sssec:galform}).

We particularly focus on critical key galaxy properties that could be affected by unexpected changes in the halo properties. In this case, we examine the results related to stellar, BH and gas properties to determine if the artefacts impact the predictions at various scales. In such a way, we track the gas disc size evolution, star formation histories (SFHs) and BH accretion histories (BHAHs) for those central galaxies hosted by problematic main branches following main progenitors back in time. 

\begin{figure*}
\centering
\captionsetup[subfigure]{labelformat=empty}
\captionsetup[subfigure]{font=Large}
\begin{subfigure}{.5\textwidth}
  \centering
  \caption{\textbf{\textsc{Shark} swapping}}
  \par\medskip
  \includegraphics[trim={0 0cm 0 0cm},clip,width=0.92\textwidth]{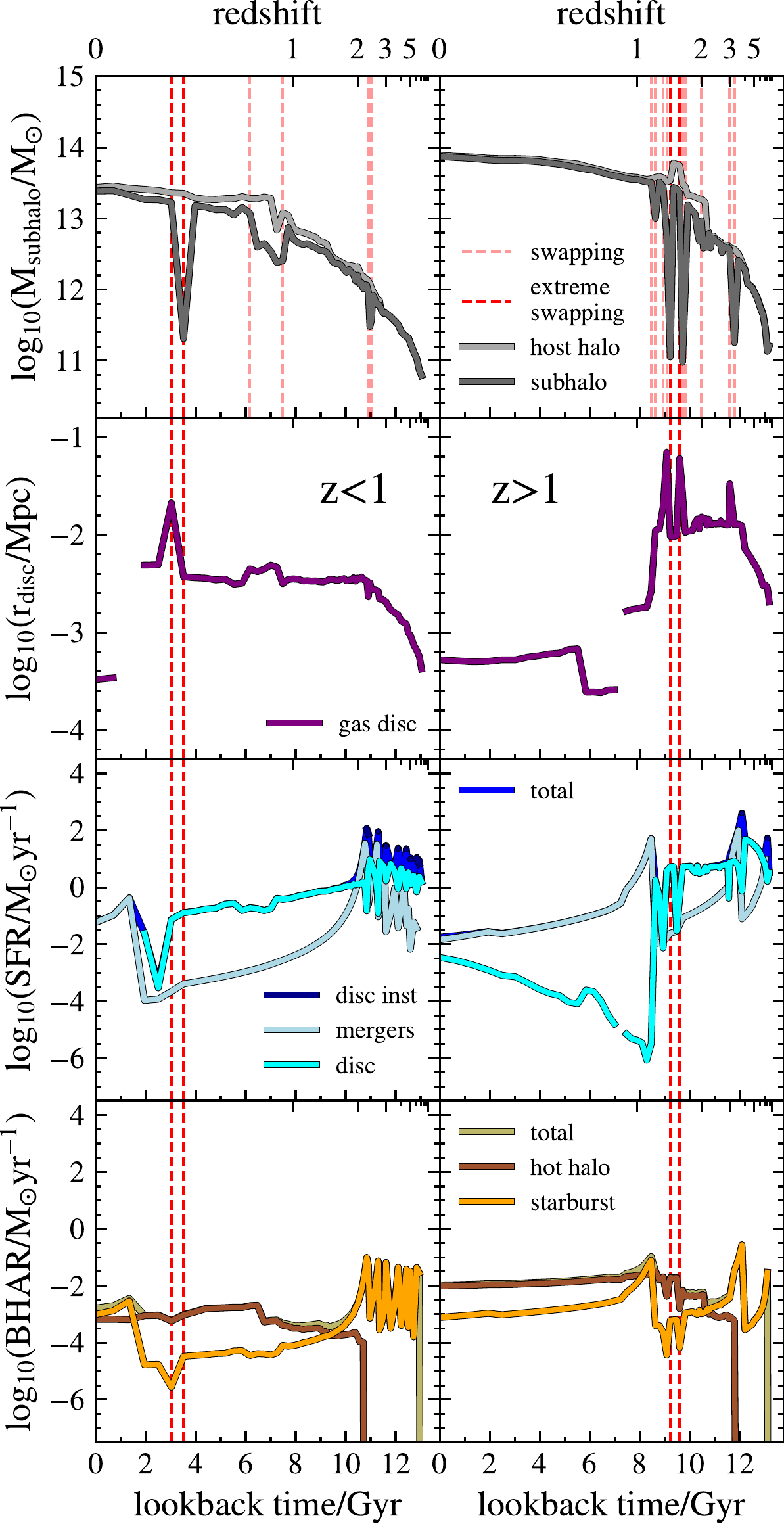}
\end{subfigure}%
\begin{subfigure}{.5\textwidth}
  \centering
  \caption{\textbf{\textsc{Galform} swapping}}
  \par\medskip
  \includegraphics[trim={0 0cm 0 0cm},clip,width=0.92\textwidth]{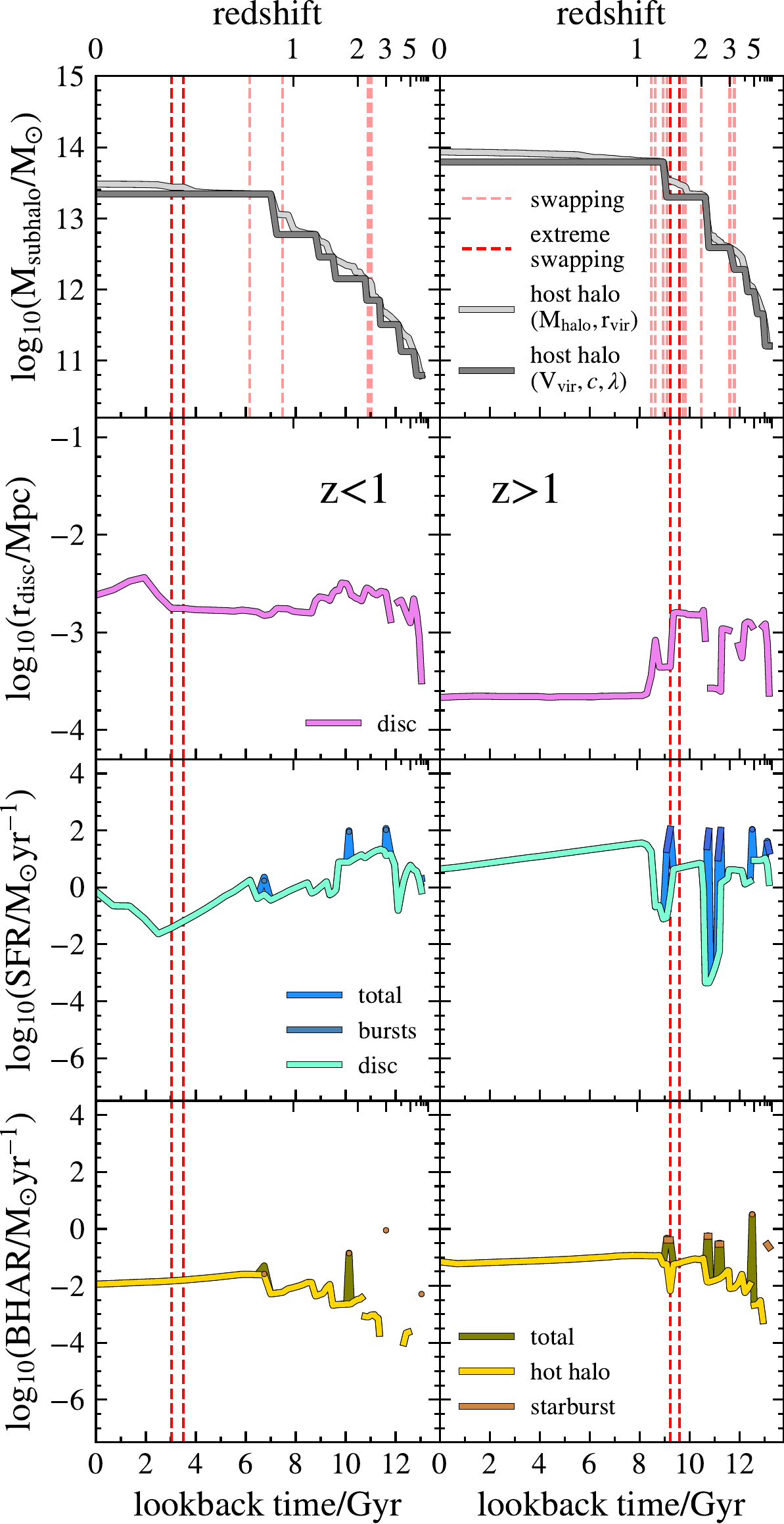}
\end{subfigure}
\caption{SAM predictions from {\sc Shark} (left panel) and {\sc Galform} (right panel) for two examples of main branches affected by extreme mass-swapping events. \textit{First row}: Subhalo (``halo formation'' event) and halo (constrained halo) MAHs in dark and light grey for {\sc Shark} ({\sc Galform}) with the specific snapshots at which the swapping events happen marked with red dashed vertical lines, with the most extreme cases in a more opaque red: in the 1st column the extreme event occurs at $z<1$, while in the 2nd column it occurs at $z>1$. \textit{Other rows}: cold gas disc size (second row), SFHs (third row) and BHAHs (fourth row) for the central galaxies associated with the main branches in the top panels.}
\label{fig:ms-sam}
\end{figure*}

\begin{figure*}
\centering
\captionsetup[subfigure]{labelformat=empty}
\captionsetup[subfigure]{font=Large}
\begin{subfigure}{.5\textwidth}
  \centering
  \caption{\textbf{\textsc{Shark} transients}}
  \par\medskip
  \includegraphics[trim={0 0cm 0 0cm},clip,width=0.92\textwidth]{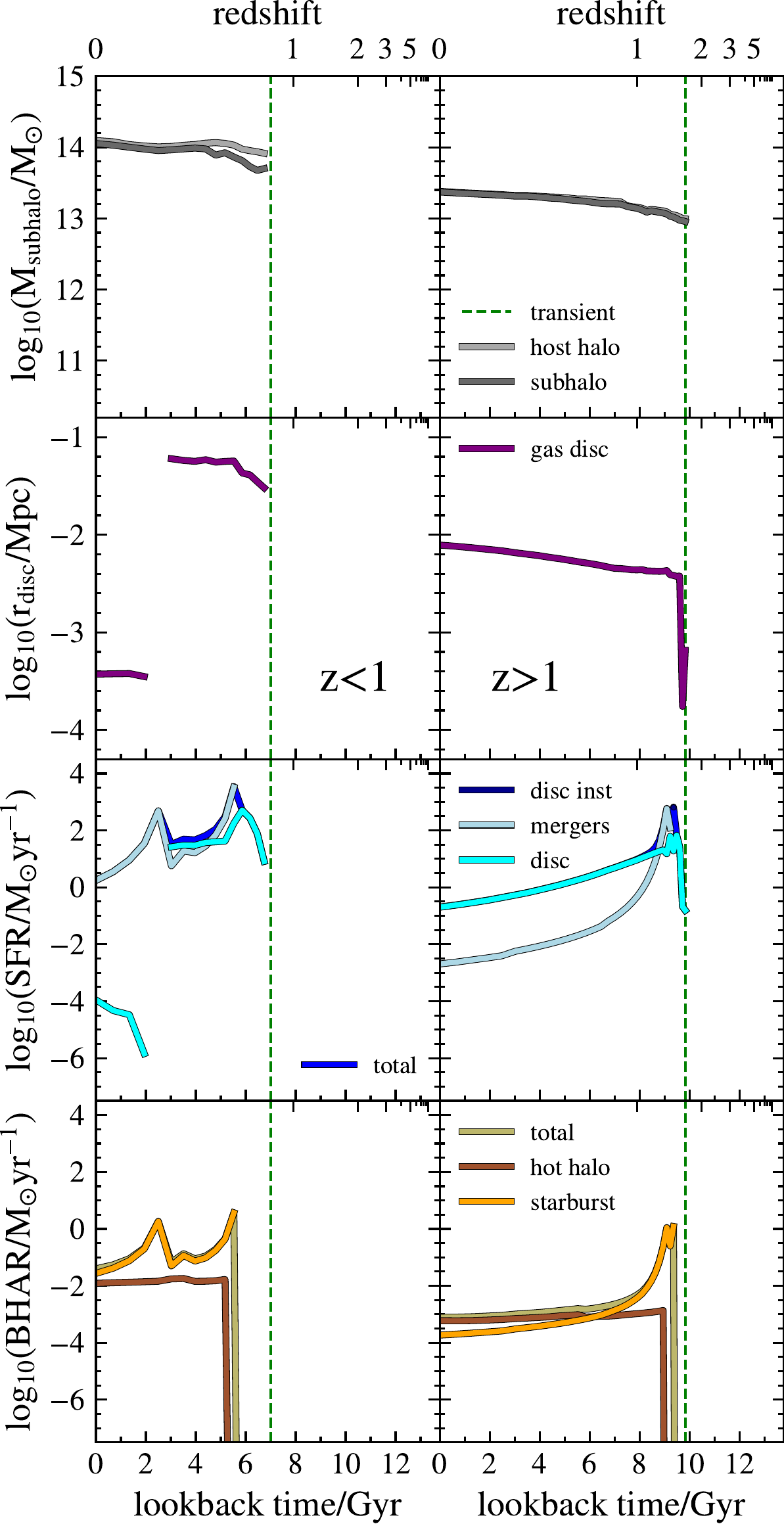}
\end{subfigure}%
\begin{subfigure}{.5\textwidth}
  \centering
  \caption{\textbf{\textsc{Galform} transients}}
  \par\medskip
  \includegraphics[trim={0 0cm 0 0cm},clip,width=0.92\textwidth]{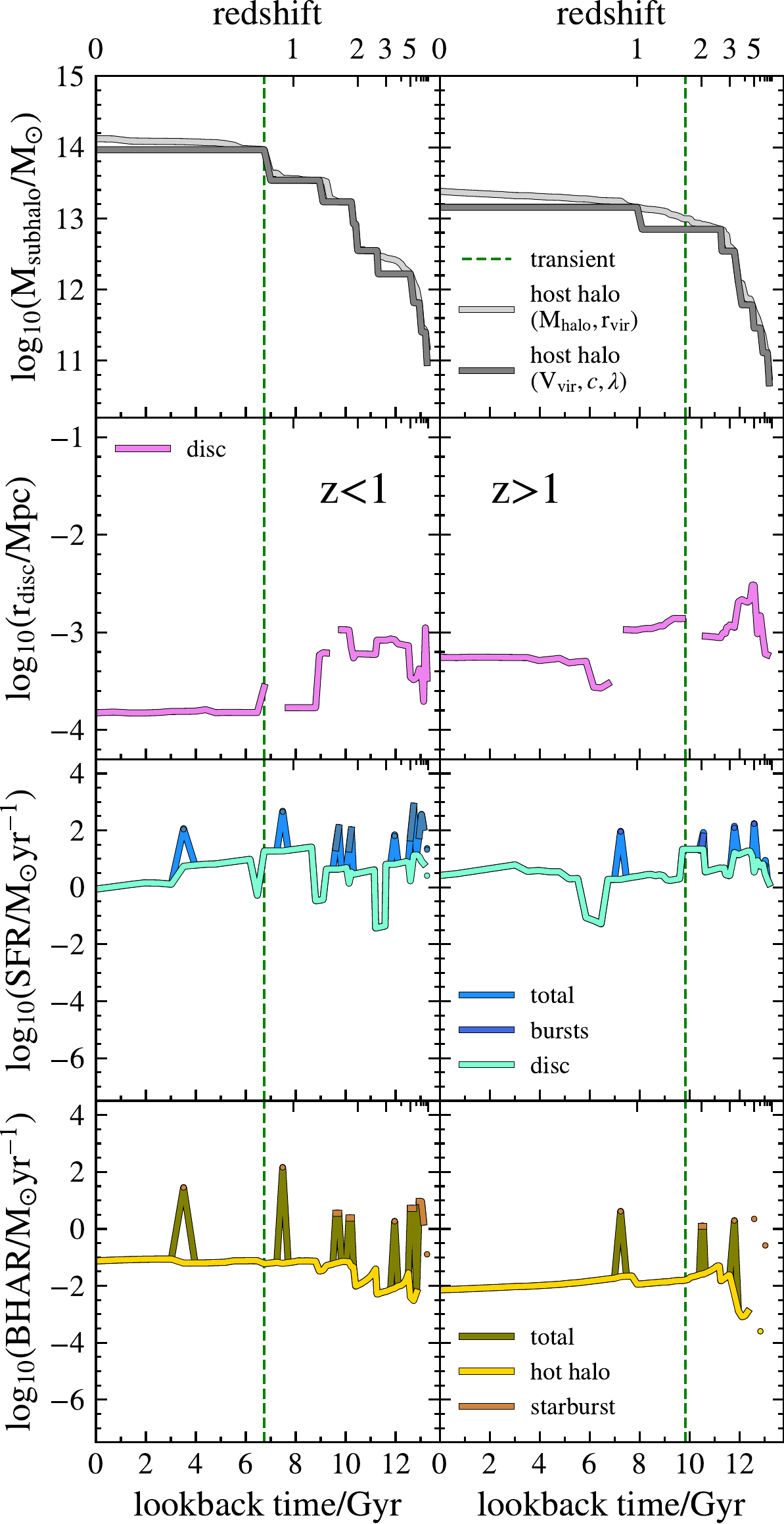}
\end{subfigure}
\caption{SAM predictions from {\sc Shark} (left panel) and {\sc Galform} (right panel) for main branches defined as extreme massive transients. \textit{First row}: Subhalo (``halo formation'' event) and halo (constrained halo) MAHs in dark and light grey for {\sc Shark} ({\sc Galform}) with the specific snapshots at which the transient are born marked with green dashed vertical lines: in the 1st column the extreme event occurs at $z<1$, while in the 2nd column it occurs at $z>1$. \textit{Other rows}: cold gas disc size (second row), SFHs (third row) and BHAHs (fourth row) for the central galaxies associated with the main branches in the top panels.}
\label{fig:mt-sam}
\end{figure*}

\subsubsection{Mass-swapping}
\label{sssec:impact-ms}

Fig.~\ref{fig:ms-sam} presents the MAHs in the first row for two example main branches affected by mass-swapping, with their predictions in the subsequent rows. The red dashed vertical lines indicate when these events occur just in the first row, with the most extreme event marked by a more opaque line in all the rows: one occurs at $z<1$ for the first column and another at $z>1$ for the second. The main conclusion is that significant subhalo mass changes (shown in red) are associated with non-physical alterations in the galaxy properties for {\sc Shark}. The properties influenced by these subhalo changes in {\sc Shark} include:

\begin{itemize}
\item \textbf{Gas disc size}: Abrupt changes in the gas disc size are seen in the second row since the subhalo maximum circular velocity, $V_{\rm max}$, used to calculate this property, is affected by the artefacts as well. A less massive subhalo results in a lower velocity profile, which leads to a more extended disc, and vice versa.
\item \textbf{SFHs}: A large decrease in subhalo mass is accompanied by a significant drop in the star formation rate (SFR) in the third row. The star formation in the disc decreases as the cold gas becomes less dense due to the disc being more extended (lower gas surface density translates to lower integrated molecular gas masses). A milder impact is observed on the SFR associated with starbursts, triggered by mergers or disc instabilities.
\item \textbf{BHAHs}: Both BH accretion modes are similarly affected in the fourth row, as they depend on the subhalo maximum circular velocity, $V_{\rm max}$. Consequently, there is a correlated drop in their values. With the gas becoming more extended, less material is accreted onto the central BH.
\end{itemize}

We emphasise that even inherently more robust subhalo properties, such as the maximum circular velocity, $V_{\rm max}$, which primarily probes the inner core of the subhalo and is less sensitive to changes in mass, are still affected by these issues. As a result, these effects propagate to the derived galaxy properties. Other galaxy properties in {\sc Shark} such as the disc and bulge gas mass, the atomic and molecular gas mass or the reheated mass from SNe are also impacted accordingly.

The right panel of Fig.~\ref{fig:ms-sam} shows the main branch evolution predicted by {\sc Galform} for central galaxies associated with the same central subhaloes selected at $z=0$. At first, we notice the MAHs in the top panels do not exhibit mass-swapping.  This is because {\sc Galform} uses host halo masses instead of subhalo masses, with the host halo mass represented constrained to grow (light grey lines). Additionally, other host halo properties, such as circular velocity $V_{\rm vir}$, halo concentration $c$ and spin $\lambda$ (except for mass $M_{\rm halo}$ and radius $r_{\rm vir}$) used to compute galaxy properties, are only updated when the host halo mass increases by a factor of 2 (dark grey lines). Therefore, it is by definition not possible to have significant mass decreases, and any non-physical mass increase typically follows a non-physical mass decrease, so they are also avoided. 

Furthermore, galaxy properties in {\sc Galform} rely on host halo properties, which are smoother and less sensitive to these artefacts (as evident from raw host halo MAHs in light grey in the first row in the {\sc Shark} panels on the left), and are constrained to increase in mass by the SAM. Therefore, galaxy properties are computed based on these smoother DM host halo properties, which effectively prevents the propagation of mass-swapping artefacts into the galaxy predictions. This behaviour aligns with the findings of \citet{gomez21}, where the authors investigated how different combinations of halo finders and merger tree builders affect {\sc Galform} predictions. On the other hand, this method can lead to situations where host halo properties remain unchanged for several Gyrs, since some are only updated when there is a factor 2 increase in host halo mass. 

\subsubsection{Massive transients}
\label{sssec:impact-mt}

The panels in Fig.~\ref{fig:mt-sam} represent SAM predictions for extreme massive transients occurring at $z<1$ and $z>1$. The dashed green vertical line marks the exact moment these events happen. In them, the correct central subhalo up until that snapshot (blue object in Fig.~\ref{fig:mt-diagram}) becomes a satellite, and thus its massive central galaxy is reclassified as a satellite galaxy (hosted by a satellite subhalo) of the new central subhalo that appears for the first time. The new central subhalo seeds a central galaxy, which rapidly develops substantial hot and cold gas components. When the former central galaxy merges with the newly defined galaxy several snapshots later, the properties return to reasonable values for a galaxy at that cosmic time but undergo a non-physical evolution. The key takeaways from the {\sc Shark} panels on the left are:

\begin{itemize}
\item \textbf{Gas disc size}: The former galaxy merging back with the newly defined main branch gives rise to a merger event, occurring at lookback times $\rm \approx6~Gyr$ in the first column and $\rm \approx9~Gyr$ in the second column for the {\sc Shark} panels, which leads to variations in galaxy properties. In some cases, these minor mergers may trigger eventually a major merger event, such as the one seen in the left panel of the second row at roughly lookback times $\rm \approx3~Gyr$, which significantly changes the cold gas disc size.
\item \textbf{SFHs}: When the former galaxy merges back with the newly defined main branch, a non-physical starburst is triggered by the merger, which produces a peak in the SFR in the third row.
\item \textbf{BHAHs}: The BH accretion activates whenever there is gas in the bulge from starbursts, which occurs only after the merger event as seen in the fourth row. Another potential channel of BH accretion is the hot halo accretion, which also activates after the merger, although its contribution is less significant as the BH is not massive enough.
\end{itemize}

\begin{figure*}
\centering
\begin{subfigure}[b]{0.305\textwidth}
    \includegraphics[trim={1.5cm 3.5cm 19.5cm 1.5cm},clip,width=\linewidth]{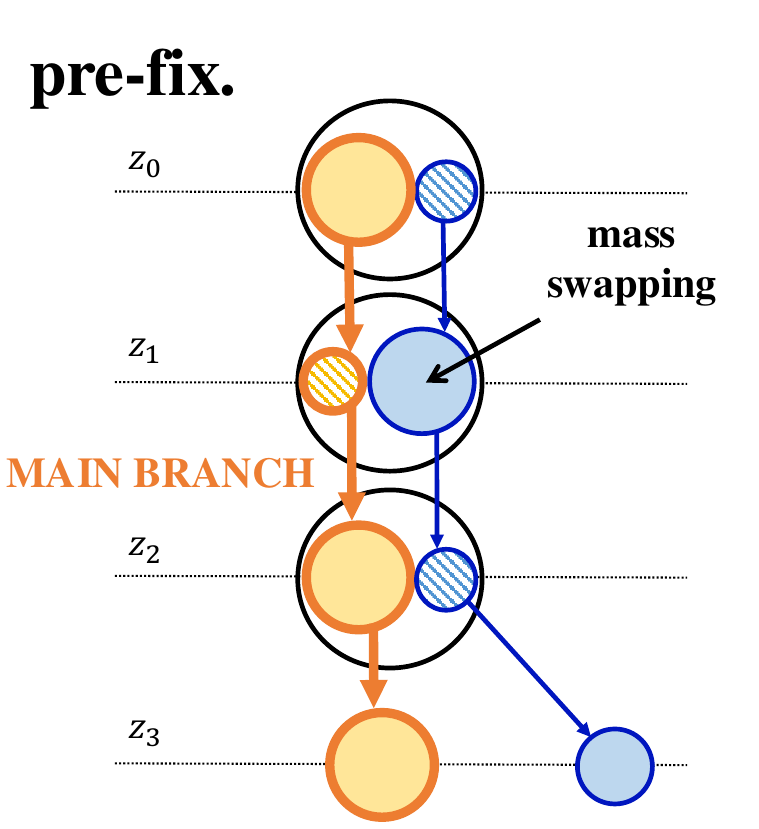}
\end{subfigure}
\begin{subfigure}[b]{0.305\textwidth}
    \includegraphics[trim={1.5cm 3.5cm 19.5cm 1.5cm},clip,width=\linewidth]{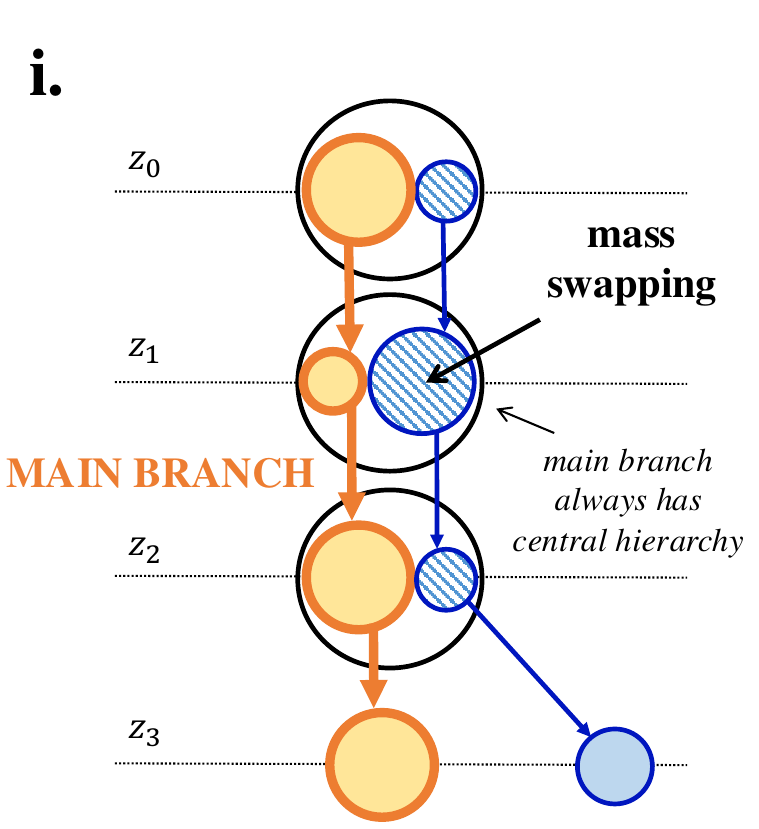}
\end{subfigure}
\begin{subfigure}[b]{0.38\textwidth}
    \includegraphics[trim={1.5cm 3.5cm 16.3cm 1.5cm},clip,width=\linewidth]{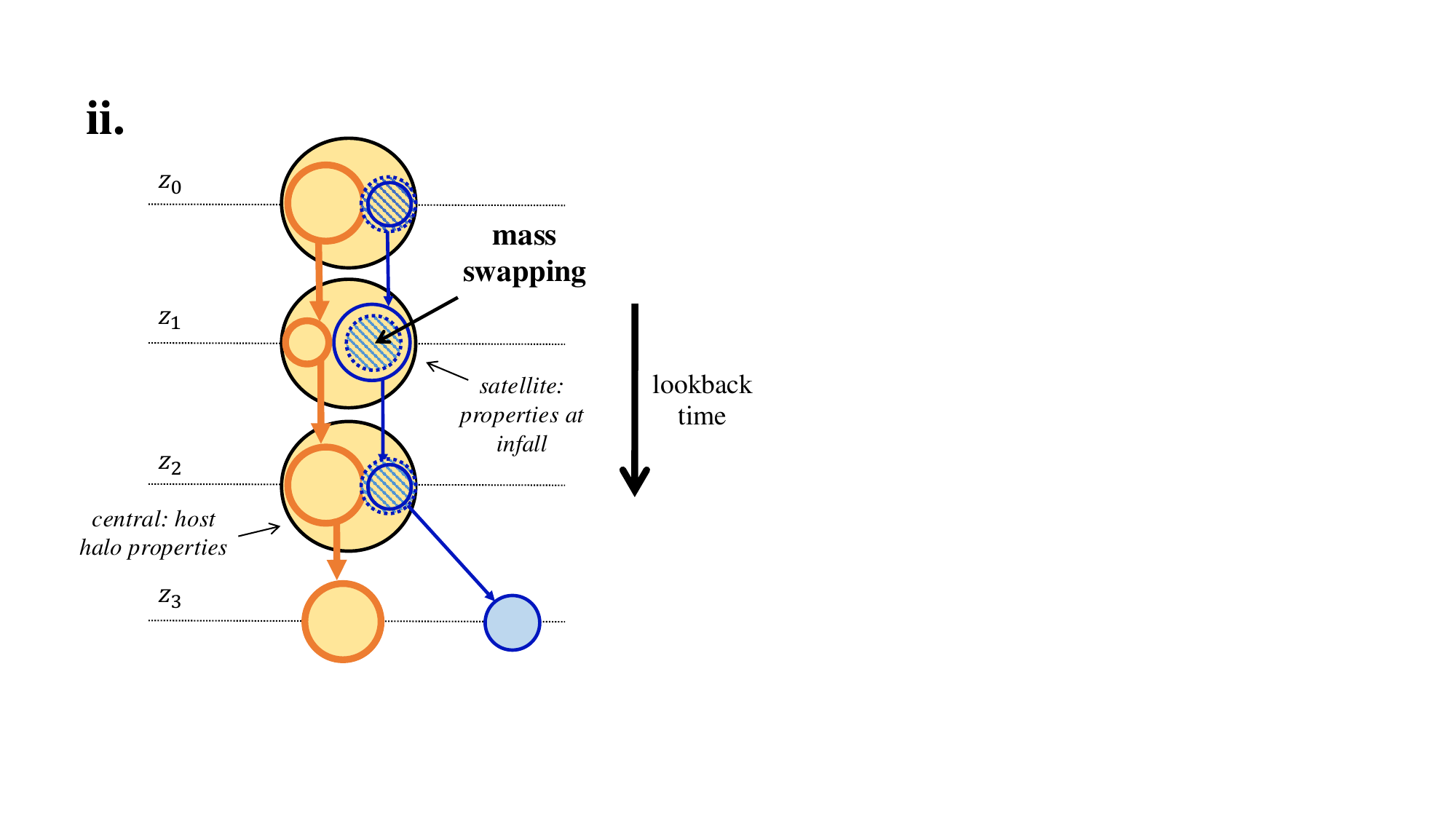}
\end{subfigure}
\caption{Schematic diagram describing the applied fixes proposed to alleviate the effects of mass-swapping events. Each colour refers to a different structure and the circle size represents qualitatively mass. The shaded area highlights the structure (host halo outlined in black, central subhalo represented in solid-coloured circles or satellite subhalo by hatched circles) used to define the galaxy properties. \textit{Pre-fix}: scenario before applying the fixes, similar to the schematic example in Fig.~\ref{fig:ms-diagram}. \textit{i}: central/satellite subhalo hierarchy fix. At $z_{1}$ the orange structure keeps its central status while the blue structure remains a satellite. \textit{ii}: the central galaxy hosted by the orange structure uses the host halo information (solid orange pattern) to define its properties, while the galaxy hosted by the blue structure (initially central and then a satellite after $z<z_{3}$) uses host halo information from the infall snapshot ($z_{3}$).}
\label{fig:ms-fix} 
\end{figure*}

Other properties such as the atomic and molecular gas mass, the hot gas mass, the cold gas disc mass or the stellar mass, also change unrealistically fast during the merger event and then decrease after the main merger, following a smooth exponential decline.

The results for {\sc Galform}, shown in the right panel of Fig.~\ref{fig:mt-sam}, demonstrate that this SAM effectively copes with transients. This is due to its host halo-based methodology, where central subhaloes without a specific main progenitor (transient definition) can be linked to progenitors from other subhaloes (satellites) within the host halo, as the code relies on the host halo merger tree structure (\S~\ref{sssec:galform}). As a result, transient events do not propagate into the predicted properties when there is a main progenitor for the host halo, as seen in the first row of the {\sc Galform} panels, where the lifetime of the central galaxies is extended to before the transient emergence, as well as in subsequent rows for the galaxy properties.

Our analysis reveals that artefacts in merger tree data are relatively common depending on the halo/subhalo-finding and tree-building codes employed and that they can affect the predicted properties of galaxies hosted by the affected main branches. The impact of these artefacts varies depending on the method used to evolve baryons. SAMs that rely on subhalo merger tree information, such as {\sc Shark}, are highly impacted. In contrast, using host halo information, as in {\sc Galform}, appears to mitigate the propagation of mass-swapping and transient events.
This approach models central galaxy properties and links galaxies across time using host halo information instead of that from the central subhalo (the gravitationally bound DM structure of the galaxy). Thus, it assumes that central subhaloes and host haloes are equivalent, with negligible influence from satellite subhaloes on the host halo.

\subsubsection{Results in the context of other galaxy formation models}
\label{sssec:impact-literature}

Widely-used SAMs, such as {\sc GAEA} \citep{delucia14,hirschmann16}, {\sc L-Galaxies} \citep{guo11,henriques15}, {\sc SAGE} \citep{croton16} or {\sc SAG} \citep{cora06,lagos08}, follow a similar subhalo-based strategy as {\sc Shark}. They also use the {\sc Subfind} code as a halo finder (along with various tree-building algorithms), which inherits the merger tree artefacts as shown in \S~\ref{sec:merger-trees-artefacts}. It is unclear whether all of these codes account for these numerical inaccuracies. There is little information available in the literature: {\sc L-Galaxies} imposes host halo mass growth \citep{ayromlou21} and corrects for gas cooling when a satellite subhalo acquires an artificially central hierarchy \citep{henriques15}; {\sc SAGE} removes hot gas from the system when dealing with host halo mass decreases \citep{croton16}; or SAG avoids using subhalo-specific angular momentum due to its noisy behavior when defining the equivalent galactic disc values \citep{padilla14}. For none of these SAMs there is any explicit reference to numerical issues in the merger tree data nor whether any corrections are made. Thus, their predictions may be affected by the same issues seen for {\sc Shark} in \S~\ref{sssec:impact-ms} and \S~\ref{sssec:impact-mt}. If that is the case, these codes would benefit from mitigation strategies, such as those introduced in \S~\ref{sec:fix-galaxy-formation-models}.

Other SAMs, such as {\sc Santa Cruz} \citep{somerville99,somerville15b} and {\sc Galacticus} \citep{benson12}, use different post-processing algorithms to identify (sub)haloes and construct merger trees, namely {\sc ROCKSTAR} \citep{rockstar} and {\sc Consistent Trees} \citep{consistenttrees}, which are not analysed in this paper. These codes detect structures based on phase-space information and link those structures through adjacent snapshots, following approaches similar to those implemented by some methods described in \S~\ref{sec:methods} \citep{behroozi15,wang16}. While the exact impact of numerical artefacts in these merger trees remains uncertain, they seem likely to be affected as well. Moreover, since it is unclear whether these SAMs primarily rely on host halo or subhalo properties, the extent of any potential biases is also unknown. If they are indeed susceptible to similar issues, then they would benefit from implementing strategies to correct for them.

\section{Fixes to galaxy formation models}
\label{sec:fix-galaxy-formation-models} 

The general conclusion from \S~\ref{ssec:impact-galaxy-formation-models} is that numerical artefacts in merger trees propagate into galaxy modelling for subhalo-based models, leading to non-physical changes in the properties of individual galaxies hosted by those affected DM structures. Given how predominant artefacts can be, especially for the more massive systems, for some combinations of halo finder and tree builder codes, we need a strategy to minimise their impact and improve the resilience of these SAMs to the numerical issues. In \S~\ref{ssec:art-treatment} we detail how to treat the artefacts at the SAM level to prevent their propagation, and in \S~\ref{ssec:new-model-predictions} we explore how their predictions are modified once these strategies are implemented.

\subsection{Merger trees artefact treatment}
\label{ssec:art-treatment}

We propose changes to the way merger trees are handled in SAMs to fix or at least minimise the mass-swapping and massive transient numerical artefacts at the SAM level. These changes are based on the findings in \S~\ref{ssec:impact-galaxy-formation-models}, particularly how the different treatments of baryons in {\sc Galform} and {\sc Shark} lead to varying effects of the artefacts, as well as on the origin of the artefacts to address them in terms of the merger tree structure. {\sc Galform} already tries to mitigate these issues through the use of host haloes that grow monotonically in mass, and a scheme for defining central subhaloes \citep{gomez21}. The proposed fixes are hence demonstrated using {\sc Shark} to exemplify how the predictions improve, but they can be adapted to any SAM based on subhalo properties.


\subsubsection{Mass-swapping}
\label{sssec:art-treatment-ms}

{\sc Shark}, as outlined in \S~\ref{sssec:shark}, primarily depends on subhalo properties to model baryonic processes. It also builds merger tree branches and provides a consistent hierarchy for them, similar to {\sc Galform}: subhaloes that are central cannot become satellites, avoiding a mixing of the central and satellite galaxies when mass-swapping events occur. This is achieved by identifying the most massive subhalo for each host halo at the final snapshot of the simulation and tracing its main progenitors backwards in time. All these branches retain central status. This allows us to implement fixes based on the subhalo hierarchy (which is well-defined after the {\sc Shark} corrections). The steps for addressing mass-swapping cases are illustrated schematically in Fig.~\ref{fig:ms-fix}, where the pattern colours represent the structures used to define galaxy properties: an orange solid pattern for the central galaxy hosted by the main branch subhalo and a blue-hatched pattern for the satellite galaxy hosted by the satellite branch. The initial conditions are depicted in Fig.~\ref{fig:ms-diagram} (pre-fix). The pre-processing applied by the code, which corrects the subhalo hierarchy, is represented in the middle panel (i), ensuring the orange subhalo retains its central status throughout its lifetime.

On the other hand, {\sc Galform} considers the host halo properties as the relevant ones in the modelling of galaxy properties (\S~\ref{sssec:galform}), with a halo mass less sensitive to the impact of mass-swapping events as these happen primarily between subhaloes hosted within the same host halo. Moreover, the code constrains host halo mass growth to be monotonic as {\sc Shark} does. This strategy makes {\sc Galform} more resilient to mass-swapping events as we stated in \S~\ref{sssec:impact-ms}.

With that in mind, we implement in {\sc Shark} the same host halo-based methodology, which is illustrated as the second step (ii) for the fix in Fig.~\ref{fig:ms-fix}. In short, we make the following adjustments: 

\begin{itemize}
    \item \textbf{Central galaxies}: information from the entire host halo, which is less affected by numerical artefacts than subhalo properties \citep{gomez21}, as we can see in the example for individual galaxies as the light grey lines in Fig.~\ref{fig:ms-sam}, is used to model the evolution of central galaxies. The smoother and strictly growing evolution is also illustrated schematically in (ii) of Fig.~\ref{fig:ms-fix}, where the orange solid pattern represents the properties computed using the entire host halo outlined in black. The mass of the satellite subhaloes can be considered negligible compared to the central subhalo mass \citep{gill04,delucia04}.
    \item \textbf{Satellite galaxies}: satellite subhaloes are also affected by mass-swapping events, as they can present abrupt mass increases due to the temporary change in the central-satellite hierarchy. To mitigate the impact of those changes on the galaxies hosted by satellite subhaloes, we rely on the host halo properties it had when it was a central subhalo as done in {\sc Galform} as well \citep{lacey16}. Specifically, we use the data from the snapshot just before the subhalo became a satellite (infall), helping avoid the numerically erroneous snapshots when the structure is misidentified as a central \citep{gomez21}. As shown in Fig.~\ref{fig:ms-fix}, the blue hatched pattern in (ii) represents the structure considered to calculate the galaxy properties for the satellite galaxy, which remains unchanged after $z<z_{3}$ as at $z_{3}$ the blue structure is a central subhalo (infall state). The assumption is that after infall the subhalo structure changes primarily on the outskirts, not affecting properties such as the maximum circular velocity ($V_{\rm max}$), concentration, etc. \citep{ghigna98}. The only instance where we choose not to use the infall properties of the haloes and instead employ the current subhalo properties is environmental processes affecting satellite galaxies, such as ram pressure and tidal stripping. 
\end{itemize}

\begin{figure*}
\centering
\begin{subfigure}[b]{0.305\textwidth}
    \includegraphics[trim={5cm 3.5cm 16.5cm 1.5cm},clip,width=\linewidth]{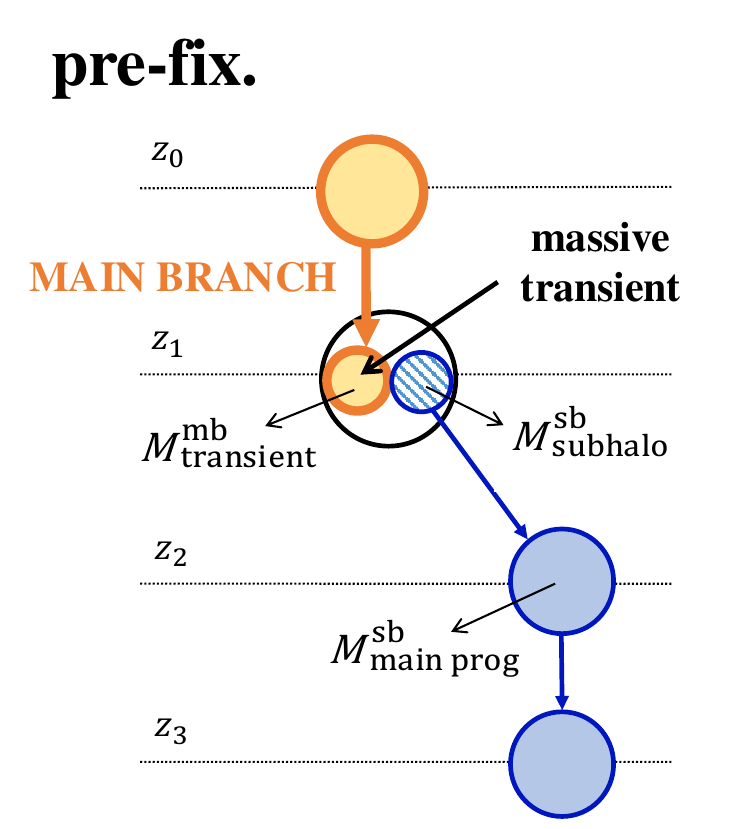}
\end{subfigure}
\begin{subfigure}[b]{0.305\textwidth}
    \includegraphics[trim={5cm 3.5cm 16.5cm 1.5cm},clip,width=\linewidth]{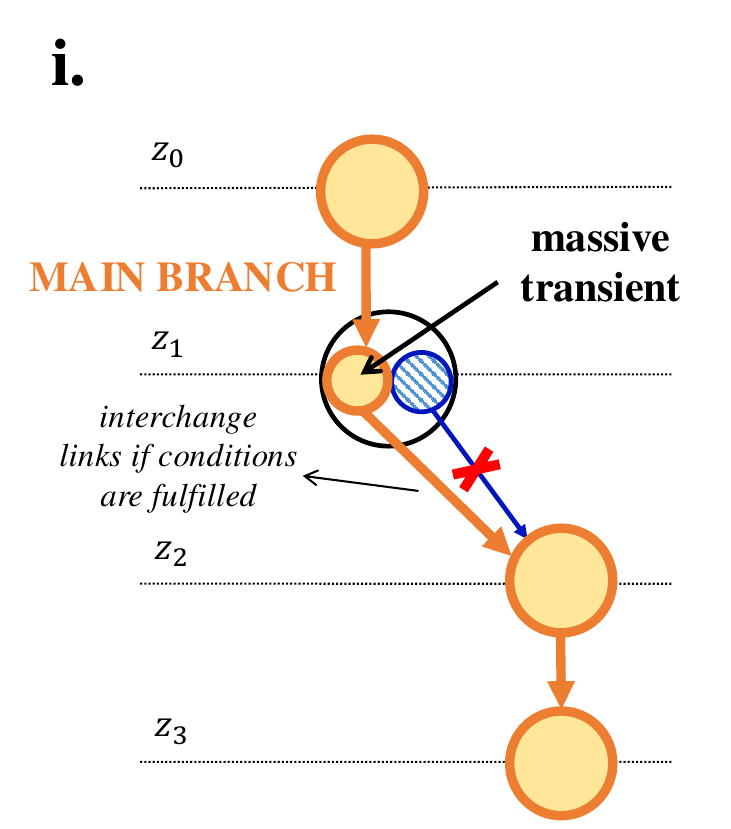}
\end{subfigure}
\begin{subfigure}[b]{0.38\textwidth}
    \includegraphics[trim={5cm 3.5cm 13.3cm 1.5cm},clip,width=\linewidth]{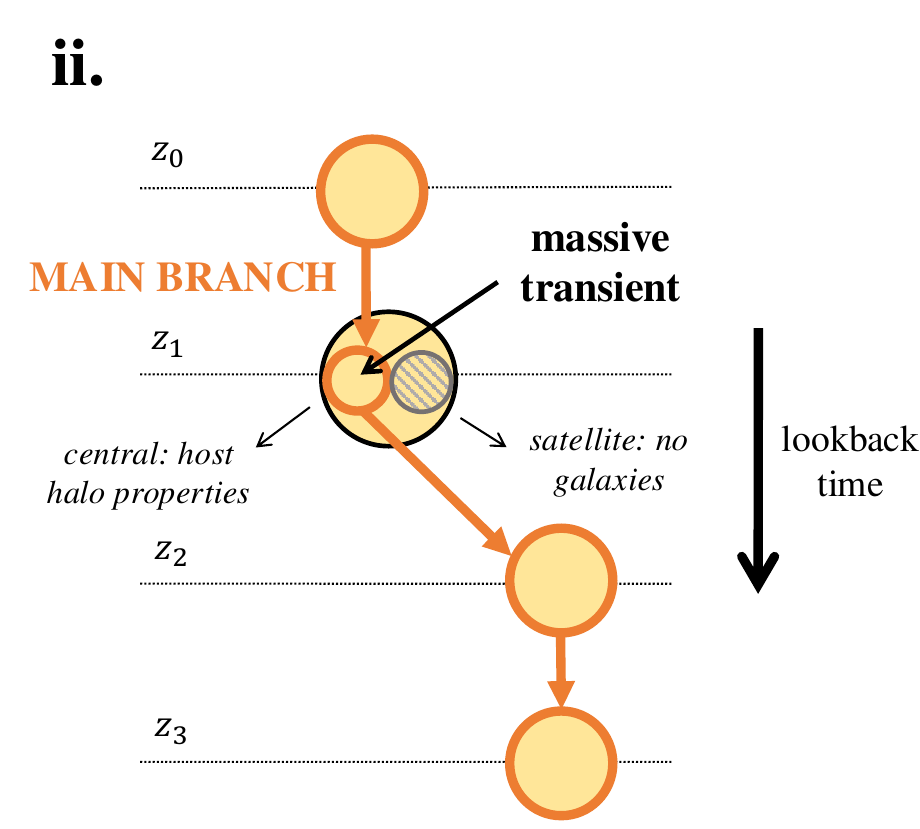}
\end{subfigure}
\caption{Schematic diagram describing the applied proposed to alleviate the effects of massive transients. Each colour refers to a different structure and the circle size represents qualitatively mass. The shaded area highlights the structure (host halo outlined in black, central subhalo represented in solid-coloured circles or satellite subhalo by hatched circles) used to define the galaxy properties. \textit{Pre-fix}: scenario before applying the fixes, similar to the schematic example in Fig.~\ref{fig:mt-diagram}, incorporating the definitions involved in equations~(\ref{eq:5})--(\ref{eq:7}). \textit{i}: possible incorrect connections are evaluated. If the link marked with a blue arrow between $z_{2}$ and $z_{1}$ meets all imposed criteria, it is broken and a new orange arrow/link is generated to extend the lifetime of the main branch. \textit{ii}: galaxy properties are well-defined thanks to the mass-swapping fixes. At $z_{1}$ the central galaxy hosted by the orange subhalo uses host halo information (solid orange pattern covering the entire black circles) to define its properties, while the other subhalo emerges directly as a satellite (grey hatched pattern) and it does not host any galaxies.} 
\label{fig:mt-fix} 
\end{figure*}

This new implementation of {\sc Shark} shares with {\sc Galform} the intermediate step of adjusting the central-satellite hierarchy to correct for numerical effects. However, in {\sc Galform} the subhalo with the highest cumulative mass in its past merger tree history, as defined at the final snapshot, is identified as central, and this designation is applied to all its main progenitors (based on the host halo structure). Specifically, both SAMs correct in step (i) the hierarchy issues caused by {\sc DHalo}, where the central identifier is assigned to the most massive structure within each group. This misassignment results in central-satellite swaps, as shown in Fig.~\ref{fig:ms-def-panel}, where main branches are incorrectly assigned as satellites (indicated by the purple regions in the colour maps). The main difference, however, is that {\sc Shark} does not define ``halo formation'' events to describe the evolution of the DM structures for computing certain galaxy properties. As a result, the halo mass (and the subhalo mass for the environmental processes of satellite galaxies) is updated more accurately at each snapshot, though it still assumes that the satellite contribution is negligible.

\begin{figure*}
\centering
\captionsetup[subfigure]{labelformat=empty}
\captionsetup[subfigure]{font=Large}
\begin{subfigure}{.5\textwidth}
  \centering
  \caption{\textbf{\textsc{Shark} swapping (fix)}}
  \par\medskip
  \includegraphics[trim={0 0cm 0 0cm},clip,width=0.92\textwidth]{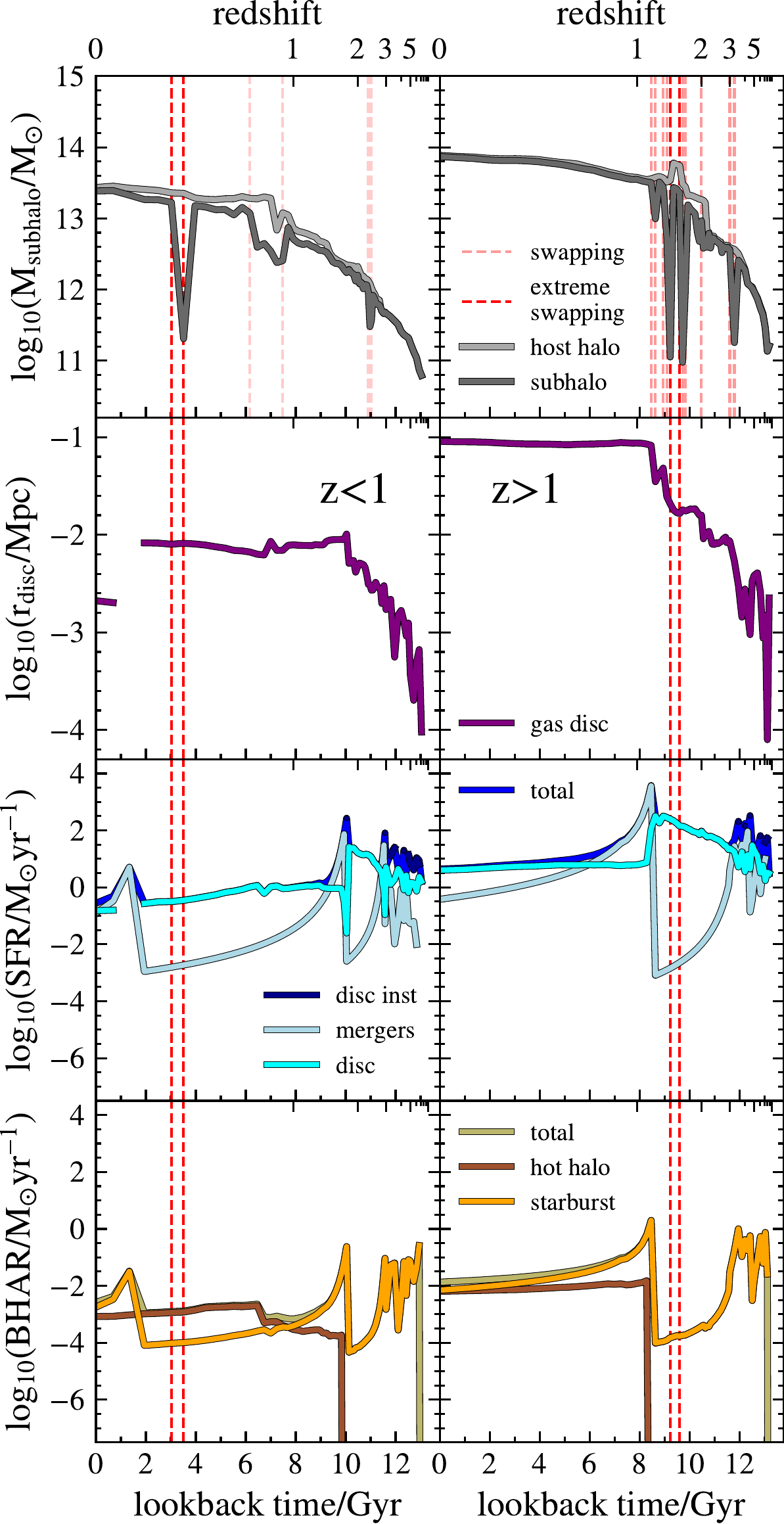}
\end{subfigure}%
\begin{subfigure}{.5\textwidth}
  \centering
  \caption{\textbf{\textsc{Shark} transients (fix)}}
  \par\medskip
  \includegraphics[trim={0 0cm 0 0cm},clip,width=0.92\textwidth]{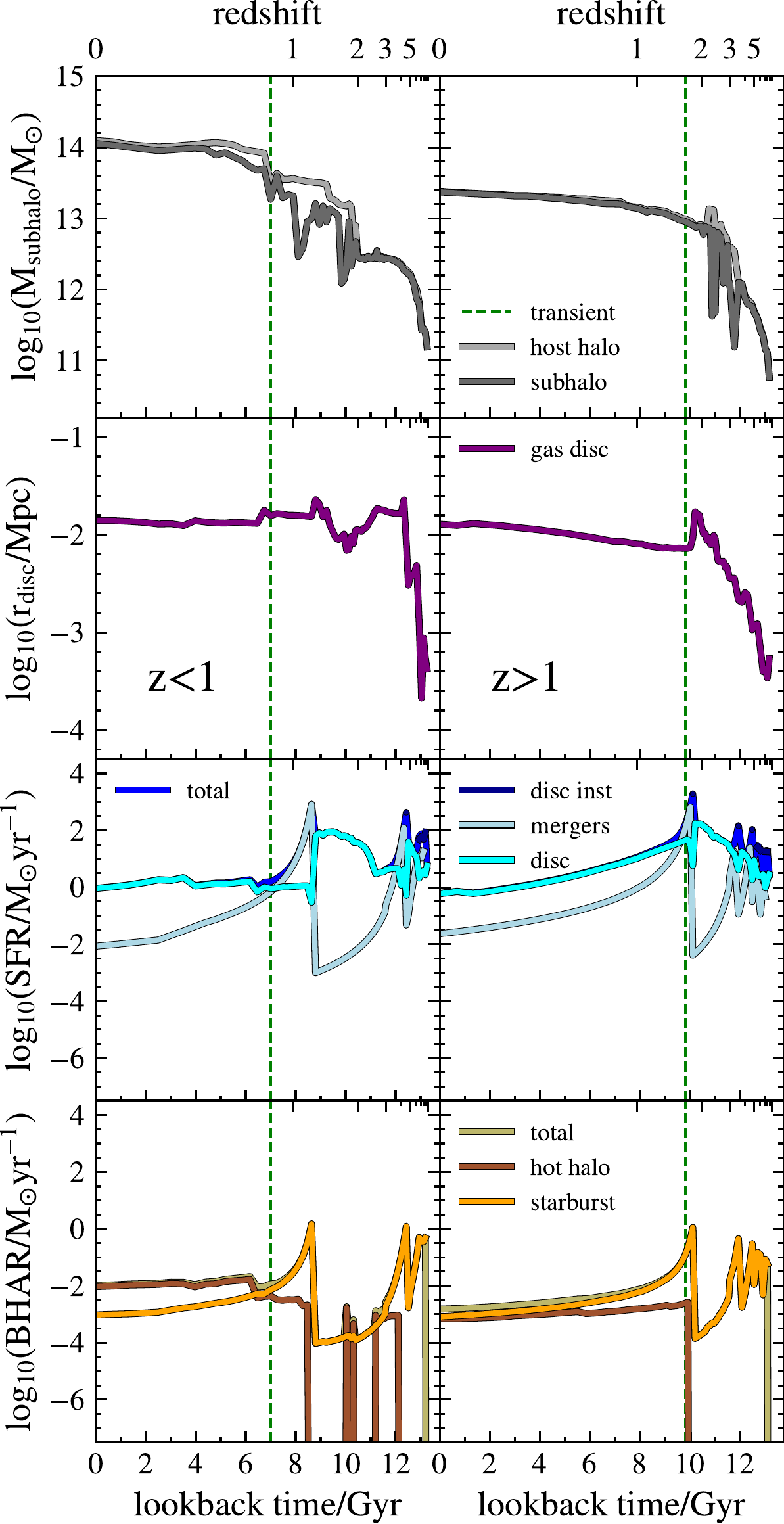}
\end{subfigure}
\caption{SAM predictions from {\sc Shark} after applying the artefact fixes at the SAM level for main branches affected by mass-swapping (left panel) and massive transients (right panel). \textit{First row}: Subhalo and halo MAHs in dark and light grey with the specific snapshots at which the swapping/transient events happen marked with red/green dashed vertical lines, with the most extreme cases of mass-swapping in a more opaque face: in the 1st column the extreme event occurs at $z<1$, while in the 2nd it occurs column at $z>1$. \textit{Other rows}: cold gas disc size (second row), SFHs (third row) and BHAHs (fourth row) for the central galaxies associated with the main branches in the top panels.}
\label{fig:sam-fix}
\end{figure*}

The implementation is activated in {\sc Shark} by default setting the newly defined parameter \verb|apply_fix_to_mass_swapping_events| to \verb|true|. After applying the mass-swapping fixes, the left panel of Fig.~\ref{fig:sam-fix} presents the updated predictions for the same individual galaxies of Fig.~\ref{fig:ms-sam}. We observe that the galaxy properties no longer exhibit the abrupt changes seen previously at the moments marked by the red dashed vertical lines, where mass-swapping events occurred. The results now appear physically consistent, with no sudden jumps in the gas disc size evolution (unless there are physically motivated starbursts) and looking at the SFHs or BHAHs independently of the considered star formation or accretion modes. For satellites, visual inspection confirms that the results are similarly reasonable, with properties usually showing declining profiles until the satellite galaxy finally merges with the central galaxy in a dynamical friction time-scale.

\subsubsection{Massive transients}
\label{sssec:art-treatment-mt}

In contrast to {\sc Galform}, {\sc Shark} does not have a built-in strategy to handle massive transients. Therefore, we need to develop a method to minimise their influence. 
As in {\sc Galform}, we propose modifying the merger tree structure directly before populating it with galaxies. However, we make these modifications for transient structures at the subhalo level, while maintaining the subhalo-based approach. When {\sc Shark} constructs the branches, it reconnects progenitor and descendant subhaloes, keeping a consistent subhalo hierarchy to define central galaxies appropriately, as explained in \S~\ref{sssec:shark}. As drawn in Fig.~\ref{fig:mt-diagram}, the link of the disappearing main branch at the moment the transient emerges (represented by the blue arrow connecting $z_{2}$ and $z_{1}$) is the critical one. By connecting the disappearing branch to the definitive main branch (shown in orange), we could solve this problem as abrupt changes in subhalo properties that can arise when both structures are linked would be assessed by the mass-swapping fix explained earlier. 
    
Fig.~\ref{fig:mt-fix} visually explains how to re-establish the links in the main branches affected by transient events. The left panel shows the initial conditions for the schematic example in Fig~\ref{fig:mt-diagram} (pre-fix). The middle panel indicates that the incorrect connection for the subhalo that ultimately disappears must be severed (marked by a red cross). The main branch would then be linked to the prior history of that disappearing branch (i), ideally extending the main branch's lifetime to ensure it reaches a reasonable particle number at birth. Furthermore, after modifying the links, the blue subhalo that ceases to have progenitors would be considered a satellite at the moment of its birth, meaning it would not host any galaxies.

This linking modification applies to all the subhaloes belonging to main branches flagged as transients. Subhaloes can be flagged as being massive transients in {\sc Shark} using one of three thresholds (with the user having to decide on one): a z-dependant $3\sigma(N^{\mathrm{birth}}_{\mathrm{part,subhalo}},z)$ (the default adopted in this paper, \verb|zdep_3sigma|), but also a constant value of $10N^{\mathrm{min}}_{\mathrm{part,subhalo}}$ (\verb|const_10minpart|) or 200 particles (\verb|const_200|). The user can choose among these options using the parameter \verb|define_transient| in {\sc Shark}, as included in Table~\ref{tab:shark-params}.

In addition, after flagging a subhalo as a massive transient, we impose several conditions that must be satisfied before breaking a link and creating a new one:

\begin{align}
    \label{eq:5}
    &M_{\mathrm{transient}}^{\mathrm{mb}}>M_{\mathrm{subhalo}}^{\mathrm{sb}} \\
    \label{eq:6}
    &M_{\mathrm{subhalo}}^{\mathrm{sb}}<f_{\mathrm{u}}^{\mathrm{sb}}\,M_{\mathrm{main}\ \mathrm{progenitor}}
    ^{\mathrm{sb}} \\
    \label{eq:7}
    &f_{\mathrm{l}}^{\mathrm{mb}}\,M_{\mathrm{main}\ \mathrm{progenitor}}^{\mathrm{sb}}<M_{\mathrm{transient}}^{\mathrm{mb}}<f_{\mathrm{u}}^{\mathrm{mb}}\,M_{\mathrm{main}\ \mathrm{progenitor}}^{\mathrm{sb}}
\end{align}

\begin{table}
	\centering
	\cprotect\caption{New parameters, as shown in the parameter file, introduced in the {\sc Shark} code for handling the numerical artefacts (mass-swapping and massive transients) from merger tree data, along with their corresponding default values. The flags \verb|apply_fix_to_mass_swapping_events| and \verb|apply_fix_to_massive_transient_events| can be set to \verb|true| or \verb|false|. The parameter \verb|define_transient| offers 3 configuration options: \verb|zdep_3sigma| (default), \verb|const_10minpart| and \verb|const_200|; while the remaining parameters accept numerical values. All of them can be modified by the user as needed.}
	\label{tab:shark-params}
	\begin{tabular}{cc} 
		\hline
		{\sc Shark} parameter & default value \\
		\hline
        \multicolumn{1}{|l|}{\textit{mass-swapping:}} \\
        \verb|apply_fix_to_mass_swapping_events| & \verb|true| \\
        \\
        \multicolumn{1}{|l|}{\textit{massive transients:}} \\
        \verb|apply_fix_to_massive_transient_events| & \verb|true| \\
        \verb|define_transient| & \verb|zdep_3sigma| \\
        \verb|transient_lostmass_ratio| & $f_{\mathrm{u}}^{\mathrm{sb}}=0.7$ \\
        \verb|transient_gainedmass_ratio_low| & $f_{\mathrm{l}}^{\mathrm{mb}}=0.1$ \\
        \verb|transient_gainedmass_ratio_up| & $f_{\mathrm{u}}^{\mathrm{mb}}=3.0$ \\
		\hline
	\end{tabular}
\end{table}

Following equation~(\ref{eq:5}), a link is severed if the transient subhalo is more massive than the subhalo belonging to the disappearing branch. Equations~(\ref{eq:6})~and~(\ref{eq:7}) introduce some free parameters to make sure the new connection is smoother in terms of mass gained/lost for both the disappearing and main branches. Specifically, equation~(\ref{eq:6}) constrains the mass loss between $z_{2}$ and $z_{1}$ for the disappearing branch to determine if it is large enough to be considered non-physical (with $f_{\mathrm{u}}^{\mathrm{sb}}$ defined as \verb|transient_lostmass_ratio| in the code). Meanwhile, equation~(\ref{eq:7}) evaluates whether the mass change in the newly created link results in an unreasonable mass decrease (lower limit $f_{\mathrm{l}}^{\mathrm{mb}}$ referred to as \verb|transient_gainedmass_ratio_low|) or increase (upper limit $f_{\mathrm{u}}^{\mathrm{mb}}$ referred to as \verb|transient_gainedmass_ratio_up|). The values assigned to the parameters involved in equations~(\ref{eq:5})--(\ref{eq:7}) are compiled in Table~\ref{tab:shark-params}. Several alternative values were tested with no significant difference found.

When these criteria are not fulfilled, the main progenitor flag for the transient is at least removed and re-evaluated for the most massive structure in the host halo. Consequently, there could still be a fix between $z_{1}$ and $z_{0}$ if there exists a more massive subhalo belonging to the host (in case the blue subhalo at $z_{1}$ is more massive than the orange transient in Fig.~\ref{fig:mt-fix}). This is because the pre-processing fixing the subhalo hierarchy in {\sc Shark} would extend the main branch's lifetime, connecting the orange main branch at $z_{0}$ to the blue satellite at $z_{1}$. The flag \verb|apply_fix_to_massive_transient_events| in {\sc Shark} is set to \verb|true|, enabling to apply the transient adjustment.

Fig.~\ref{fig:sam-fix} displays in the right panel the updated predictions for the same individual galaxies shown in Fig.~\ref{fig:mt-sam} once the changes to alleviate the presence of massive transients are applied. By linking transient structures using a subhalo-based method that follows subhalo branches — rather than relying on host halo information, as in {\sc Galform} — we successfully extend the lifetimes of the main branches beyond the previous lookback times/redshifts, as indicated by the dashed vertical green lines. The subhaloes now emerge at the initial snapshots of the simulation with a much lower number of particles. As a result, there are no artificial starbursts or abrupt build-ups of other properties for them. Plus, due to the mass-swapping fix, the galaxy properties in these transient branches are now computed using a host halo-based approach.

\begin{figure}
\includegraphics[trim={0 0.2cm 0 0cm},clip,width=0.98\linewidth]{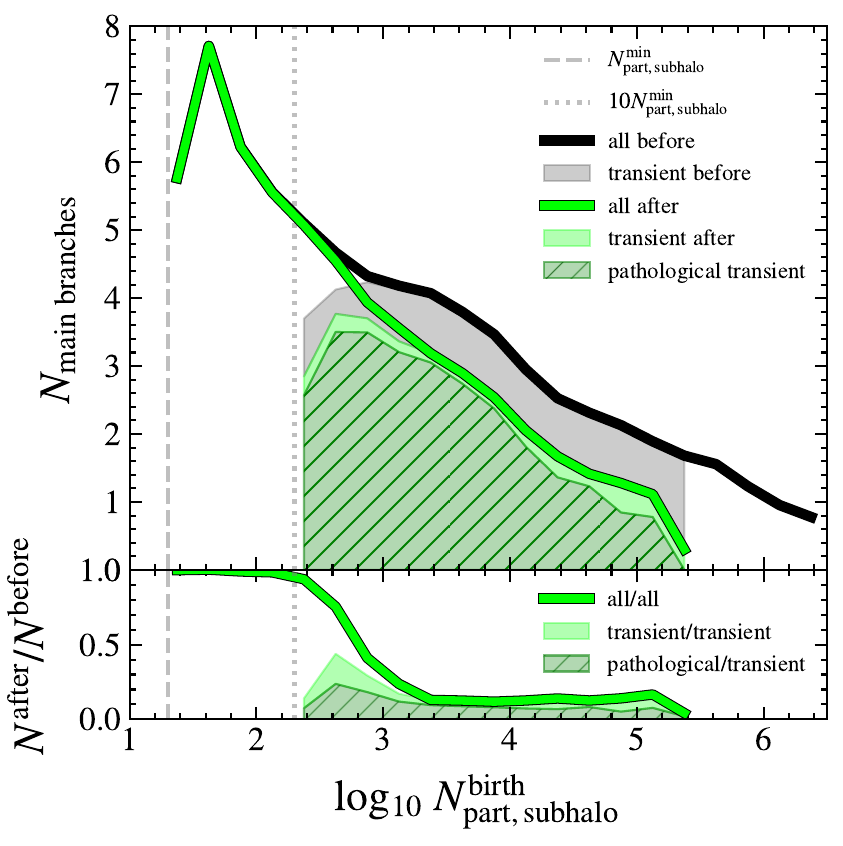}
\caption{1D histogram showing the particle number at birth for all main branches for the {\sc FLAM-DM-VR} catalogue. The black line represents the pre-fixed results before running {\sc Shark}, while the lime-green line shows results after running the code with the transient fixes, as explained in \S~\ref{sssec:art-treatment-mt}. The shaded areas indicate subhaloes identified as massive transient artefacts, both before (grey) and after (lime-green) the implementations, while the hatched pattern region in green points out pathological cases that are very challenging to address.} 
\label{fig:mt-fix-stats} 
\end{figure}

To statistically quantify how many transients are fixed, we analyse the {\sc FLAM-DM-VR} merger tree catalogue post-processed by {\sc Shark}. Fig.~\ref{fig:mt-fix-stats} presents a histogram showing the particle number at birth for subhaloes belonging to main branches, comparing the results before (black line) and after (limegreen line) the fixes were applied. The total number of main branches remains identical for both cases ($\approx10^{7}$ from Table~\ref{tab:statistics}). The shaded areas cover subhaloes classified as transients in each of the runs, with the pre-fix version in grey and the updated version in a solid lime-green pattern. We observe at least a 1~dex improvement in each mass bin populated by transients, falling within the shaded regions. These regions are delimited by a dotted vertical line indicating the minimum particle number for a transient, as defined in equation~(\ref{eq:4}).

The improvement is particularly noticeable in the bottom panel, where the ratio between the new and old results is plotted, highlighting the $\approx10\%$ values in the high-mass regime (particle number above $10^3$). Between the vertical dotted — indicating the minimum particle number required to detect a structure — and dashed lines, there can not be improvements by definition, but also as transients that are fixed now emerge with a particle number in that range, the ratio can exceed one. The ratio of transients between the shaded areas is also illustrated by the lime-green solid pattern in the bottom panel. 

\begin{table}
	\centering
	\caption{Statistics for main branches before (pre-fix) and after applying the fix for massive transients for the {\sc FLAM-DM-VR} catalogue. $N_{\rm t}$: number of massive transients; $f_{\rm unfixed}$: fraction of massive transients that remain unfixed compared to the pre-fixed values (and when excluding non-pathological cases) in $\%$; $\tilde{x}(N_{\rm p,subh,t}^{\rm birth})$: median particle number at birth for massive transients; and $3\sigma(N_{\rm p,subh}^{\rm birth})$: 3 times the standard deviation of the particle number for all main branches at birth.}
	\label{tab:mt-fix}
	\begin{tabular}{ccccc} 
		\hline
		\multirow{2}{*}{catalogue} & \multirow{2}{*}{$N_{\rm t}$} & $f_{\rm unfixed}$ & \multirow{2}{*}{$\tilde{x}(N_{\rm p,subh,t}^{\rm birth})$} & \multirow{2}{*}{$3\sigma(N_{\rm p,subh}^{\rm birth})$} \\
        & & /\% & & \\
		\hline
        pre-fix & 71826 & - & 1015 & 402 \\
        fix & 16821 & 23 & 638 & 254 \\
        (non-pathol.) & (6505) & (9) & - & - \\
		\hline
	\end{tabular}
\end{table}

\begin{figure*}
\vspace{0.1cm}
\centering
\begin{subfigure}[b]{0.4\textwidth}
   \includegraphics[width=\textwidth]{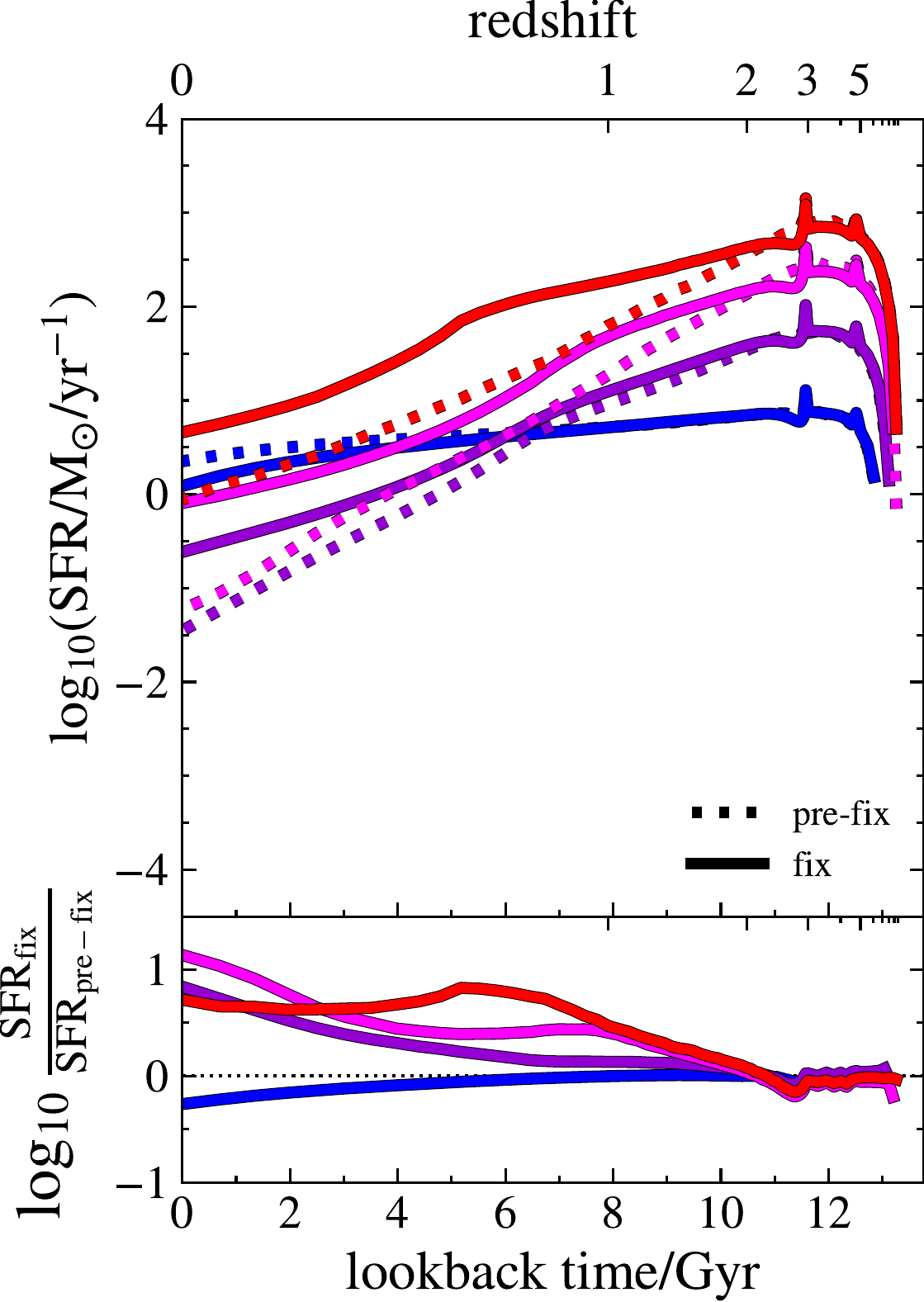}
\end{subfigure} \hfil
\begin{subfigure}[b]{0.4\textwidth}
   \includegraphics[width=\textwidth]{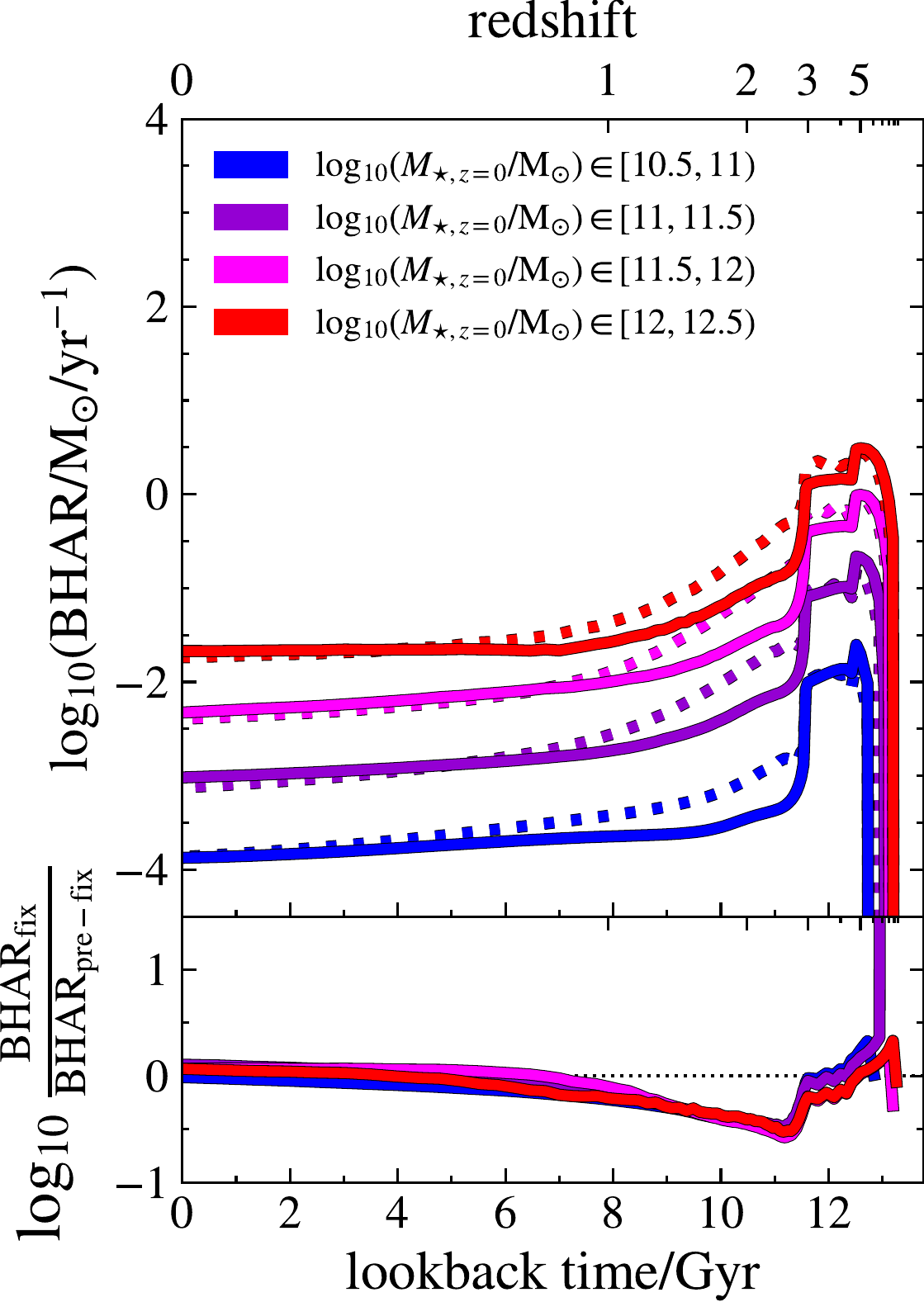}
\end{subfigure}
\caption{\textit{Left panel}: median SFHs for several stellar mass bins at $z=0$ (in different colours) considering all the galaxies modelled by {\sc Shark} over the {\sc FLAM-DM-VR} catalogue, indicating the predictions before (dotted lines) and after (solid lines) applying the numerical artefact fixes. The ratio between both runs is shown in the bottom panel. \textit{Right panel}: same but for the median BHAHs.} 
\label{fig:shark-histories} 
\end{figure*}

Additionally, the green hatched pattern region in Fig.~\ref{fig:mt-fix-stats} highlights transient subhaloes that are particularly challenging to address due to their pathological nature. These cases can be divided into 2 main categories to deal with: (i) massive transients showing up as isolated haloes, making it difficult to connect them to any previous structure using the approach described in \S~\ref{sssec:art-treatment-mt} unless we take into account spatial information— an aspect not addressed by the {\sc Galform} approach either; or (ii) transient subhaloes closely associated with interpolated subhaloes — artificially created by the merger tree builder algorithms to link structures separated by multiple snapshots in time, often due to the misidentification of nearby structures. This second category collects transients' descendants which share the host halo with structures that, at the snapshot when the transient emerges, are either interpolated subhaloes or are hosted by them. Most of the transients that remain unfixed by our implementations lie in this pathological regime as shown in both the top and bottom panels.

Table~\ref{tab:mt-fix} provides the statistics for massive transients in the specific case of the {\sc FLAM-DM-VR} merger tree catalogue, comparing numbers before and after these fixes are applied. We can observe that the total number of massive transients is successfully reduced to $23\%$, or $9\%$ when excluding pathological cases. This success is further highlighted by the reduction in the median particle number at birth for these transient objects, indicating that the high-mass end of subhaloes emerging as transients has been effectively addressed. In some cases, a subhalo’s lifetime may have been extended, but not enough to avoid classification as numerically incorrect. Plus, the value for 3 times the standard deviation of the particle number at birth, $3\sigma(N_{\rm part,subh}^{\rm birth})$, for all main branches (not just transients) has decreased, even though the total number of main branches is significantly larger than the transient ones (3 dex). 

In short, after applying the fixes for both the mass-swapping and the massive transients inherited by the merger tree catalogues due to the structure-finding and tree-building post-processing algorithms the galaxy properties evolve smoothly and are significantly less affected by them. While the merger tree data used as input may not be entirely accurate, we have proposed an approach to generate reliable predictions for galaxy properties when running SAMs. This strategy has been applied to these types of models, but it could be generalised for any kind of galaxy formation and evolution models.  To facilitate its adoption, a pseudo-code for each implementation is provided in Appendix~\ref{asec:pseudo-code}, which can be used by other subhalo-based SAMs.

\subsection{New model predictions}
\label{ssec:new-model-predictions} 

Using {\sc Shark} we compared the predictions of the SAMs before and after applying the fixes described in \S~\ref{ssec:art-treatment} for individual galaxies affected by numerical artefacts. However, SAMs are most effective when used to generate statistical results for comparison with observations or testing physics models describing baryonic processes. Thus, we aim to assess how the statistical observables are impacted by the artefacts characterised in \S~\ref{sec:merger-trees-artefacts}. We present the combined impact of both artefacts, as the massive transients alone do not cause a substantial effect, given that only a few branches are affected by them. This also serves as a sanity check to validate the results for both central and satellite subhaloes, as well as central and satellite galaxies. Since the parameter values in {\sc Shark} are calibrated with the fixes enabled (Appendix~\ref{asec:shark-calibration}), we run the code with and without the flags in Table~\ref{tab:shark-params}. 

\begin{figure*}
\vspace{0.2cm}
\centering
\includegraphics[width=0.98\linewidth]{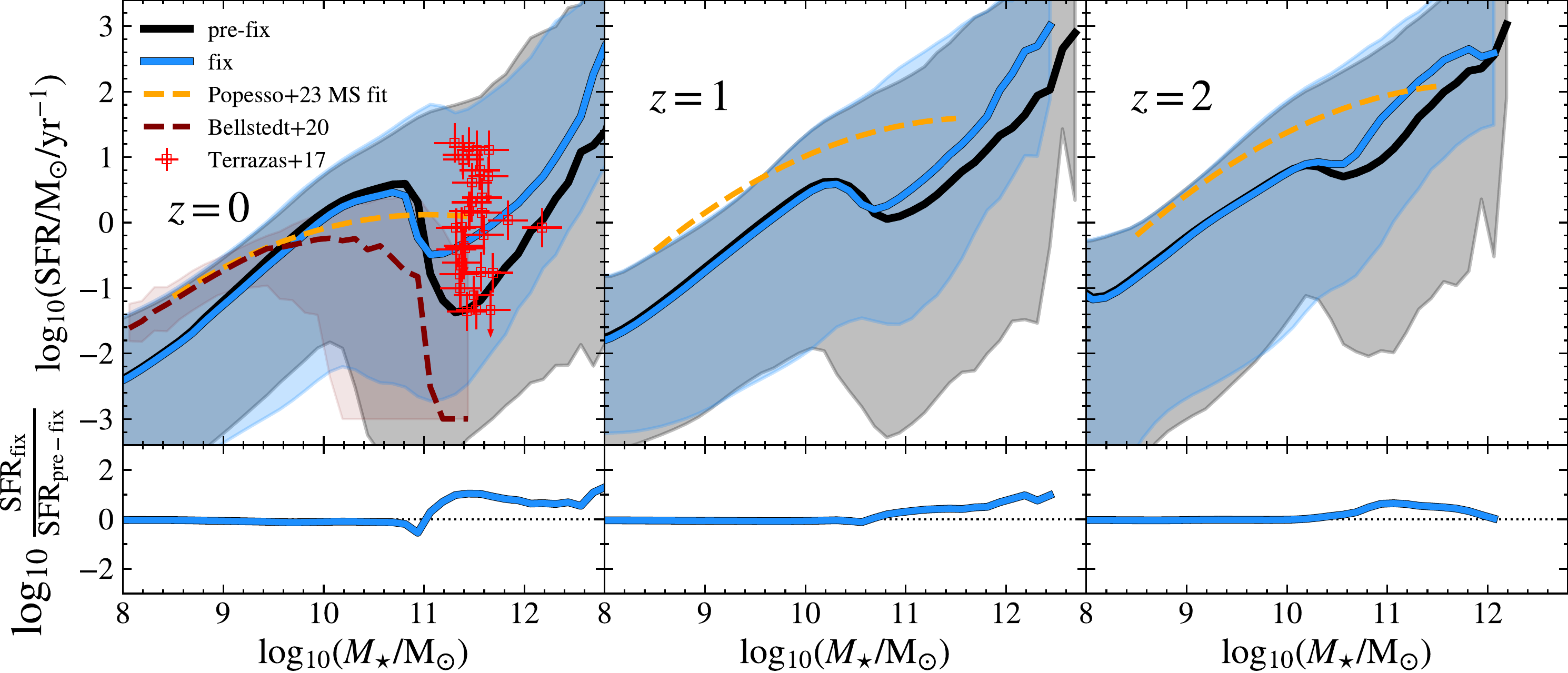}
\caption{SFR versus stellar mass plane at different redshifts ($z=0,\,1,\,2$) predicted by {\sc Shark} over the {\sc FLAM-DM-VR} catalogue before (black line) and after (sky-blue lines) applying the numerical artefact fixes. Solid lines indicate median values, while shaded regions represent the standard deviation (16th-84th percentiles). In the corresponding bottom panels, the ratios of the median values between both runs (pre-fix and fix) are shown. Observational data from \citet{bellstedt20,popesso23,terrazas17} are included.} 
\label{fig:sfms} 
\end{figure*}

We run {\sc Shark} on the {\sc FLAM-DM-VR} catalogue and show how the predictions change for it since this catalogue is highly affected by the issues. Simultaneously, we run {\sc Shark} over the more accurate {\sc FLAM-DM-HBT} catalogue. We maintain the same calibration of {\sc Shark} parameters in all cases, despite using different catalogues, in order to completely isolate the effect of the artefacts at the merger tree level. This ensures that no potential degeneracies are introduced due to differences in the modelled physics at the SAM level, aligning with the paper's goal of providing as robust predictions as possible. Nonetheless, we tested whether keeping the same calibration affected the results by visually inspecting how much the $z=0$ SMF (which is the reference to calibrate {\sc Shark}) changed when using different subhalo and merger tree catalogues, and found negligible differences.

\begin{figure*}
\centering
\includegraphics[width=0.98\linewidth]{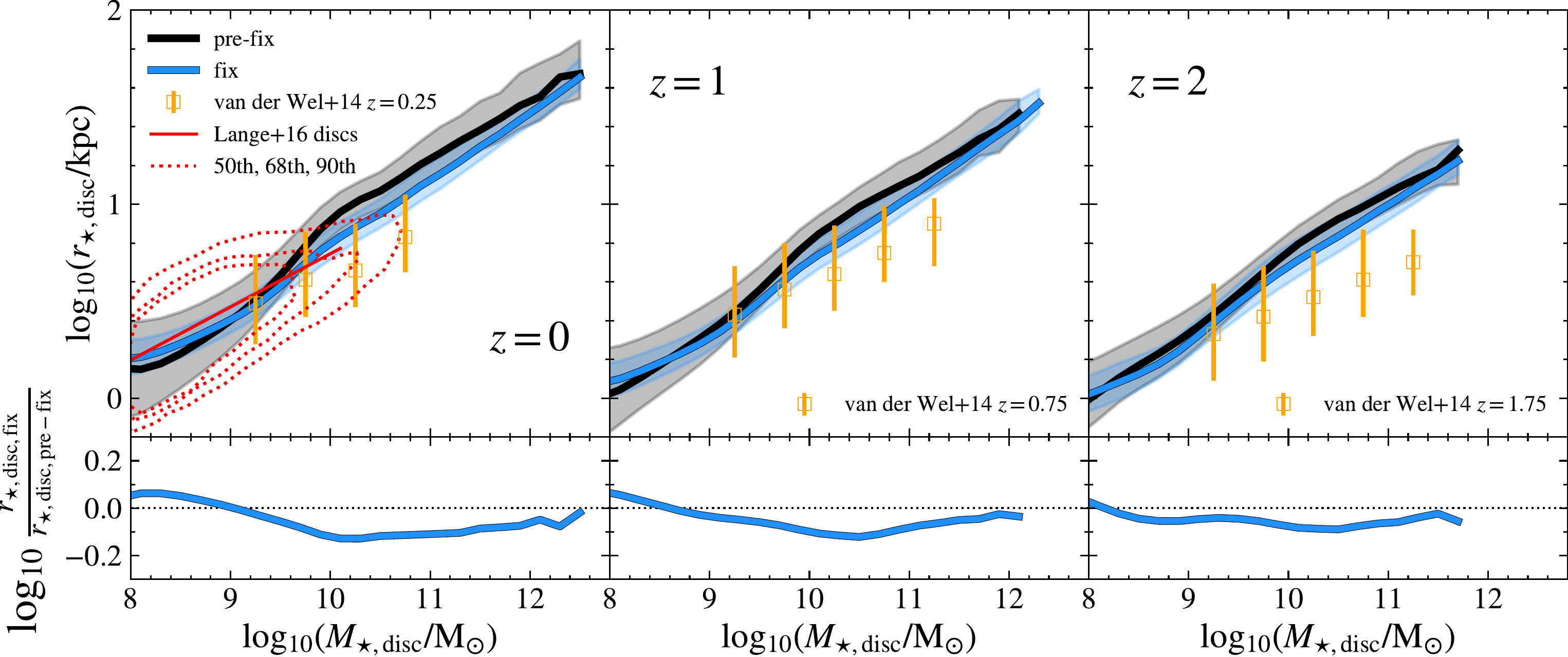}
\caption{Stellar disc size versus stellar mass relation at different redshifts ($z=0,1,2$) predicted by {\sc Shark} for galaxies with $M_{\star\mathrm{,disc}}/M_{\star}>0.5$ over the {\sc FLAM-DM-VR} catalogue before (black line) and after (sky-blue lines) applying the numerical artefact fixes. Solid lines indicate median values, while shaded regions represent the standard deviation (16th-84th percentiles). In the corresponding bottom panels, the ratios of the median values between both runs (pre-fix and fix) are shown. Observational data from \citet{lange16,vanderwel14} are included.} 
\label{fig:sizes} 
\end{figure*}

The comparison between the {\sc FLAM-DM-VR} and {\sc FLAM-DM-HBT} runs is presented in Appendix~\ref{asec:shark-hbt-predictions}, where we provide a fair comparison between the predictions for both catalogues and assess the magnitudes of their difference. We prove the modelled galaxy properties are very consistent despite variations in the underlying merger tree catalogues once the catalogues are fixed, which is in agreement with \citet{gomez21}. This supports the argument that the fixes should always be applied, regardless of the merger tree catalogue used, even for less-affected catalogues, as noticeable differences still exist that need to be addressed.

We look particularly into observables related to the galaxy properties most affected by these artefacts in \S~\ref{ssec:impact-galaxy-formation-models} (star formation, BH accretion and disc size) over the full range of galaxy scales. First, we examine how the numerical fixes statistically propagate through cosmic time by analysing the evolution of the SFRs and BH accretion rates. In Fig.~\ref{fig:shark-histories}, we show the median SFHs and BHAHs for all progenitors across different stellar mass bins at $z=0$, regardless of whether the galaxies are central or satellite. Dotted lines represent the predictions without the fixes, while solid lines show the new results. The ratio between them is computed in the bottom panels. The discontinuities that appear in both panels at $z=3$ and $z=5$ are due to the change in cadence between snapshots in the {\sc FLAMINGO} simulation.

On the left panel, we observe minimal changes in the SFH for the lowest mass bins, while more massive bins show an increase in SFRs at $z\lesssim 2$ (the higher the stellar mass, the higher the difference between the two runs). The fixes address non-physical drops (previously shown in Fig.~\ref{fig:ms-sam}) caused by the mass-swapping events, as well as remove artificial starbursts from massive transients (as seen in Fig.~\ref{fig:mt-sam}). However, since transient events are relatively rare, the net effect is an increase in SFR, except for a slight decrease around lookback times $\rm \approx11~Gyr$, where star formation activity peaks for all mass bins. As expected, more massive systems, which reside in larger subhaloes, exhibit the most significant differences.

The right panel of Fig.~\ref{fig:shark-histories} shows the median BHAHs, which are similarly affected regardless of the stellar mass bin. This suggests an interesting global reduction in accretion via the starburst mode between lookback times $\rm =6-12~Gyr$, consistent with the decrease in starburst activity at lookback times $\rm \approx11~Gyr$. Although the massive transients are not fully resolved with our implementations, they no longer drive so many artificial starbursts, leading to a noticeable reduction in accretion. The lower BH accretion rates are likely the cause of the slightly elevated SFRs in massive galaxies as AGN feedback becomes weaker.

Next, we evaluate predictions for the entire galaxy population across different redshifts. Fig.~\ref{fig:sfms} shows the SFR-stellar mass plane at $z=0,\,1$~and~$2$ with median values represented by solid lines and standard deviations as shaded areas, comparing the predictions from both runs (pre-fix and fix) alongside observational data. The ratio of the median values is shown in the bottom panels. The corrected predictions indicate higher SFRs for more massive galaxies, as the fixes eliminate the non-physical drops caused by numerical artefacts — particularly in systems with stellar masses above $10^{11} \mathrm{M}_{\odot}$. The updated predictions align more closely with observations at various cosmic times, particularly with the data points for massive galaxies from \citet{terrazas17}. We notice, however, that even with the higher SFRs, these massive galaxies are well below the main sequence (by $>1$~dex), and would be safely considered as being passive. At higher redshifts, both the corrected and uncorrected predictions converge, as numerical artefacts are more prominent at lower redshifts. Dense environments, which challenge structure-finding and tree-building algorithms, are more common in the low-redshift Universe. Notably, the scatter in the SFR-stellar mass plane is reduced after applying the fixes, particularly at stellar masses $\gtrsim 10^{10}\,\rm M_{\odot}$, as shown by the shaded regions representing the standard deviation.

Fig.~\ref{fig:sizes} presents the stellar disc size-stellar mass relation for disc-dominated galaxies ($M_{\star\mathrm{,disc}}/M_{\star}>0.5$) at different cosmic times compared to observational data. Similarly, solid lines represent median values, while the shaded areas indicate the scatter. The bottom panels display the ratio between both median values. The corrected predictions reveal that the most massive discs (above $\approx10^{9}\mathrm{M}_{\odot}$) are now relatively smaller because the fixes eliminate non-physical size jumps caused by mass-swapping events. This improvement holds across all redshifts and results in a better match with the observations from \citet{vanderwel14}. Furthermore, the scatter is significantly reduced across the entire mass range, resulting in a decrease in the stochasticity of the predictions and enhancing the ability to constrain galaxy properties.

To sum up, the variations produced by the artefacts exceed a factor of 3 at the high-mass end — a discrepancy larger than the offsets in galaxy properties caused by different DM merger trees \citep{gomez21} or variations in the implemented physics of galaxy formation models \citep{nifty}. The paper aims to produce robust predictions, thus our primary concern is not how well the observables are reproduced. The key point is that, regardless of the simulation’s resolution, the statistical predictions for observables in the most massive systems (galaxies whose DM-counterparts contain at least $10^{3}$ particles) will inevitably inherit the numerical artefacts discussed in this paper. Therefore, addressing these numerical issues is essential to obtain reliable predictions from SAMs, especially at the high-mass end, irrespective of the specific mass range involved. Achieving this level of reliability is crucial for making robust interpretations from a computational perspective. This is particularly relevant given the growing focus on massive galaxies following recent observations with the James Webb Space Telescope (JWST), e.g. the detection of massive galaxies at $z>7$ \citep{labbe23} and massive-quenched galaxies at $z>3$ \citep{carnall23, glazebrook24}, which seem to be in tension with the $\Lambda$CDM model, as these objects likely formed at $z>10$ \citep{boylan-kolchin23}.

\section{Conclusions}
\label{sec:conclusions}

In this paper, we conduct a detailed analysis of the numerical issues that affect merger tree data, which arise from the difficulties faced by post-processing tools that convert simulation raw data into products that allow comparison with observations. Specifically, we examine several merger tree catalogues generated using different combinations of these post-processing codes that detect bound structures (\textit{``halo finder''}) and link them across the various snapshots of a simulation (\textit{``tree builder''}). We demonstrate that inaccuracies in the merger trees propagate into predictions made by semi-analytic galaxy formation and evolution models, particularly those whose baryonic treatment relies on subhalo information and that lack specific strategies to address these inaccuracies. In our case, we propose possible implementations to mitigate these effects and produce more reliable results. Below we summarise our key findings:

\begin{itemize}
    \item First, we characterise 2 typical numerical artefacts affecting all the catalogues, regardless of the combination of halo finder and merger tree builder codes used: \textit{``mass-swapping''} and \textit{``massive transients''}. Mass-swapping occurs when the hierarchy between a central and a satellite subhalo reverses during certain simulation snapshots, causing large mass fluctuations as material on the outskirts is swapped between them (Fig.~\ref{fig:ms-diagram}). Massive transients refer to subhaloes that emerge late in the simulation with a very high number of particles, as they are not correctly linked to the previous substructures. We give physical definitions of these artefacts in \S~\ref{ssec:mass-swapping}~and~\S~\ref{ssec:massive-transients}, respectively. 
    \item Mass-swapping affects around $10\%$ of main branches in configuration- and phase-space finder codes, as well as tree-builder codes that rely on adjacent snapshot information (Table~\ref{tab:statistics}). The impact is particularly severe in the intermediate-to-high-mass range ($\rm >10^{3}$ particles), where almost all subhaloes are affected (left panel of Fig.~\ref{fig:diagnostics-part}). However, catalogues processed by {\sc HBT-HERONS} perform significantly better with roughly $2\%$ of impacted main branches.
    \item Massive transients are less frequent than the mass-swapping events (affecting $\lesssim0.1\%$ of main branches), though they are still highly prominent among the most massive subhaloes ($\rm >10^{4}-10^{5}$ particles), where their occurrence reaches $\approx 10-20\%$ in configuration-, phase-space or adjacent snapshot approaches (middle panel of Fig.~\ref{fig:diagnostics-part}). Again, the {\sc HBT}-related catalogues show minimal impact from these transients ($\approx10^{-4}\%$ of main branches).
    \item These artefacts arise due to difficulties in assigning particles to structures in dense environments, thus being more likely to emerge for the more massive systems. All the analysed codes struggle with them, independently of their structure-finding (configuration-, phase-space-, or history-based) and tree-building approach (adjacent snapshots, or history-based). However, we conclude that the history-based method used by the algorithm {\sc HBT-HERONS}, relying on temporal information to track which particles belong to which subhalo, is highly effective at mitigating these problems.
    \item The inaccuracies in the merger trees can lead to non-physical effects in the predictions of SAMs, depending on how the modelling of galaxies is addressed. This is exemplified using  {\sc Shark} (subhalo-based) and {\sc Galform} (host halo-based) in \S~\ref{ssec:impact-galaxy-formation-models}, and specifically placed in a broader context in \S~\ref{sssec:impact-literature}. As shown in Figs.~\ref{fig:ms-sam}~and~\ref{fig:mt-sam}, artefacts have significant consequences in subhalo-based models for various galaxy properties (e.g. star formation, BH information, gas sizes), making the predictions less reliable.  In contrast, a host-halo approach appears to mitigate the propagation of these artefacts, as host halo masses are less affected, assuming that satellite subhalo properties are negligible.
    \item We introduce fixes to {\sc Shark} to alleviate the artefacts at the SAM level in \S~\ref{ssec:art-treatment}. Our results show that these fixes successfully address the treatment for the affected structures (Fig.~\ref{fig:sam-fix}), keeping their impact as minimal as possible. In addition, the fixes allow us to make more reliable predictions for the high-mass end, which can be highly affected by the artefacts, as shown in \S~\ref{ssec:new-model-predictions}, and they are consistent across different underlying merger tree catalogues.
    \item {\sc HBT-HERONS} should be the preferred choice for generating accurate merger tree data since it displays a significantly better performance in terms of numerical effects, with fewer and less extreme artefacts. However, even these catalogues still exhibit some artefacts, which need to be addressed when modelling galaxy properties via SAMs. The implementations introduced in this paper provide a solution for this.
\end{itemize}

Despite having quantified the numerical artefacts mentioned above, some uncertainties remain. These artefacts present a persistent challenge for algorithms at both the halo finder and tree builder levels. Attempts to resolve these issues have been carried out — by tweaking parameters in various codes and imposing additional conditions when making tree connections — yet they remain unresolved. Both halo/subhalo-finding and tree-building processes likely play a role in the appearance of these artefacts, but it remains unclear whether one type of post-processing code is predominantly responsible. However, combining the subhalo-finding and the tree-building into a single process avoids most of these issues as is the case for {\sc HBT-HERONS}. They arise always in dense environments, with bound structures containing above $10^3$ particles, as shown in Appendix~\ref{asec:small-box-tests}. Thus, larger simulation boxes and/or higher resolution simulations, which produce a larger number of $>10^3$ particle haloes and at earlier epochs, will be more affected by these artefacts.


In summary, this paper explains that history-based codes seem to resolve the underlying issues at the merger tree level. It provides valuable insight into their frequency across different structure-finding and tree-building approaches, demonstrates how they impact galaxy predictions and offers a solution that can be applied to SAMs before solving for the evolution of galaxies. This awareness is crucial when interpreting observations based on these models. While the study focuses on SAMs, and more particularly {\sc Shark}, similar implementations could be adapted for a range of models of galaxy formation and evolution in the future.

\section*{Acknowledgments}

We thank the anonymous referee for their constructive feedback. ACG thanks Alexander Knebe for productive discussions. ACG acknowledges Research Training Program and ICRAR scholarships. This work was supported by resources provided by The Pawsey Supercomputing Centre with funding from the Australian Government and the Government of Western Australia. VJFM acknowledges support by NWO through the Dark Universe Science Collaboration (OCENW.XL21.XL21.025). CGL and JCH acknowledge support from STFC grants ST/T000244/1 and ST/X001075/1. This work used the DiRAC@Durham facility managed by the Institute for Computational Cosmology on behalf of the STFC DiRAC HPC Facility (\url{www.dirac.ac.uk}). The equipment was funded by BEIS capital funding via STFC capital grants ST/K00042X/1, ST/P002293/1, ST/R002371/1 and ST/S002502/1, Durham University and STFC operations grant ST/R000832/1. DiRAC is part of the National e-Infrastructure. Minor typos, grammar and spelling mistakes were identified with the assistance of ChatGPT-4o\footnote{\url{openai.com}} when preparing this document. No passages of text or structural outlines for this paper were created with the help of any large language models.

\section*{Data Availability}

The codes referenced throughout this paper are publicly available, except for {\sc D-Trees}+{\sc DHalo} and {\sc Galform}.

\begin{itemize}
    \item {\sc Subfind}: the code can be found at \url{http://gitlab.mpcdf.mpg.de/vrs/gadget4}.
    \item {\sc VELOCIraptor}: the code can be found at \url{https://github.com/pelahi/VELOCIraptor-STF}.
    \item {\sc HBT-HERONS}: the code can be found at \url{https://github.com/SWIFTSIM/HBT-HERONS}.
    \item {\sc TreeFrog}: the code can be found at \url{https://github.com/pelahi/TreeFrog/tree/master}.
    \item {\sc Shark}: the code can be found at \url{https://github.com/ICRAR/shark/}.
\end{itemize}




\bibliographystyle{mnras}
\bibliography{tree-issues} 




\appendix

\section{Small simulation box tests}
\label{asec:small-box-tests} 

\begin{figure*}
\centering
\includegraphics[width=1\linewidth]{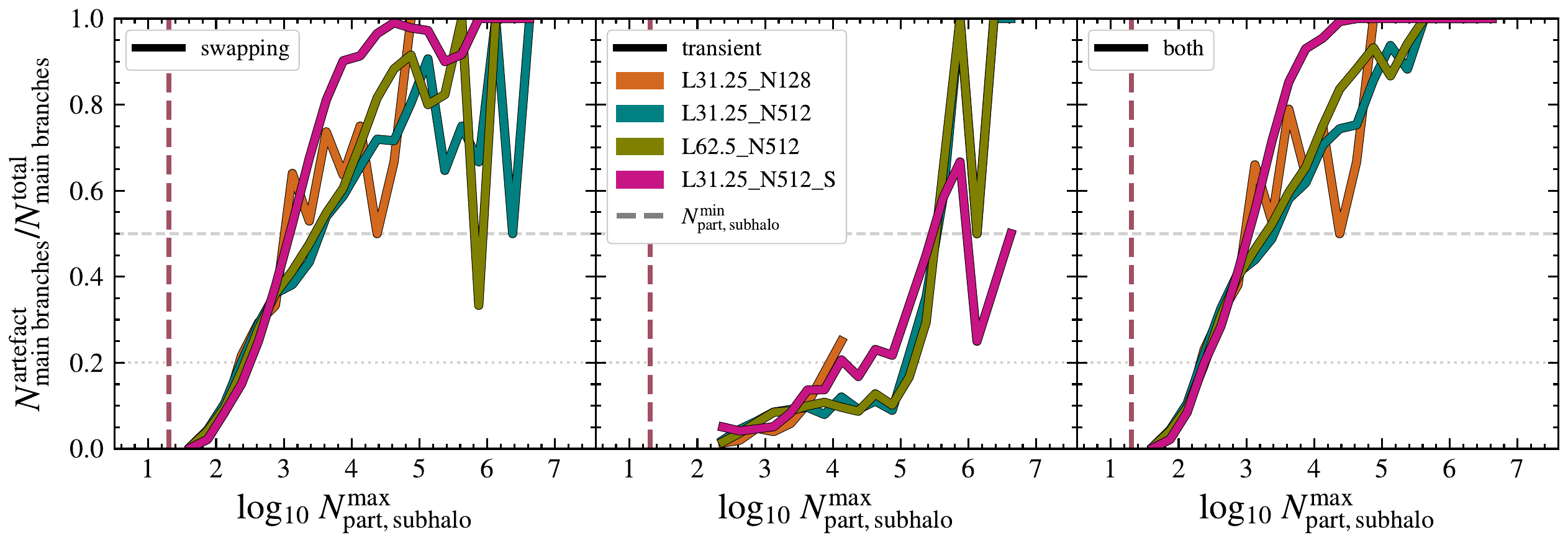}
\caption{\textit{Left panel}: fraction of main branches affected at least by 1 mass-swapping event during their lifetime as a function of the maximum subhalo particle number along the main branch. Each colour represents a different small box merger tree catalogue, as labelled, while the vertical lines mark the minimum particle number for a structure in each catalogue (usually around $20$ particles). \textit{Middle panel}: fraction of main branches defined as massive transients as a function of the maximum subhalo particle number along the main branch. \textit{Right panel}: fraction of main branches affected either by mass-swapping or massive transients as a function of the maximum subhalo particle number along the main branch.} 
\label{fig:diagnostics-part-appendix} 
\end{figure*}

\begin{figure*}
\centering
\includegraphics[width=1\linewidth]{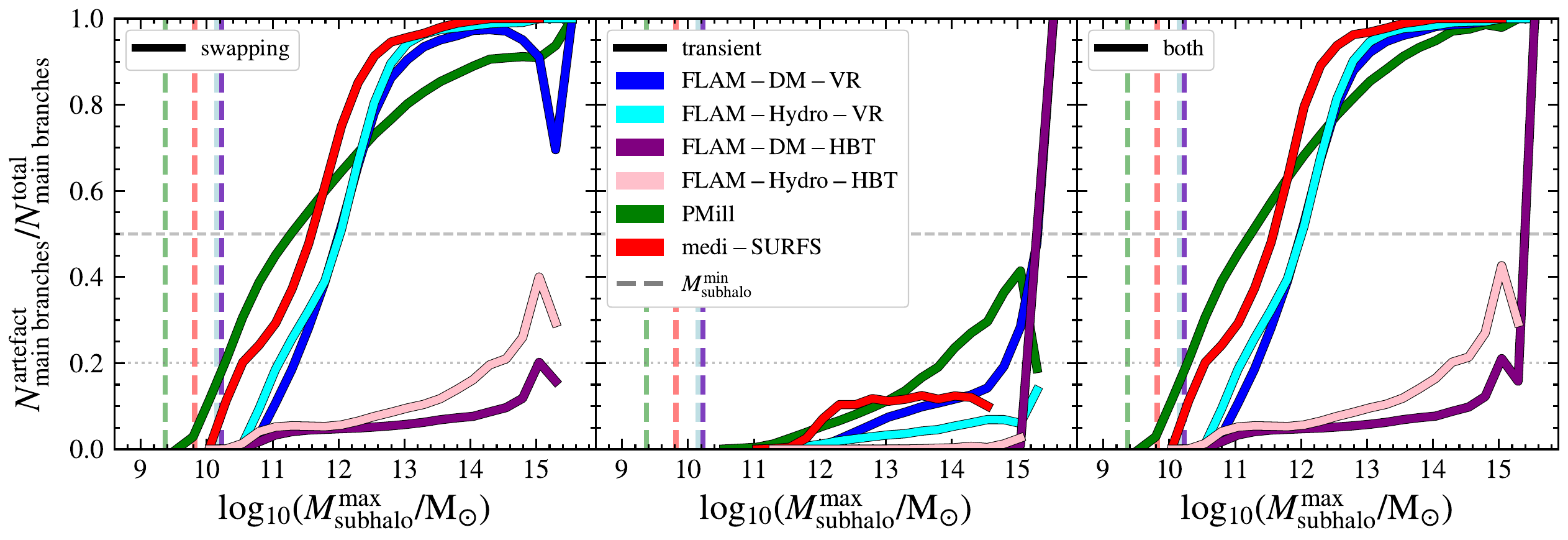}
\caption{Similar to Fig.~\ref{fig:diagnostics-part}, but replacing the x-axis subhalo particle number with subhalo mass. \textit{Left panel}: fraction of main branches affected at least by 1 mass-swapping event during their lifetime as a function of the maximum subhalo mass along the main branch. Each colour represents a different merger tree catalogue, as labelled, while the vertical lines mark the minimum particle number for a structure in each catalogue (usually around $20$ particles). \textit{Middle panel}: fraction of main branches defined as massive transients as a function of the maximum subhalo mass along the main branch. \textit{Right panel}: fraction of main branches affected either by mass-swapping or massive transients as a function of the maximum subhalo mass along the main branch} 
\label{fig:diagnostics} 
\end{figure*}

We run small DM-only cosmological boxes to understand better the numerical artefacts we identified — mass-swapping and massive transients. These simulations share the same cosmology as the {\sc FLAMINGO} catalogues previously analysed, and they use the same combination of post-processing tools: {\sc VELOCIraptor} as the halo finder and {\sc D-Trees}+{\sc DHalo} as the tree builder. As a result, the merger tree catalogues generated from these simulations are equivalent to the {\sc FLAM-DM-VR} catalogue, which has been the default catalogue analysed throughout this paper since it shows being extremely affected by the numerical effects. Table~\ref{atab:small-box} details these small simulation boxes.  

These small boxes allow us to investigate various aspects of mass-swapping and the massive transients. First, we examine how the mass resolution and the number of snapshots affect these numerical artefacts, as discussed in \S~\ref{assec:small-box-mass}. Additionally, we use them to explore the environmental and merger history features that characterise the subhaloes affected by these artefacts in \S~\ref{assec:small-box-analysis}.

\subsection{Mass and snapshot number dependency}
\label{assec:small-box-mass} 


As in \S~\ref{ssec:mass-swapping} and \S~\ref{ssec:massive-transients}, we apply the definitions of mass-swapping and massive transients to the catalogues generated from the small cosmological runs.
Next, we quantify the fraction of main branches affected by each artefact (left and middle panel) and by either of them (right panel) in Fig.~\ref{fig:diagnostics-part-appendix}. We show in these diagnostics plots that the trends for each curve closely match those observed in the {\sc FLAM-DM-VR} catalogue (blue curve) in Fig.~\ref{fig:diagnostics-part} since the combination of post-processing tools are the same ({\sc VELOCIraptor} and {\sc D-Trees}+{\sc DHalo}) and they are all DM-only runs. The frequency of subhaloes with a high maximum number of particles (above $10^{3}$ particles) varies across the boxes, reflected in the different noise levels in the curves. However, the key takeaway is that, regardless of mass resolution, the shapes of the curves in each panel resemble each other. In other words, mass-swapping primarily impacts main branches containing more than $10^{3}$ particles, while massive transients affect those with around $10^{4}$ particles most significantly. When analysed by snapshot number, mass-swapping and massive transients tend to have a slightly stronger impact on massive branches, though the fraction of transients decreases at the very high-mass end due to low statistics.

\begin{table}
	\centering
	\caption{Small DM-only cosmological simulations. $L$: the periodic box size in comoving Mpc; $N_{\rm part}$: the number of DM particles in the volume; $m_{\rm part}$: the particle mass in M$_{\odot}$; $N_{\rm out}$: the number of snapshots; and $N^{\rm min}_{\rm p,subhalo}$: the minimum particle number for a subhalo to be detected in the simulation. The merger tree catalogues obtained from these simulations are processed with {\sc VELOCIraptor}+{\sc D-Trees}+{\sc DHalo.}}
	\label{atab:small-box}
	\begin{tabular}{cccccccc} 
		\hline
		simulation & $L$/Mpc & $N_{\rm part}$ & $m_{\rm part}$/M$_{\odot}$ & $N_{\rm out}$ & $N^{\rm min}_{\rm p,subhalo}$\\
		\hline
        {\sc L31.25\_N128} & 31.25 & 128$^3$ & 5.75$\times$10$^8$ & 77 & 20\\
        {\sc L31.25\_N512} & 31.25 & 512$^3$ & 8.98$\times$10$^6$ & 77 & 20\\
        {\sc L62.5\_N512} & 62.5 & 512$^3$ & 7.18$\times$10$^7$ & 77 & 20\\
        {\sc L31.25\_N512\_S} & 31.25 & 512$^3$ & 8.98$\times$10$^6$ & 201 & 20\\
		\hline
	\end{tabular}
\end{table}

\begin{figure*}
\centering 
\captionsetup[subfigure]{labelfont=bf,font=large}
\begin{subfigure}{0.355\textwidth}
  \includegraphics[trim={0.1cm 0.1cm 0.1cm 0.1cm},clip,width=\linewidth]{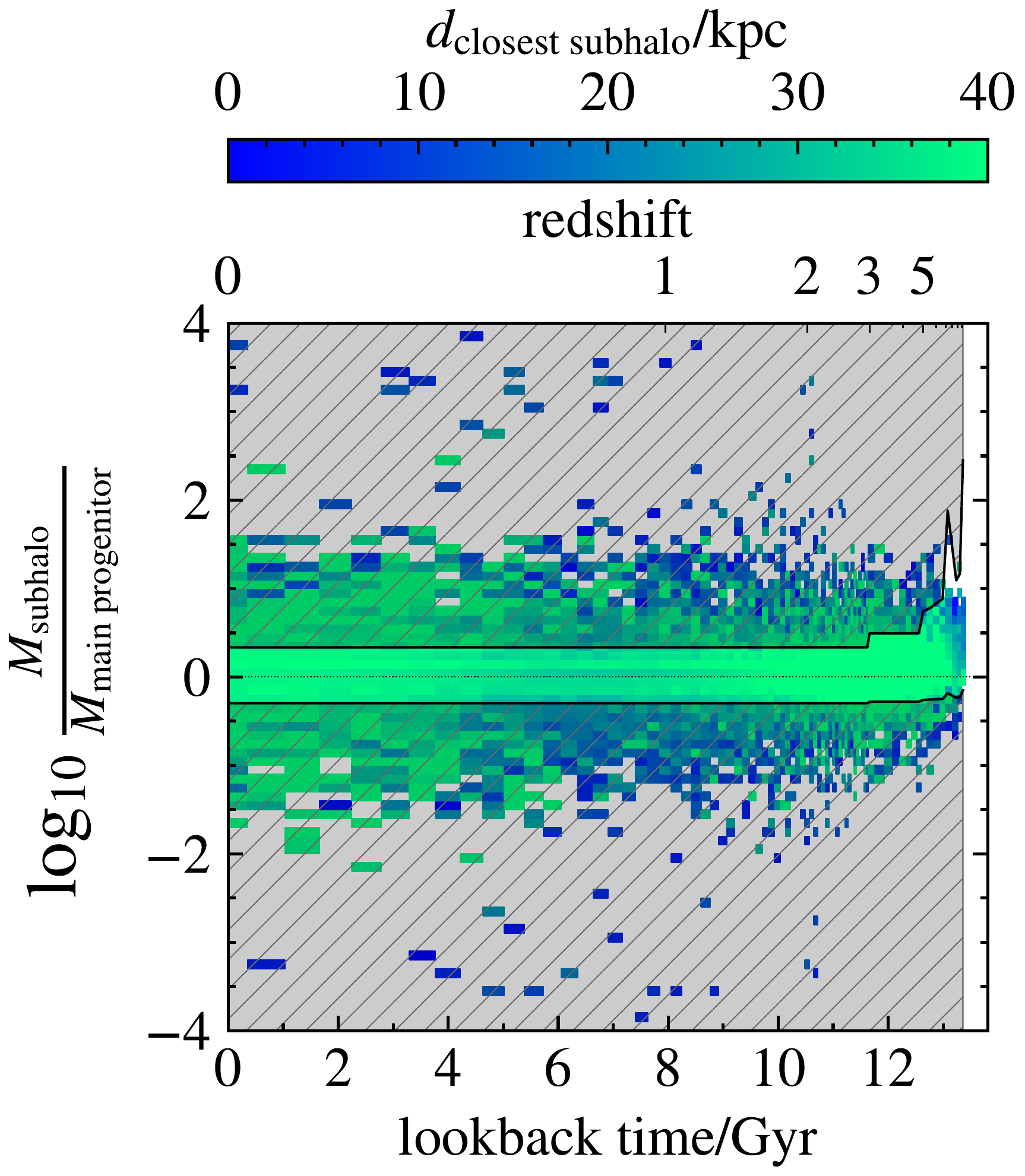}
  \caption{\textbf{Distance to closest subhalo (environment)}}
  \label{fig:msfeat1}
\end{subfigure}\hfil 
\begin{subfigure}{0.3025\textwidth}
  \includegraphics[trim={0.1cm 0.1cm 0.1cm 0.1cm},clip,width=\linewidth]{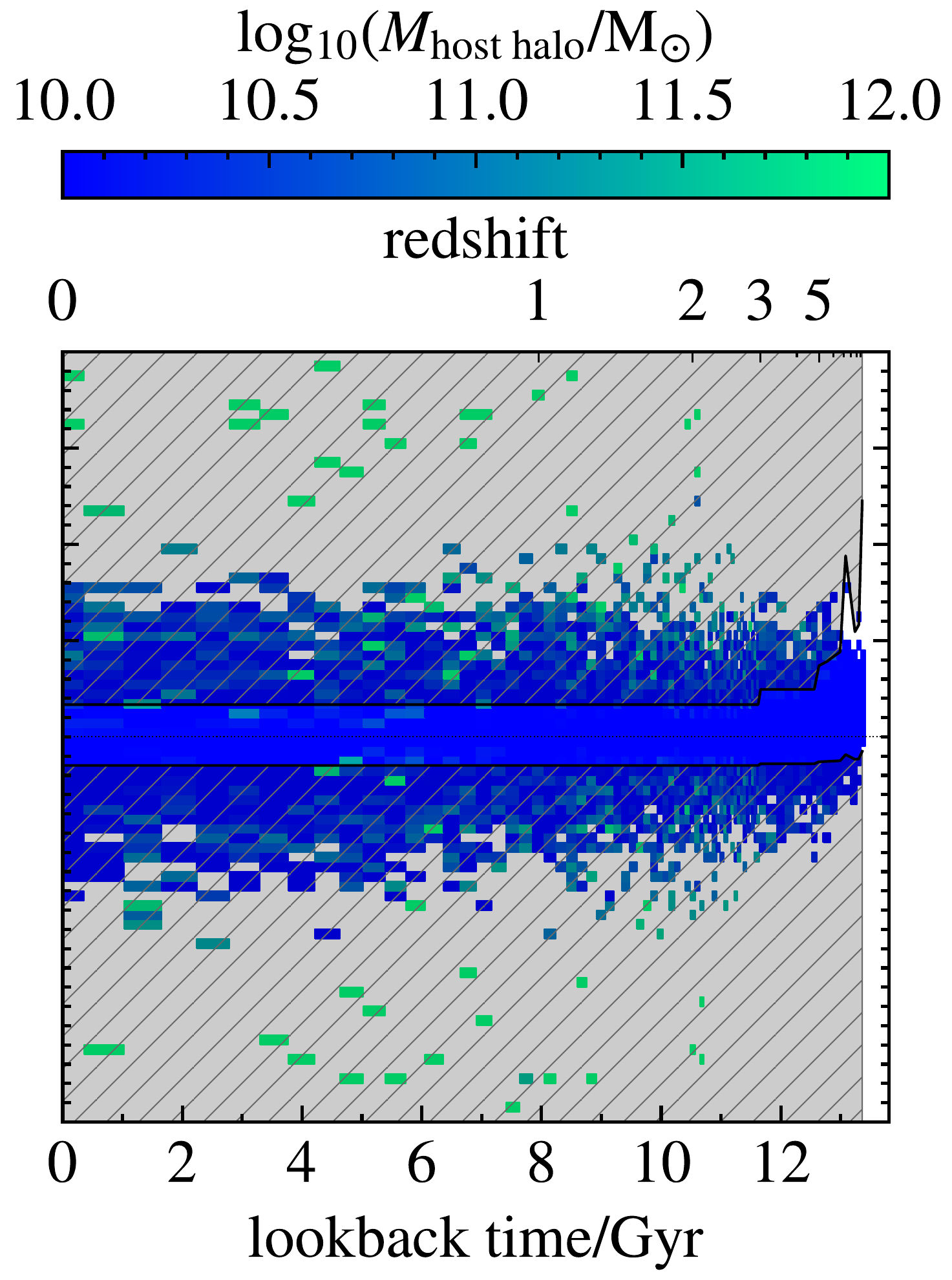}
  \caption{\textbf{Mass of the host halo (environment)}}
  \label{fig:msfeat2}
\end{subfigure}\hfil 
\begin{subfigure}{0.278\textwidth}
  \includegraphics[trim={0.1cm 0.1cm 0.1cm 0.1cm},clip,width=\linewidth]{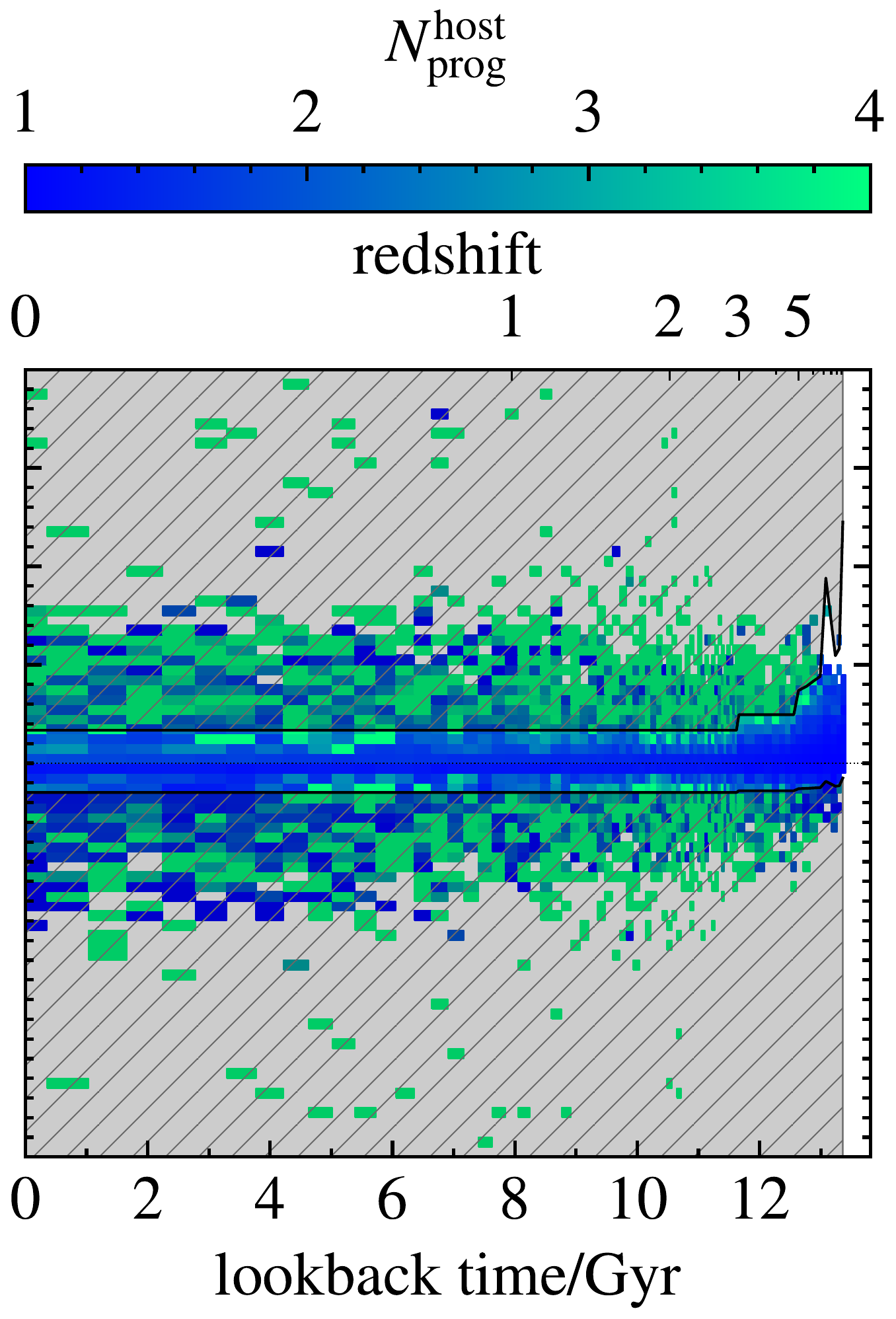}
  \caption{\textbf{Progenitor number for host halo (merger history)}}
  \label{fig:msfeat3}
\end{subfigure}
\\[0.3cm]
\caption{2D maps of the relative change in mass when there is a connection between a main progenitor and a descendant subhalo for a main branch as a function of the lookback time/redshift at which that event is happening, for the {\sc L31.25\_N512} merger tree catalogue described in Table~\ref{atab:small-box}. The panels are colour-coded by the average values within each 2D bin of different environmental and merger history tracers, following the colour bars at the top. The shaded area marks the numerical artefacts based on the thresholds defined in equation~(\ref{eq:3}) (solid line). A dotted line represents cases where there is no mass change ($M_{\mathrm{subhalo}}=M_{\mathrm{main}\ \mathrm{progenitor}}$).}
\label{fig:ms-feat-panel}
\end{figure*}

\begin{figure*}
\captionsetup[subfigure]{labelfont=bf,font=large}
\vspace{0.5cm}
    \centering 
\begin{subfigure}{0.349\textwidth}
  \includegraphics[trim={0.1cm 0.1cm 0.1cm 0.1cm},clip,width=\linewidth]{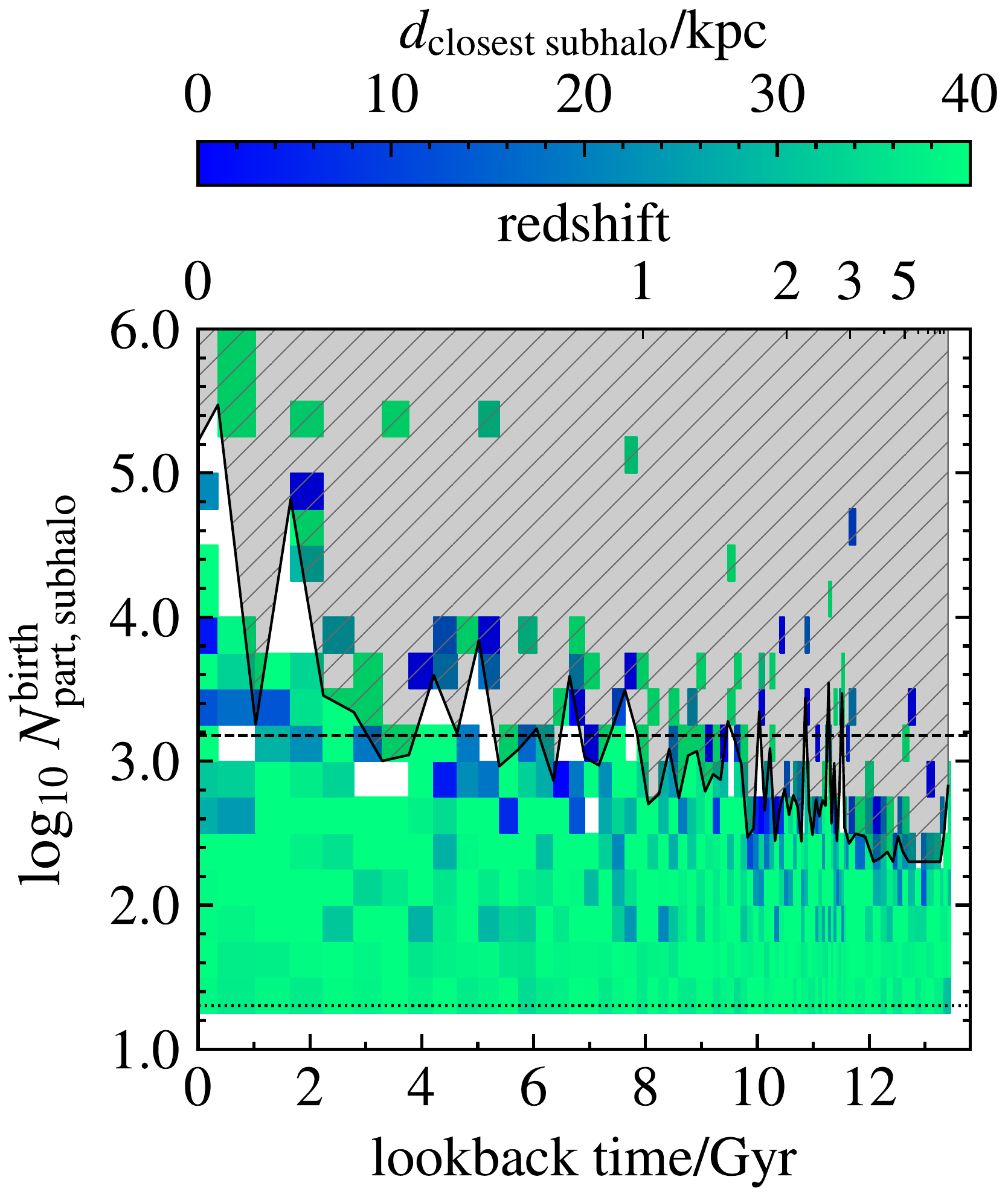}
  \caption{\textbf{Distance to closest subhalo (environment)}}
  \label{fig:mtfeat1}
\end{subfigure}\hfil 
\begin{subfigure}{0.318\textwidth}
  \includegraphics[trim={0.1cm 0.1cm 0.1cm 0.1cm},clip,width=\linewidth]{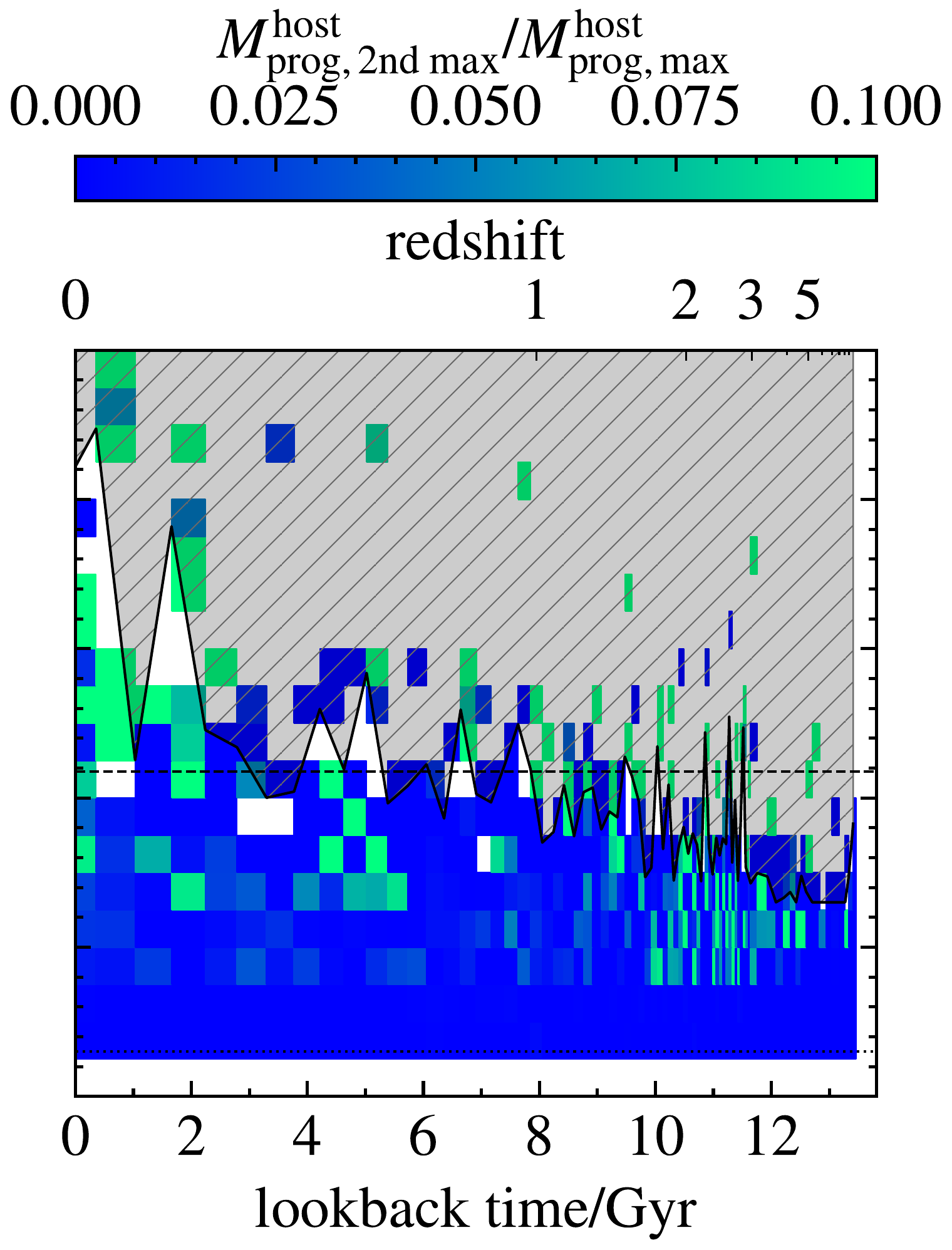}
  \caption{\textbf{Mass ratio massive prog. for the host (merger history)}}
  \label{fig:mtfeat2}
\end{subfigure}\hfil 
\begin{subfigure}{0.317\textwidth}
  \includegraphics[trim={0.1cm 0.1cm 0.1cm 0.1cm},clip,width=\linewidth]{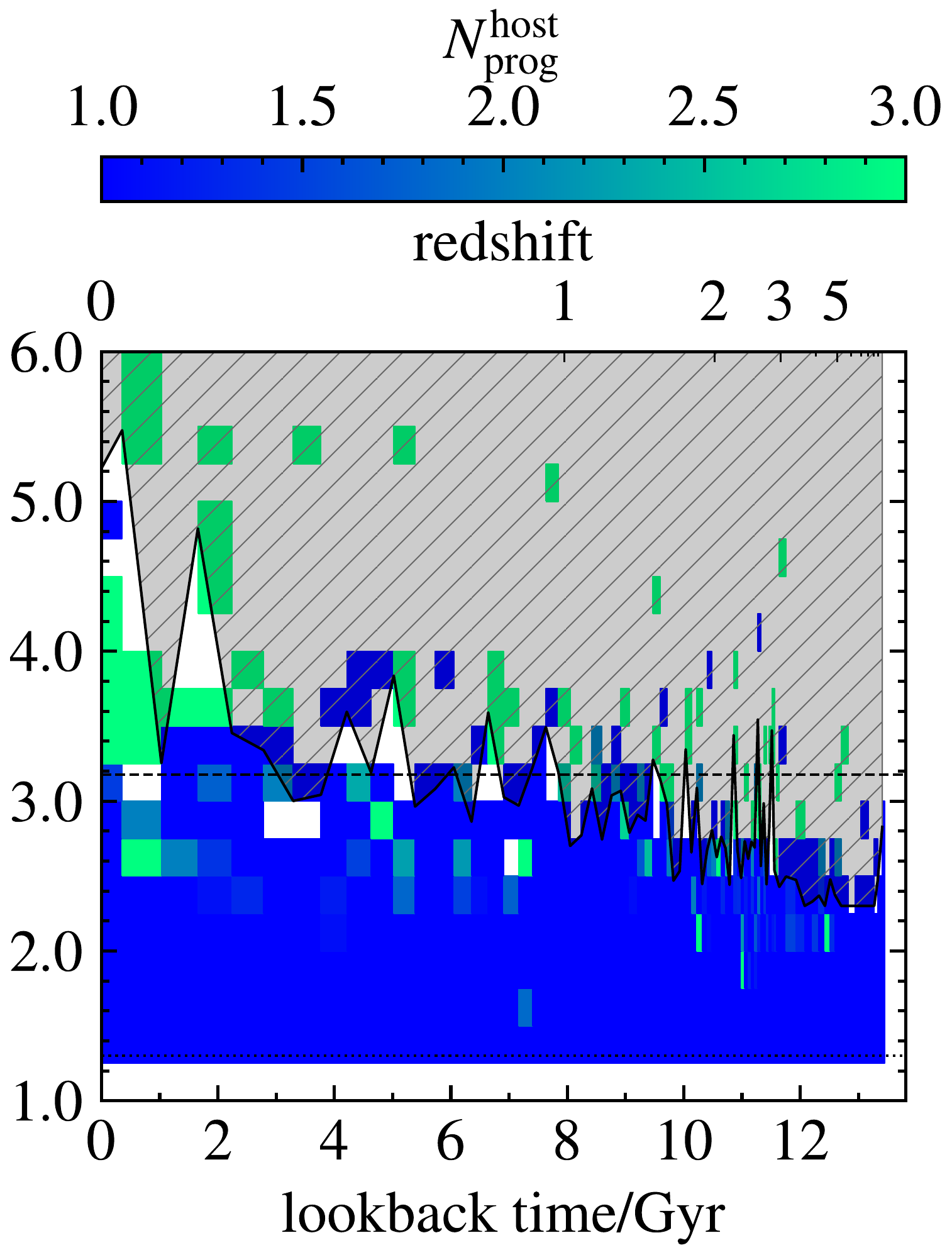}
  \caption{\textbf{Progenitor number for host halo (merger history)}}
  \label{fig:mtfeat3}
\end{subfigure}
\\[0.3cm]
\caption{2D maps of the particle number with which main branches are born in the simulation as a function of the lookback time/redshift at which the birth is happening, for the {\sc L31.25\_N512} merger tree catalogue described in Table~\ref{atab:small-box}. The panels are colour-coded by the average values within each 2D bin of different environmental and merger history tracers, following the colour bars at the top. The shaded area marks the numerical artefacts based on the threshold $3\sigma(N^{\mathrm{birth}}_{\mathrm{part,subhalo}},z)$ defined in equation~(\ref{eq:4}) (solid line). A dotted line represents the minimum particle number for a subhalo to be detected ($N^{\mathrm{min}}_{\mathrm{part,subhalo}}$), whereas a dotted line shows $3\sigma(N^{\mathrm{birth}}_{\mathrm{part,subhalo}})$ globally.}
\label{fig:mt-feat-panel}
\end{figure*}

In the right panel, the fraction of main branches affected by both numerical effects increases slightly when the number of snapshots is higher. This is because a greater number of snapshots raises the likelihood of these artefacts occurring, which can degrade the quality of the merger trees \citep{wang16}. Nevertheless, the overall impact on the fraction of main branches affected by numerical effects remains similar, regardless of box size, particle number or snapshot number. The issues consistently emerge when halo/subhalo-finding and tree-building codes encounter sufficiently dense environments with at least $10^{3}$ particles.

The features of the simulation box are irrelevant: the issues always arise when systems with $10^{3}$ particles are formed. This demonstrates that our analysis, despite using several merger tree catalogues based on cosmological simulations with different features in Table~\ref{tab:sim}, is not biased by the simulation box characteristics. As a complementary material, Fig.~\ref{fig:diagnostics} presents the same diagnostic results as Fig.~\ref{fig:diagnostics-part}, but now as a function of the maximum mass for the subhalo along the main branch in the x-axis. This allows us to examine the structures that inherit numerical artefacts across all the catalogues in Table~\ref{tab:cat}.

\subsection{Environment and merger history analysis}
\label{assec:small-box-analysis}

Using the merger tree catalogue from the {\sc L31.25\_N512} simulation, we demonstrate that these issues arise from very dense environments and their connection to the merger history. We analyse the 2D maps used to define the numerical artefacts (both mass-swapping with the mass ratio between snapshots in the y-axis, as well as massive transients using the particle number for an emerged main branch) as in Figs.~\ref{fig:ms-def-panel}~and~\ref{fig:mt-def-panel}. In this case, the maps are colour-coded by the average values within each 2D bin, incorporating different tracers of the environment and past merger history for the subhalo. In such a way, we can detect any particular trends arising in the shaded areas that define numerically incorrect structures. We look into several environmental properties (e.g., the distance to the closest subhalo, the host halo mass, the number/mass/maximum mass of subhaloes in a particular aperture surrounding the analysed structure) and past merger-related quantities (e.g., the number of progenitors, the mass ratio between the 2 most massive progenitors, the number of major mergers). However, we only present the properties that yield the most interesting results in the figures below.

In Fig.~\ref{fig:ms-feat-panel}, the top row shows that subhaloes affected by mass-swapping events tend to be closer to other subhaloes (left panel), reside in more massive host haloes (middle panel) and have a higher number of progenitors (right panel) as indicated by the bluer or greener colours in the artefact shaded areas. This confirms the hypothesis that these events occur between subhaloes that live in really dense environments, where they are in close proximity to each other, involve a large number of particles, and have more complex merger histories, making them more susceptible to these numerical artefacts.

Similarly, Fig.~\ref{fig:mt-feat-panel} presents the same analysis for the massive transients. Once again, bluer or greener colours dominate the artefact-shaded areas, indicating that subhaloes that are closer to one another (left panel), have a higher number of progenitors (for the host halo since there is no direct progenitor for these transients) (middle panel) and exhibit larger mass ratios for those progenitors (right panel) are more prone to this second numerical artefact. To summarise, we confirm that the numerical artefacts analysed in the main text primarily arise in massive systems undergoing major mergers, where particle assignment becomes more challenging, increasing the susceptibility of these structures to numerical inaccuracies. This is evidenced by the environmental and merger history features studied.

\begin{figure}
\centering
\includegraphics[width=0.98\linewidth]{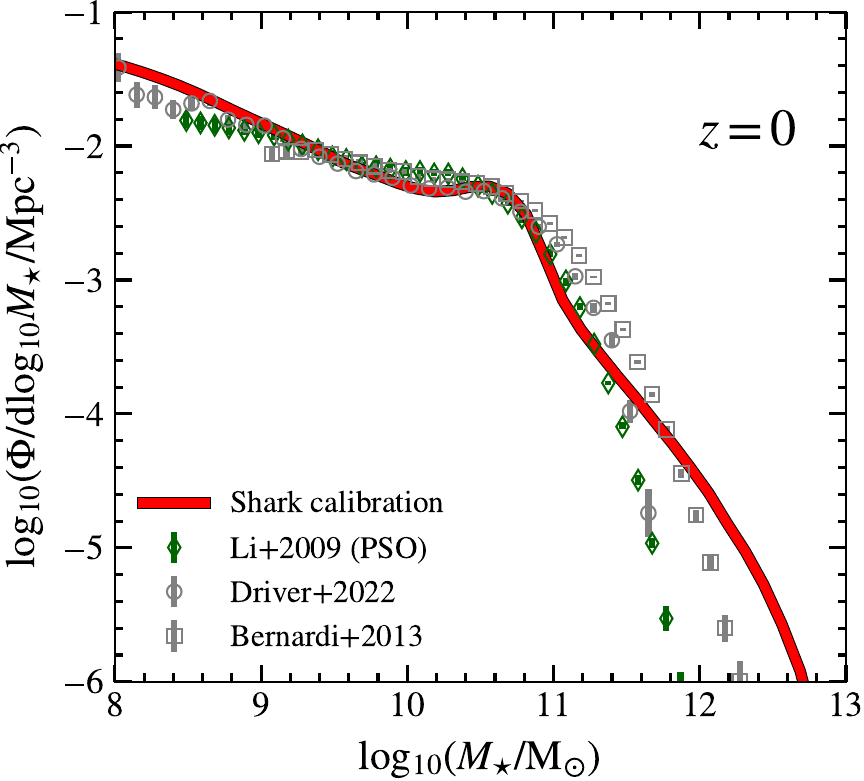}
\caption{SMF at $z=0$ predicted by the new {\sc Shark} calibration when running the code on top of the {\sc FLAM-DM-HBT} merger tree catalogue. Observational data from \citet{li09} (input for the automatic parameter exploration), and \citet{bernardi13,driver22} for comparison.} 
\label{fig:shark-cal-smf0} 
\end{figure}

\begin{table}
	\centering
	\cprotect\caption{Parameters, as shown in the parameter file, tuned to calibrate the {\sc Shark} code to reproduce the $z=0$ SMF for the {\sc FLAM-DM-HBT} merger tree catalogue. The rest of the parameters keep the same values as detailed in table~2 in \citet{sharkv2}. $\kappa$ balances the BH accretion rate due to the hot-halo mode (equation~11 in \citet{sharkv2}); while $\Gamma_{\rm thresh}$ sets the threshold for the cooling-to-heating specific energy ratio, above which the AGN jet-mode feedback is activated \citep[section~3.3.2 in][]{sharkv2}.}
	\label{tab:shark-cal}
	\begin{tabular}{ccc} 
		\hline
		{\sc Shark} parameter & {\sc Shark}.v2 value & recalibration value \\
		\hline
        \multicolumn{1}{l}{\textit{AGN feedback and BH growth:}} \\
        $\kappa$ (equation~11) & 10.31 & 0.18957 \\
        $\Gamma_{\rm thresh}$ (section~3.3.2) & 10 & 6.03 \\
		\hline
	\end{tabular}
\end{table}

\section{New {\sc Shark} calibration}
\label{asec:shark-calibration}

\begin{figure*}
\centering
\begin{subfigure}{0.4\textwidth}
   \includegraphics[width=\textwidth]{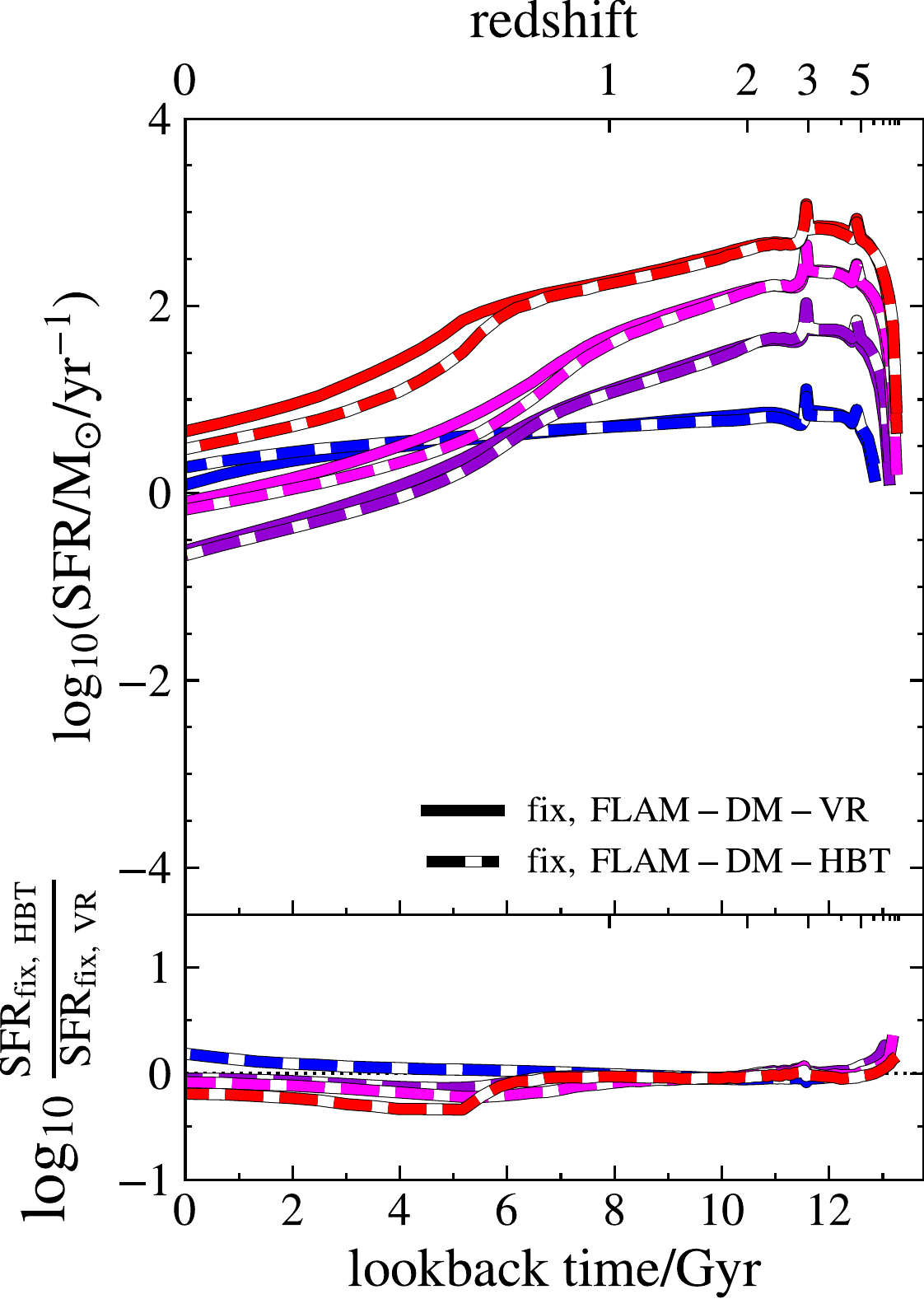}
\end{subfigure} \hfil
\begin{subfigure}{0.4\textwidth}
   \includegraphics[width=\textwidth]{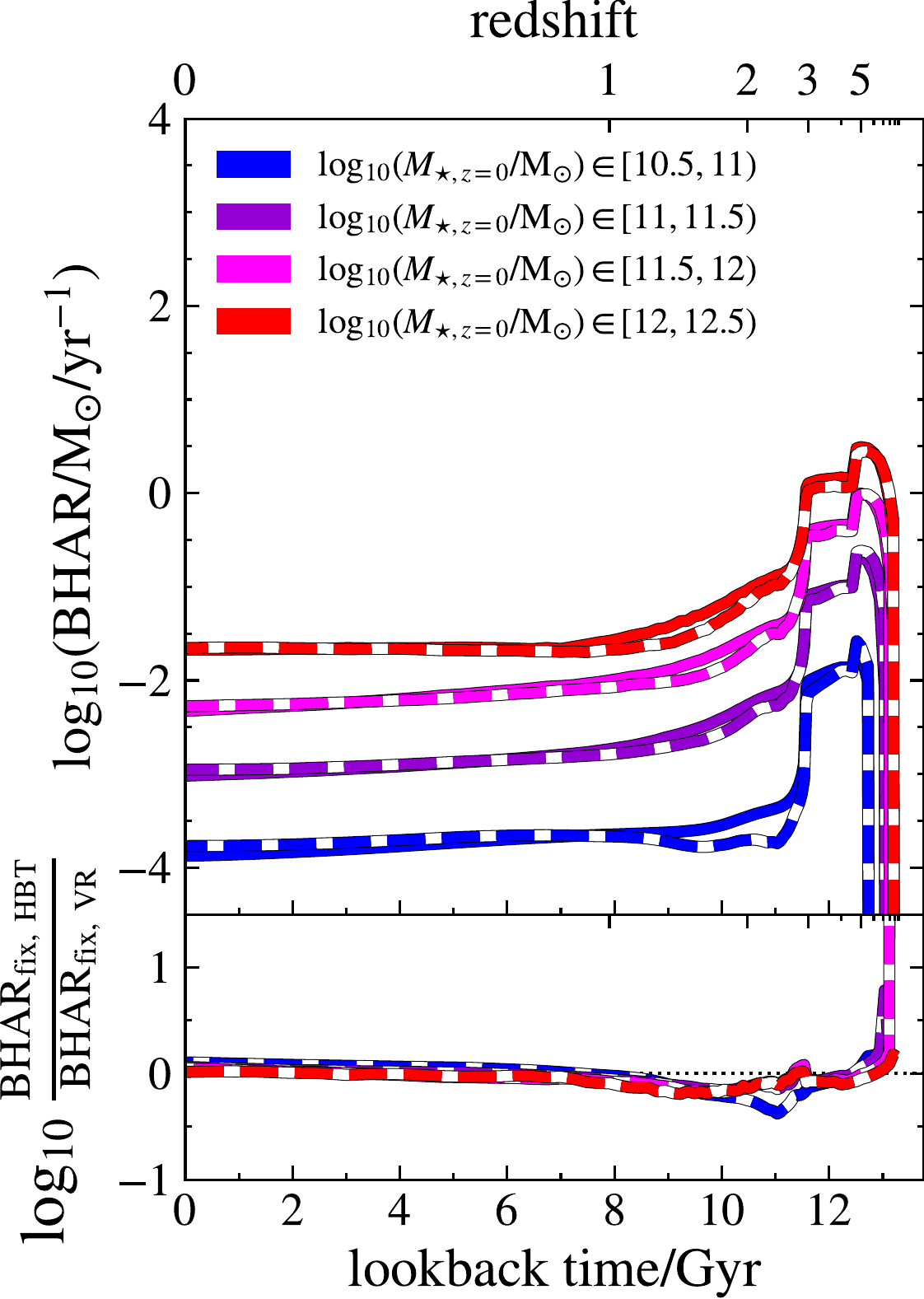}
\end{subfigure}
\\[0.3cm]
\caption{Similar to Fig.~\ref{fig:shark-histories}, but comparing the {\sc Shark} predictions for the {\sc FLAM-DM-VR} catalogue to the ones for the more accurate {\sc FLAM-DM-HBT} catalogue. \textit{Left panel}: median SFHs for several stellar mass bins at $z=0$ (in different colours) considering all the galaxies modelled by {\sc Shark} over the previously mentioned merger tree catalogues, indicating the predictions for {\sc FLAM-DM-VR} (solid lines) and {\sc FLAM-DM-HBT} (solid bicoloured lines). The ratio between both runs is shown in the bottom panel. \textit{Right panel}: same but for the median BHAHs.} 
\label{fig:shark-histories-hbtvr} 
\end{figure*}

We describe the parameters in {\sc Shark} that are adjusted from the default values in version 2.0, as presented in table~2 in \citep{sharkv2}, to better fit the $z=0$ SMF using the {\sc FLAM-DM-HBT} merger tree catalogue (detailed in Table~\ref{tab:cat}, with {\sc HBT-HERONS} as the halo finder and tree builder codes). To run {\sc Shark} the catalogue is processed by {\sc DHalo}. The SMF at $z=0$ is shown in Fig.~\ref{fig:shark-cal-smf0}, against the observational dataset used for calibration \citep{li09} and additional observations \citep{bernardi13,driver22}. This calibration is performed once the implementations introduced in this paper to address the numerical artefacts in merger tree data are activated. It helps assess the impact of numerical artefacts on SAM predictions (see \S\ref{ssec:impact-galaxy-formation-models}), evaluate the improvements made by applying the proposed fixes (\S~\ref{ssec:art-treatment}) and analyse how these changes affect the statistical predictions (\S~\ref{ssec:new-model-predictions}). Table~\ref{tab:shark-cal} lists the two parameters related to AGN activity that were tuned via an automatic parameter exploration algorithm, which results in a less efficient hot-halo accretion by the BH and relaxes the threshold for its activation.

\section{{\sc Shark} predictions for the {\sc FLAM-DM-HBT} catalogue}
\label{asec:shark-hbt-predictions} 

We present the predictions for the {\sc FLAM-DM-HBT} catalogue after processing the data through {\sc Shark}. As explained previously, the catalogue is first processed by {\sc DHalo} to ensure the correct format, similar to the {\sc medi-SURFS} catalogue. This step modifies the subhalo hierarchy and groups subhaloes into different host halo structures, as discussed in \S~\ref{sssec:dhalos}. We verified that the {\sc DHalo} corrections introduce only minimal variations in the diagnostics plots in Fig.~\ref{fig:diagnostics-part}.

We show the median values and the standard deviations for the SFHs and BHAHs in different stellar mass bins in Fig.~\ref{fig:shark-histories-hbtvr} comparing the results after applying the fixes for both the {\sc FLAM-DM-VR} catalogue (used throughout the main text of the paper and heavily affected by artefacts, shown with solid bicoloured lines) and the {\sc FLAM-DM-HBT} catalogue (minimally affected by artefacts, shown with solid lines). The predictions are highly consistent across the different stellar mass bins at $z=0$ for both properties. Similar results are observed for the SFR-stellar mass plane and the galaxy stellar disc size-stellar mass relation. The main conclusion is that, regardless of the merger tree catalogue employed (which may be more or less affected by artefacts), once processed through the new implementations introduced in this paper, the predictions remain consistent, with a maximum variation of about a factor of 2, typically lower. This aligns with the findings of \citet{gomez21}, which showed that the underlying halo/subhalo-finding and tree-building codes do not impact predictions in {\sc Galform}. We demonstrate that the same holds for {\sc Shark}, despite its different approach to modelling galaxy properties. Furthermore, the absolute magnitude of the differences in this paper appears less significant than the ones shown in fig.~15 in \citet{gomez21}, but we focus here on median values.

Although some catalogues are more accurate, and the fixes applied have less impact when the data contain fewer artefacts, they still cause slight variations in the predictions. This underscores that fixes must be implemented to ensure reliable predictions from galaxy formation and evolution models, even when dealing with more accurate data generated by history-space algorithms, such as {\sc HBT-HERONS}. This highlights the importance of the work presented in this paper.

\section{Pseudo-code for numerical artefact fixes}
\label{asec:pseudo-code} 

This appendix presents the algorithms we implement to correct the numerical artefacts (mass-swapping and massive transients) in the merger tree data, using pseudo-code. The approach is designed to be adaptable to any galaxy formation model, particularly subhalo-based SAMs that lack strategies to mitigate these artefacts.



\subsection{Mass-swapping}
\label{asec:pseudo-code-ms} 

\begin{algorithmic}
\State \textit{read merger tree catalogue}
\While{\textit{snapshot from max to min}} \Comment{Central subhaloes}
    \For{\textit{host haloes in snapshot}}
        \If{\textit{central subhalo not defined}}
            \State \textit{most massive subhalo is central}  \Comment{Central flag}
        \EndIf
        \State \textit{central subhalo props using host halo} \Comment{Defined props}
    \EndFor
    \While{\textit{ascendant not defined}}
        \State \textit{1. find the main progenitor for the subhalo and its host halo}
        \If{\textit{main progenitor not defined}}
            \State \textit{most massive progenitor is the main progenitor}
        \EndIf
        \State \textit{2. central subhalo is the main prog} \Comment{Central flag}
        \State \textit{3. central subhalo props using host halo} \Comment{Defined props}
    \EndWhile
\EndWhile
\While{\textit{snapshot from max to min}} \Comment{Satellite subhaloes}
    \For{\textit{satellite subhaloes in snapshot}}
        \State \textit{1. find infall snapshot (when central hierarchy)}
        \State \textit{2. satellite subhalo properties at infall} \Comment{Defined props}
    \EndFor
\EndWhile
\State \textit{galaxy properties defined using subhalo properties}
\State \textit{when satellite subhalo properties used, consider the infall properties for all cases except for environmental processes}
\end{algorithmic}

\subsection{Massive transients}
\label{asec:pseudo-code-mt} 

\begin{algorithmic}
\While{\textit{snapshot from max-1 to min}}
    \color{teal}
    \For{\textit{subhaloes in snapshot}} \Comment{Compute $3\sigma(N,z)$}
        \State \textit{find main branch subhalo at birth \& its particle number}
    \EndFor
    \State \textit{compute the threshold for transients}
    \color{black}
    \For{\textit{subhaloes in snapshot}} \Comment{Transient flags}
        \If{\textit{particle number > threshold}}
            \State \textit{flag transient subhalo and host halo}
        \EndIf
    \EndFor
    \For{\textit{subhaloes in snapshot}} \Comment{Fix transients}
        \If{\textit{subhalo descendant's host halo is a transient}}
            \For{\textit{subhaloes in subhalo descendant's host halo}}
                \If{\textit{transient subhalo}}
                    \If{\textit{conditions equations~(\ref{eq:5})--(\ref{eq:7})}}
                        \State \textit{link subhalo and transient}
                    \Else
                        \State \textit{remove transient main progenitor flag}
                    \EndIf
                \EndIf
            \EndFor
            \If{\textit{subhalo was not linked}}
                \State \textit{apply the default link}
            \EndIf
        \EndIf
    \EndFor
\EndWhile
\end{algorithmic}



\bsp	
\label{lastpage}
\end{document}